\documentclass{amsart}
\usepackage{amsmath}
\usepackage{amssymb}
\usepackage{amscd}

\input tcilatex
\theoremstyle{definition}
\theoremstyle{remark}
\numberwithin{equation}{section}

\begin{document}
\title[Quantum Stochastic CP Evolutions]{Quantum Stochastic Positive
Evolutions:\\
Characterization, Construction, Dilation}
\author{V.\thinspace P.\thinspace Belavkin}
\address{Mathematics Department, University of Nottingham\\
NG7 2RD, UK.}
\date{November 10, 1995}
\thanks{Published in: \textit{Commun. Math. Phys.} \textbf{184} (1997)
533-566.}
\maketitle

\begin{abstract}
A characterization of the unbounded stochastic generators of quantum
completely positive flows is given. This suggests the general form of
quantum stochastic adapted evolutions with respect to the Wiener
(diffusion), Poisson (jumps), or general Quantum Noise. The corresponding
irreversible Heisenberg evolution in terms of stochastic completely positive
(CP) maps is constructed. The general form and the dilation of the
stochastic completely dissipative (CD) equation over the algebra $\mathcal{L}%
\left( \mathcal{H}\right) $ is discovered, as well as the unitary quantum
stochastic dilation of the subfiltering and contractive flows with unbounded
generators. A unitary quantum stochastic cocycle, dilating the subfiltering
CP flows over $\mathcal{L}\left( \mathcal{H}\right) $, is reconstructed.
\end{abstract}

\section*{Introduction}

In quantum theory of open systems there is a well known Lindblad's form \cite%
{19} of the quantum Markovian master equation, satisfied by the
one-parameter semigroup of completely positive (CP) maps. This
nonstochastical equation is obtained by averaging the stochastic Langevin
equation for quantum diffusion \cite{EvH} over the driving quantum noises.
On the other hand the quantum Langevin equation is satisfied by a quantum
stochastic process of dynamical representations, which are obviously
completely positive due to *-multiplicativity of the homomorphisms,
describing these representations. The homomorphisms give the examples of
pure, i.e. extreme point CP maps, but among the extreme points of the convex
cone of all CP maps there are not only the homomorphisms. This means a
possibility to construct the dynamical semigroups by averaging of pure, i.e.
non-mixing irreversible quantum stochastic CP dynamics, which is not driven
by a Langevin equation.

The examples of such dynamics having recently been found in many physical
applications, will be considered in the first section. The rest of the paper
will be devoted to the mathematical derivation of the general structure for
the quantum stochastic CP evolutions and the corresponding equations. The
results of the paper not only generalize the Evans-Hudson (EH) flows \cite%
{EvH} from homomorphism-valued maps to the general CP maps, but also prove
the existence of the homomorphic dilations for the subfiltering and
contractive CP flows. Here in the introduction we would like to outline this
structure on the formal level.

The initial purpose of this paper was to extend the Evans--Lewis
differential analog \cite{17} of the Stinespring dilation \cite{Stn} for the
CP semigroups to the stochastic differentials, generating an It\^{o} $*$%
--algebra 
\begin{equation}
\mathrm{d}\Lambda \left( a\right) ^{*}\mathrm{d}\Lambda \left( a\right) =%
\mathrm{d}\Lambda \left( a^{\star }a\right) ,\quad \sum \lambda _i\mathrm{d}%
\Lambda \left( a_i\right) =\mathrm{d}\Lambda \left( \sum \lambda
_ia_i\right) ,\quad \mathrm{d}\Lambda \left( a\right) ^{*}=\mathrm{d}\Lambda
\left( a^{\star }\right)  \label{0.1}
\end{equation}
with given mean values $\langle \mathrm{d}\Lambda \left( t,a\right) \rangle
=l\left( a\right) \mathrm{d}t$, $a\in \mathfrak{a}$. Here $\mathfrak{a}$ is
in general a noncommutative $\star $-algebra with a self-adjoint annihilator
(death) $d=d^{\star }$, $\mathfrak{a}d=0$, corresponding to $\mathrm{d}t=%
\mathrm{d}\Lambda \left( t,d\right) $, and $l:\mathfrak{a}\rightarrow 
\mathbb{C}$ is a positive $l\left( a^{\star }a\right) \geq 0$ linear
functional, normalized as $l\left( d\right) =1$, corresponding to the
determinism $\left\langle \mathrm{d}t\right\rangle =\mathrm{d}t$. The
functional $l$ defines the GNS representation $a\mapsto \boldsymbol{a}%
=\left( a_\nu ^\mu \right) _{\nu =+,\bullet }^{\mu =-,\bullet }$ of $%
\mathfrak{a}$ in terms of the quadruples 
\begin{equation}
a_{\bullet }^{\bullet }=j\left( a\right) ,\quad a_{+}^{\bullet }=k\left(
a\right) ,\quad a_{\bullet }^{-}=k^{*}\left( a\right) ,\quad
a_{+}^{-}=l\left( a\right) ,  \label{0.2}
\end{equation}
where $j\left( a^{\star }a\right) =j\left( a\right) ^{*}j\left( a\right) $
is the operator representation $j\left( a\right) ^{*}k\left( a\right)
=k\left( a^{\star }a\right) $ on the pre-Hilbert space $\mathcal{E}=k\left( 
\mathfrak{a}\right) $ of the Kolmogorov decomposition $l\left( a^{\star
}a\right) =k\left( a\right) ^{*}k\left( a\right) $, and $k^{*}\left(
a\right) =k\left( a^{\star }\right) ^{*}$.

As was proved in \cite{Bcs}, a quantum stochastic stationary processes $t\in 
\mathbb{R}_{+}\mapsto \Lambda \left( t,a\right) ,a\in \mathfrak{a}$ with $%
\Lambda \left( 0,a\right) =0$ and independent increments $\mathrm{d}\Lambda
\left( t,a\right) =\Lambda \left( t+\mathrm{d}t,a\right) -\Lambda \left(
t,a\right) $, forming an It\^{o} $\star $-algebra, can be represented in the
Fock space $\mathfrak{F}$ over the space of $\mathcal{E}$ -valued
square-integrable functions on $\mathbb{R}_{+}$ as $\Lambda _\mu ^\nu \left(
t,a_\nu ^\mu \right) =a_\nu ^\mu \Lambda _\mu ^\nu \left( t\right) $. Here 
\begin{equation}
a_\nu ^\mu \Lambda _\mu ^\nu \left( t\right) =a_{\bullet }^{\bullet }\Lambda
_{\bullet }^{\bullet }\left( t\right) +a_{+}^{\bullet }\Lambda _{\bullet
}^{+}\left( t\right) +a_{\bullet }^{-}\Lambda _{-}^{\bullet }\left( t\right)
+a_{+}^{-}\Lambda _{-}^{+}\left( t\right) ,  \label{0.3}
\end{equation}
is the canonical decomposition of $\Lambda $ into the exchange $\Lambda
_{\bullet }^{\bullet }$, creation $\Lambda _{\bullet }^{+}$, annihilation $%
\Lambda _{-}^{\bullet }$ and preservation (time) $\Lambda _{-}^{+}=t\mathrm{I%
}$ processes of quantum stochastic calculus \cite{Par}, \cite{Mey} having
the mean values $\left\langle \Lambda _\mu ^\nu \left( t\right)
\right\rangle =t\delta _{+}^\nu \delta _\mu ^{-}$ with respect to the vacuum
state in $\mathfrak{F}$. Thus the parametrizing algebra $\mathfrak{a}$ can
be always identified with a $\star $-subalgebra of the algebra $\mathcal{Q}%
\left( \mathcal{E}\right) $ of all quadruples $\boldsymbol{a}=\left( a_\nu
^\mu \right) _{\nu =+,\bullet }^{\mu =-,\bullet }$, where $a_\nu ^\mu :%
\mathcal{E}_\nu \rightarrow \mathcal{E}_\mu $ are the linear operators on $%
\mathcal{E}_{\bullet }=\mathcal{E},\mathcal{E}_{+}=\mathbb{C=}\mathcal{E}%
_{-} $, having the adjoints $a_\nu ^{\mu *}\mathcal{E}_\mu \subseteq 
\mathcal{E}_\nu $, with the Hudson--Parthasarathy (HP) multiplication table 
\cite{16} 
\begin{equation}
\boldsymbol{a}\bullet \boldsymbol{b}=\left( a_{\bullet }^\mu b_\nu ^{\bullet
}\right) _{\nu =+,\bullet }^{\mu =-,\bullet },  \label{0.4}
\end{equation}
the unique death $\boldsymbol{d}=\left( \delta _{-}^\mu \delta _\nu
^{+}\right) _{\nu =+,\bullet }^{\mu =-,\bullet }$, and the involution $%
a_{-\nu }^{\star \mu }=a_{-\mu }^{\nu *}$, where $-(-)=+$, $-\bullet
=\bullet $, $-(+)=-$.

The stochastic differential of a CP flow $\phi =\left( \phi _t\right) _{t>0}$
over an operator algebra $\mathcal{B}$ is written in terms of the quantum
canonical differentials as $\mathrm{d}\phi =\phi \circ \lambda _\nu ^\mu 
\mathrm{d}\Lambda _\mu ^\nu $ with $\phi _0=\imath $ at $t=0$, where $\imath
\left( B\right) =B$ is the identical representation of $\mathcal{B}$. The
main result of this paper is the construction of CP flows and their
filtering dilation to the HP flows, based on the linear quantum stochastic
evolution equation of the form 
\begin{equation*}
\mathrm{d}\phi _t\left( B\right) +\phi _t\left( K^{*}B+BK-L^{*}\jmath \left(
B\right) L\right) \mathrm{d}t=\phi _t\left( L^{\bullet }\jmath \left(
B\right) L_{\bullet }-B\otimes \delta _{\bullet }^{\bullet }\right) \mathrm{d%
}\Lambda _{\bullet }^{\bullet }
\end{equation*}
\begin{equation}
+\phi _t\left( L^{\bullet }\jmath \left( B\right) L-K^{\bullet }B\right) 
\mathrm{d}\Lambda _{\bullet }^{+}+\phi _t\left( L^{*}\jmath \left( B\right)
L_{\bullet }-BK_{\bullet }\right) \mathrm{d}\Lambda _{-}^{\bullet },
\label{0.5}
\end{equation}
where $\jmath $ is an operator representation of $\mathcal{B}$, $\delta
_{\bullet }^{\bullet }$ is the identity operator in $\mathcal{E}$, and the
operator $K$ satisfies the conservativity condition $K+K^{*}=L^{*}L$ for the
deterministic generator $\lambda =\lambda _{+}^{-}$. This form of the CP
evolution equation was discovered in \cite{15} as a result of the general CP
differential structure 
\begin{equation*}
\boldsymbol{\lambda }\left( B\right) =\boldsymbol{L}^{*}\jmath (B)%
\boldsymbol{L}-\boldsymbol{K}^{*}B-B\boldsymbol{K}
\end{equation*}
of the bounded quantum stochastic generators $\boldsymbol{\lambda }=\left(
\lambda _\nu ^\mu \right) _{\nu =+,\bullet }^{\mu =-,\bullet }$ over a von
Neumann algebra $\mathcal{B}$ even in the nonlinear case. The dilation of
the stochastic differentials for CP processes over arbitrary $*$-algebras,
giving this structure for the bounded generators as a consequence of the
Christensen-Evans theorem \cite{CrE}, was constructed in \cite{Bge}. Here we
shall prove that such a quantum stochastic extension of Lindblad's structure 
$\lambda \left( B\right) =L^{*}\jmath \left( B\right) L-K^{*}B-BK$, can be
always used for the construction and the dilation of the CP flows also in
the case of the unbounded maps $\lambda _\nu ^\mu :\mathcal{B}\rightarrow 
\mathcal{B}$ over the algebra $\mathcal{B}=\mathcal{L}\left( \mathcal{H}%
\right) $ of all operators in a Hilbert space $\mathcal{H}$. We shall prove
that this structure is necessary at least in the case of the w*-continuous
generators, which are extendable to the covariant ones over the algebra of
all bounded operators $\mathcal{L}\left( \mathcal{H}\right) $. The existence
of a minimal CP\ solution which is constructed under certain continuity
conditions proves that this structure is also sufficient for the CP property
of any solution to this stochastic equation. The construction of the
differential dilations and the CP solutions of such quantum stochastic
differential equations with the bounded generators over the simple
finite-dimensional It\^{o} algebra $\mathfrak{a}=\mathcal{Q}\left( \mathcal{E%
}\right) $ and the arbitrary $\mathcal{B}\subseteq \mathcal{L}\left( 
\mathcal{H}\right) $ was recently discussed in \cite{Be96,LiP} (the latter
paper contains also a characterization of the bounded generators for the
contractive CP flows.)

The Evans--Lewis case $\Lambda \left( t,a\right) =\alpha t\mathrm{I}$ is
described by the simplest one-dimensional It\^{o} algebra $\mathfrak{a}=%
\mathbb{C}d$ with $l\left( a\right) =\alpha \in \mathbb{C}$, $\alpha ^{\star
}=\bar{\alpha}$, and the nilpotent multiplication $\alpha ^{\star }\alpha =0$
corresponding to the non-stochastic (Newton) calculus $\left( \mathrm{d}%
t\right) ^2=0$ in $\mathcal{E}=0$. The standard Wiener process $\mathrm{Q}%
=\Lambda _{-}^{\bullet }+\Lambda _{\bullet }^{+}$ in Fock space is described
by the second order nilpotent algebra $\mathfrak{a}$ of pairs $a=\left(
\alpha ,\xi \right) $ with $d=\left( 1,0\right) $, $\xi \in \mathbb{C}$,
represented by the quadruples $a_{+}^{-}=\alpha ,\quad a_{\bullet }^{-}=\xi
=a_{+}^{\bullet },\quad a_{\bullet }^{\bullet }=0$ in $\mathcal{E}=\mathbb{C}
$, corresponding to $\Lambda \left( t,a\right) =\alpha t\mathrm{I}+\xi 
\mathrm{Q}\left( t\right) $. The unital $\star $-algebra $\mathbb{C}$ with
the usual multiplication $\zeta ^{\star }\zeta =\left| \zeta \right| ^2$ can
be embedded into the two-dimensional It\^{o} algebra $\mathfrak{a}$ of $%
a=\left( \alpha ,\zeta \right) $, $\alpha =l\left( a\right) $, $\zeta \in 
\mathbb{C}$ as $a_{\bullet }^{\bullet }=\zeta $, $a_{+}^{\bullet }=+i\zeta $%
, $a_{\bullet }^{-}=-i\zeta $, $a_{+}^{-}=\zeta $. It corresponds to $%
\Lambda \left( t,a\right) =\alpha t\mathrm{I}+\zeta \mathrm{P}\left(
t\right) $, where $\mathrm{P}=\Lambda _{\bullet }^{\bullet }+i\left( \Lambda
_{\bullet }^{+}-\Lambda _{-}^{\bullet }\right) $ is the representation of
the standard Poisson process, compensated by its mean value $t$. Thus our
results are applicable also to the classical stochastic differentials of
completely positive processes, corresponding to the commutative It\^{o}
algebras, which are decomposable into the Wiener, Poisson and Newton
orthogonal components.

\section{Quantum filtering dynamics}

The quantum filtering theory, which was outlined in \cite{1, 2} and
developed then since \cite{3}, provides the derivations for new types of
irreversible stochastic equations for quantum states, giving the dynamical
solution for the well-known quantum measurement problem. Some particular
types of such equations have been considered recently in the
phenomenological theories of quantum permanent reduction \cite{4,5},
continuous measurement collapse \cite{6,7}, spontaneous jumps \cite{8,9},
diffusions and localizations \cite{10,11}. The main feature of such dynamics
is that the reduced irreversible evolution can be described in terms of a
linear dissipative stochastic wave equation, the solution to which is
normalized only in the mean square sense.

The simplest dynamics of this kind is described by the continuous filtering
wave propagators $V_t\left( \omega \right) $, defined on the space $\Omega $
of all Brownian trajectories as an adapted operator-valued stochastic
process in the system Hilbert space $\mathcal{H}$, satisfying the stochastic
diffusion equation 
\begin{equation}
\mathrm{d}V_t+KV_t\mathrm{d}t=LV_t\mathrm{dQ},\quad V_0=I  \label{1.3}
\end{equation}
in the It\^{o} sense. Here $\mathrm{Q}\left( t,\omega \right) $ is the
standard Wiener process, which is described by the independent increments $%
\mathrm{dQ}\left( t\right) =\mathrm{Q}\left( t+\mathrm{d}t\right) -\mathrm{Q}%
\left( t\right) $, having the zero mean values $\langle \mathrm{dQ}\rangle
=0 $ and the multiplication property $(\mathrm{dQ})^2=\mathrm{d}t$, $K$ is
an accretive operator, $K+K^{\dagger }\geq L^{*}L$, defined on a dense
domain $\mathcal{D\subseteq H}$, with $K^{\dagger }=K^{*}|\mathcal{D}$, and $%
L$ is a linear operator $\mathcal{D}\rightarrow \mathcal{H}$. This
stochastic wave equation was first derived \cite{13} from a unitary cocycle
evolution by a quantum filtering procedure. A sufficient analyticity
condition, under which it has the unique solution in the form of a
stochastic multiple integral even in the case of unbounded $K$ and $L$ is
given in the Appendix. Using the It\^{o} formula 
\begin{equation}
\mathrm{d}\left( V_t^{*}V_t\right) =\mathrm{d}V_t^{*}V_t+V_t^{*}\mathrm{d}%
V_t+\mathrm{d}V_t^{*}\mathrm{d}V_t,  \label{1.2}
\end{equation}
and averaging $\left\langle \cdot \right\rangle $ over the trajectories of $%
\mathrm{Q}$, one obtains $\langle V_t^{*}V_t\rangle \leq I$ as a consequence
of $\mathrm{d}\langle V_t^{*}V_t\rangle \leq 0$. Note that the process $V_t$
is not necessarily unitary if the filtering condition $K^{\dagger }+K=L^{*}L$
holds, and even if $L^{\dagger }=-L$, it might be only isometric, $%
V_t^{*}V_t=I$, in the unbounded case.

Another type of the filtering wave propagator $V_t\left( \omega \right)
:\psi _0\in \mathcal{H}\mapsto \psi _t\left( \omega \right) $ in $\mathcal{H}
$ is given by the stochastic jump equation 
\begin{equation}
\mathrm{d}V_t+KV_t\mathrm{d}t=LV_t\mathrm{dP},\quad V_0=I  \label{1.1}
\end{equation}
at the random time instants $\omega =\left\{ t_1,t_2,...\right\} $. Here $%
L=J-I$ is the jump operator, corresponding to the stationary discontinuous
evolutions $\psi _{t+}=J\psi $ at $t\in \omega $, and $\mathrm{P}\left(
t,\omega \right) $ is the standard Poisson process, counting the number $%
\left| \omega \cap [0,t)\right| $ compensated by its mean value $t$. It is
described as the process with independent increments $\mathrm{dP}\left(
t\right) =\mathrm{P}\left( t+\mathrm{d}t\right) -\mathrm{P}\left( t\right) $%
, having the values $\left\{ 0,1\right\} $ at $\mathrm{d}t\rightarrow 0$,
with zero mean $\langle \mathrm{dP}\rangle =0$, and the multiplication
property $\left( \mathrm{dP}\right) ^2=\mathrm{dP}+\mathrm{d}t$. This
stochastic wave equation was first derived in \cite{12} by the conditioning
with respect to the spontaneous reductions $J:\psi _t\mapsto \psi _{t+}$. An
analyticity condition under which it has the unique solution in the form of
the multiple stochastic integral even in the case of unbounded $K$ and $L$
is also given in the Appendix. Using the It\^{o} formula (\ref{1.2}) with $%
\mathrm{d}V_t^{*}\mathrm{d}V_t=V_t^{*}L^{*}LV_t(\mathrm{dP}\mathbf{+}\mathrm{%
d}t\mathrm{)}$, one can obtain 
\begin{equation*}
\mathrm{d}\left( V_t^{*}V_t\right) =V_t^{*}\left( L^{*}L-K-K^{\dagger
}\right) V_t\mathrm{d}t+V_t^{*}\left( L^{\dagger }+L+L^{*}L\right) V_t%
\mathrm{dP}.
\end{equation*}
Averaging $\left\langle \cdot \right\rangle $ over the trajectories of $%
\mathrm{P}$, one can easily find that $\mathrm{d}\langle V_t^{*}V_t\rangle
\leq 0$ under the sub-filtering condition $L^{*}L\leq K+K^{\dagger }$. Such
evolution is not needed to be unitary even if $L^{*}L=K+K^{\dagger }$, but
it might be isometric, $V_t^{*}V_t=I$ if the jumps are isometric, $J^{*}J=I$.

This proves in both cases that the stochastic wave function $\psi _t\left(
\omega \right) =V_t\left( \omega \right) \psi _0$ is not normalized for each 
$\omega $, but it is normalized in the mean square sense to the survival
probability $\langle ||\psi _t||^2\rangle \leq ||\psi _0||^2=1$ for the
quantum system not to be demolished during its observation up to the time $t$%
. If $\left\langle ||\psi _t||^2\right\rangle =1$, then the positive
stochastic function $||\psi _t\left( \omega \right) ||^2$ is the probability
density of a diffusive $\widehat{\mathrm{Q}}$ or counting $\widehat{\mathrm{P%
}}$ output process up to the given $t$ with respect to the standard Wiener $%
\mathrm{Q}$ or Poisson $\mathrm{P}$ input processes.

Using the It\^{o} formula for $\phi _t\left( B\right) =V_t^{*}BV_t$, one can
obtain the stochastic equations 
\begin{equation}
\mathrm{d}\phi _t\left( B\right) +\phi _t\left( K^{*}B+BK-L^{*}BL\right) 
\mathrm{d}t=\phi _t\left( L^{*}B+BL\right) \mathrm{dQ},  \label{1.5}
\end{equation}
\begin{equation}
\mathrm{d}\phi _t\left( B\right) +\phi _t\left( K^{*}B+BK-L^{*}BL\right) 
\mathrm{d}t=\phi _t\left( J^{*}BJ-B\right) \mathrm{dP},  \label{1.4}
\end{equation}
describing the stochastic evolution $Y_t=\phi _t\left( B\right) $ of a
bounded system operator $B\in \mathcal{L}\left( \mathcal{H}\right) $ as $%
Y_t\left( \omega \right) =V_t\left( \omega \right) ^{*}BV_t\left( \omega
\right) $. The maps $\phi _t:B\mapsto Y_t$ are Hermitian in the sense that $%
Y_t^{*}=Y_t$ if $B^{*}=B$, but in contrast to the usual Hamiltonian
dynamics, are not multiplicative in general, $\phi _t\left( B^{*}C\right)
\neq \phi _t\left( B\right) ^{*}\phi _t\left( C\right) $, even if they are
not averaged with respect to $\omega $. Moreover, they are usually not
normalized, $R_t\left( \omega \right) :=\phi _t\left( \omega ,I\right) \neq
I $, although the stochastic positive operators $R_t=V_t^{*}V_t$ under the
filtering condition are usually normalized in the mean, $\langle R_t\rangle
=I$, and satisfy the martingale property $\epsilon _t\left[ R_s\right] =R_t$
for all $s>t$, where $\epsilon _t$ is the conditional expectation with
respect to the history of the processes $\mathrm{P}$ or $\mathrm{Q}$ up to
time $t$.

Although the filtering equations (\ref{1.1}), (\ref{1.3}) look very
different, they can be unified in the form of the quantum stochastic
equation 
\begin{equation}
\mathrm{d}V_t+KV_t\mathrm{d}t+K^{-}V_t\mathrm{d}\Lambda _{-}=\left(
J-I\right) V_t\mathrm{d}\Lambda +L_{+}V_t\mathrm{d}\Lambda ^{+},  \label{1.6}
\end{equation}
where $\Lambda ^{+}\left( t\right) $ is the creation process, corresponding
to the annihilation $\Lambda _{-}\left( t\right) $ on the interval $[0,t)$,
and $\Lambda \left( t\right) $ is the number of quantums on this interval.
These canonical quantum stochastic processes, representing the quantum noise
with respect to the vacuum state $|0\rangle $ of the Fock space $\mathcal{F}$
over the single-quantum Hilbert space $L^2\left( \mathbb{R}_{+}\right) $ of
square-integrable functions of $t\in [0,\infty )$, are formally given in 
\cite{14} by the integrals 
\begin{equation*}
\Lambda _{-}\left( t\right) =\int_0^t\Lambda _{-}^r\mathrm{d}r,\quad \Lambda
^{+}\left( t\right) =\int_0^t\Lambda _r^{+}\mathrm{d}r,\quad \Lambda \left(
t\right) =\int_0^t\Lambda _r^{+}\Lambda _{-}^r\mathrm{d}r,
\end{equation*}
where $\Lambda _{-}^r,\Lambda _r^{+}$ are the generalized quantum
one-dimensional fields in $\mathcal{F}$, satisfying the canonical
commutation relations 
\begin{equation*}
\left[ \Lambda _{-}^r,\Lambda _s^{+}\right] =\delta \left( s-r\right)
I,\quad \left[ \Lambda _{-}^r,\Lambda _{-}^s\right] =0=\left[ \Lambda
_r^{+},\Lambda _s^{+}\right] .
\end{equation*}
They can be defined by the independent increments with 
\begin{equation}
\langle 0|\mathrm{d}\Lambda _{-}|0\rangle =0,\quad \langle 0|\mathrm{d}%
\Lambda ^{+}|0\rangle =0,\quad \langle 0|\mathrm{d}\Lambda |0\rangle =0
\label{1.7}
\end{equation}
and the noncommutative multiplication table 
\begin{equation}
\mathrm{d}\Lambda \mathrm{d}\Lambda =\mathrm{d}\Lambda ,\quad \mathrm{d}%
\Lambda _{-}\mathrm{d}\Lambda =\mathrm{d}\Lambda _{-},\quad \mathrm{d}%
\Lambda \mathrm{d}\Lambda ^{+}=\mathrm{d}\Lambda ^{+},\quad \mathrm{d}%
\Lambda _{-}\mathrm{d}\Lambda ^{+}=\mathrm{d}tI  \label{1.8}
\end{equation}
with all other products being zero: $\mathrm{d}\Lambda \mathrm{d}\Lambda
_{-}=\mathrm{d}\Lambda ^{+}\mathrm{d}\Lambda =\mathrm{d}\Lambda ^{+}\mathrm{d%
}\Lambda _{-}=0$. The standard Poisson process $\mathrm{P}$ as well as the
Wiener process $\mathrm{Q}$ can be represented in $\mathfrak{F}$ by the
linear combinations \cite{16} 
\begin{equation}
\mathrm{P}\left( t\right) =\Lambda \left( t\right) +i\left( \Lambda
^{+}\left( t\right) -\Lambda _{-}\left( t\right) \right) ,\quad \mathrm{Q}%
\left( t\right) =\Lambda ^{+}\left( t\right) +\Lambda _{-}\left( t\right) ,
\label{1.9}
\end{equation}
so Eq. (\ref{1.6}) corresponds to the stochastic diffusion equation (\ref%
{1.3}) if $J=I$, $L_{+}=L=-K^{-}$, and it corresponds to the stochastic jump
equation (\ref{1.1}) if $J=I+L$, $L_{+}=iL=K^{-}$. The quantum stochastic
equation for $\phi _t\left( B\right) =V_t^{*}BV_t$ has the following general
form 
\begin{equation*}
\mathrm{d}\phi _t\left( B\right) +\phi _t\left( K^{*}B+BK-L^{-}BL_{+}\right) 
\mathrm{d}t=\phi _t\left( J^{*}BJ-B\right) \mathrm{d}\Lambda
\end{equation*}
\begin{equation}
+\phi _t\left( J^{*}BL_{+}-K_{+}B\right) \mathrm{d}\Lambda ^{+}+\phi
_t\left( L^{-}BJ-BK^{-}\right) \mathrm{d}\Lambda _{-},  \label{1.10}
\end{equation}
where $L^{-}=L_{+}^{*},K_{+}^{*}=K^{-}$, coinciding with either (\ref{1.5})
or with (\ref{1.4}) in the particular cases. Equation (\ref{1.10}) is
obtained from (\ref{1.6}) by using the It\^{o} formula (\ref{1.2}) with the
multiplication table (\ref{1.8}). The sub-filtering condition $K+K^{\dagger
}\geq L^{-}L_{+}$ for Eq. (\ref{1.6}) defines in both cases the positive
operator-valued process $R_t=\phi _t\left( I\right) $ as a sub-martingale
with $R_0=I$, or a martingale in the case $K+K^{\dagger }=L^{-}L_{+}$. In
the particular case 
\begin{equation*}
J=S,\quad K^{-}=L^{-}S,\quad L_{+}=SK_{+},\quad S^{*}S=I,
\end{equation*}
corresponding to the Hudson--Evans flow \cite{EvH} if $S^{*}=S^{-1}$, the
evolution is isometric, and identity preserving, $\phi _t\left( I\right) =I$%
, at least in the case of bounded $K$ and $L$.

In the next sections we define a multidimensional analog of the quantum
stochastic equation (\ref{1.10}) and will show that the suggested general
structure of its generator indeed follows just from the property of complete
positivity of the map $\phi _t$ for all $t>0$ and the normalization
condition $\phi _t\left( I\right) =R_t$ to a form-valued sub-martingale with
respect to the natural filtration of the quantum noise in the Fock space $%
\mathfrak{F}$ .

\section{Quantum completely positive flows}

Throughout the complex pre-Hilbert space $\mathcal{D}\subseteq \mathcal{H}$
is a reflexive Fr\'{e}chet space, $\mathcal{E}\otimes \mathcal{D}$ denotes
the projective tensor product ($\pi $-product) with another such space $%
\mathcal{E}$, $\mathcal{D}^{\prime }\supseteq \mathcal{H}$ denotes the dual
space of continuous antilinear functionals $\eta ^{\prime }:\eta \in 
\mathcal{D}\mapsto \langle \eta |\eta ^{\prime }\rangle $, with respect to
the canonical pairing $\left\langle \eta |\eta ^{\prime }\right\rangle $
given by $\left\| \eta \right\| ^2$ for $\eta ^{\prime }=\eta \in \mathcal{H}
$, $\mathcal{B}\left( \mathcal{D}\right) $ denotes the linear space of all
continuous sesquilinear forms $\langle \eta |B\eta \rangle $ on $\mathcal{D}$%
, identified with the continuous linear operators $B:\mathcal{D}\rightarrow 
\mathcal{D}^{\prime }$ (kernels), $B^{\dagger }\in \mathcal{B}\left( 
\mathcal{D}\right) $ is the Hermit conjugated form (kernel) $\langle \eta
|B^{\dagger }\eta \rangle =\langle \eta |B\eta \rangle ^{*}$, and $\mathcal{L%
}\left( \mathcal{D}\right) \subseteq \mathcal{B}\left( \mathcal{D}\right) $
denotes the algebra of all strongly continuous operators $B:\mathcal{D}%
\rightarrow \mathcal{D}$. Any such space $\mathcal{D}$ can be considered as
a projective limit with respect to an increasing sequence of Hilbertian
norms $\left\| \cdot \right\| _p>\left\| \cdot \right\| $ on $\mathcal{D}$;
for the definitions and properties of this standard topological notions see
for example \cite{Ob94}. The space $\mathcal{D}^{\prime }$ will be equipped
with weak topology induced by its predual (= dual) $\mathcal{D}$, and $%
\mathcal{B}\left( \mathcal{D}\right) $ will be equipped with w*-topology
(induced by the predual $\mathcal{B}_{*}\left( \mathcal{D}\right) =\mathcal{D%
}\otimes \mathcal{D}$), coinciding with the weak topology on each bounded
subset with respect to a norm $\left\| \cdot \right\| _p$. Any operator $%
A\in \mathcal{L}\left( \mathcal{D}\right) $ with $A^{\dagger }\in \mathcal{L}%
\left( \mathcal{D}\right) $ can be uniquely extended to a weakly continuous
operator onto $\mathcal{D}^{\prime }$ as $A^{\dagger *}$, denoted again as $%
A $, where $A^{*}$ is the dual operator $\mathcal{D}^{\prime }\rightarrow 
\mathcal{D}^{\prime }$, $\langle \eta |A^{*}\eta ^{\prime }\rangle
=\left\langle A\eta |\eta ^{\prime }\right\rangle $, defining the involution 
$A\mapsto A^{*}$ for such continuations $A:\mathcal{D}^{\prime }\rightarrow 
\mathcal{D}^{\prime }$. We say that the operator $A$ commutes with a
sesquilinear form, $BA=AB$ if $\left\langle \eta |BA\eta \right\rangle
=\left\langle A^{\dagger }\eta |B\eta \right\rangle $ for all $\eta \in 
\mathcal{D}$. The commutant $\mathcal{A}^c=\left\{ B\in \mathcal{B}\left( 
\mathcal{D}\right) :\left[ A,B\right] =0,\forall A\in \mathcal{A}\right\} $
of an operator $*$-algebra $\mathcal{A}\subseteq \mathcal{L}\left( \mathcal{D%
}\right) $ is weakly closed in $\mathcal{B}\left( \mathcal{D}\right) $, so
that the weak closure $\overline{\mathcal{B}}\subseteq \mathcal{B}\left( 
\mathcal{D}\right) $ of any $\mathcal{B}\subseteq \mathcal{A}^c$ also
commutes with $\mathcal{A}$.

1. Let $\mathcal{B}\subseteq \mathcal{L}\left( \mathcal{H}\right) $ be a
unital $*$-algebra of bounded operators $B:\mathcal{H}\rightarrow \mathcal{H}
$, $\left\| B\right\| <\infty $, and $\left( \Omega ,\mathfrak{A},P\right) $
be a probability space with a filtration $\left( \mathfrak{A}_t\right)
_{t>0},$ $\mathfrak{A}_t\subseteq \mathfrak{A}$ of $\sigma $-algebras on $%
\Omega $. One can assume that the filtration $\mathfrak{A}_t\subseteq 
\mathfrak{A}_s,\forall t<s$ is generated by $x_t=\left\{ r\mapsto x\left(
r\right) :r<t\right\} $ of a stochastic process $x\left( t,\omega \right) $
with independent increments $\mathrm{d}x\left( t\right) =x\left( t+\Delta
\right) -x\left( t\right) $, and the probability measure $P$ is invariant
under the measurable representations $\omega \mapsto \omega _s\in \Omega $, $%
A_s^{-1}=\left\{ \omega :\omega _s\in A\right\} \in \mathfrak{A}$, $\forall
A\in \mathfrak{A}$ on $\Omega \ni \omega $ of the time shifts $t\mapsto
t+s,s>0$, corresponding to the shifts of the random increments 
\begin{equation*}
\mathrm{d}x\left( t,\omega _s\right) =\mathrm{d}x\left( t+s,\omega \right)
,\quad \forall \omega \in \Omega ,t\in \mathbb{R}_{+}.
\end{equation*}
The\emph{\ filtering dynamics} over $\mathcal{B}$ with respect to the
process $x\left( t\right) $ is described by a cocycle flow $\phi =\left(
\phi _t\right) _{t>0}$ of linear completely positive \cite{Stn}
w*-continuous stochastic adapted maps $\phi _t\left( \omega \right) :%
\mathcal{B}\rightarrow \overline{\mathcal{B}}$, $\omega \in \Omega $ such
that the stochastic process $y_t\left( \omega \right) =\left\langle \eta
|\phi _t\left( \omega ,B\right) \eta \right\rangle $ is causally measurable
for each $\eta \in \mathcal{D}$, $B\in \mathcal{B}$ in the sense that $%
y_t^{-1}\left( B\right) \in \mathfrak{A}_t$, $\forall t>0$ and any Borel $%
B\subseteq \mathbb{C}$. The maps $\phi _t$ can be extended on the $\mathfrak{%
A}$-measurable functions $Y:\omega \mapsto Y\left( \omega \right) $ with
values $Y\left( \omega \right) \in \overline{\mathcal{B}}$ as the normal
maps $\phi _t\left[ Y\right] \left( \omega \right) =\overline{\phi }_t\left(
\omega ,Y\left( \omega _t\right) \right) $, defined for each $\omega \in
\Omega $ by the normal extension $\overline{\phi }_t\left( \omega \right) $
onto $\overline{\mathcal{B}}$, and the cocycle condition $\phi _r\left(
\omega \right) \circ \phi _s\left( \omega _r\right) =\phi _{r+s}\left(
\omega \right) $, $\forall r,s>0$ reads as the semigroup condition $\phi _r%
\left[ \phi _s\left[ Y\right] \right] =\phi _{r+s}\left[ Y\right] $ of the
extended maps. As it was noted in the previous section, the maps $\phi
_t\left( \omega \right) $ are not considered to be normalized to the
identity, and can be even unbounded, but they are supposed to be normalized, 
$\phi _t\left( \omega ,I\right) =R_t\left( \omega \right) $, to an
operator-valued martingale $R_t=\epsilon _t\left[ R_s\right] \geq 0$ with $%
R_0\left( \omega \right) =I$, or to a positive submartingale, $R_t\geq
\epsilon _t\left[ R_s\right] ,\forall s>t$ in the subfiltering case, where $%
\epsilon _t$ is the conditional expectation over $\omega $ with respect to $%
\mathfrak{A}_t$.

2. Now we give a noncommutative generalization of the filtering
(subfiltering) CP flows for an arbitrary It\^{o} algebra, which was
suggested in \cite{Be91} for a Gaussian It\^{o} algebra of finite
dimensional quantum thermal noise, and in \cite{15} for the simple quantum It%
\^{o} algebra $\mathcal{Q}\left( \mathbb{C}^d\right) $ even in the nonlinear
case.

The role of the classical process $x\left( t\right) $ will play the quantum
stochastic process 
\begin{equation*}
X\left( t\right) =A\otimes I+I\otimes \Lambda \left( t,a\right) ,\quad A\in 
\mathcal{A},a\in \mathfrak{a}
\end{equation*}
indexed by an operator algebra $\mathcal{A}\subset \mathcal{L}\left( 
\mathcal{D}\right) $ and a noncommutative It\^{o} algebra $\mathfrak{a}$.
Here $\Lambda \left( t,a\right) $ is the process with independent increment
on a dense subspace $\mathfrak{F}\subset \Gamma \left( \mathfrak{E}\right) $
of the Fock space $\Gamma \left( \mathfrak{E}\right) $ over the space $%
\mathfrak{E}=L_{\mathcal{E}}^2\left( \mathbb{R}_{+}\right) $ of all
square-norm integrable $\mathcal{E} $-valued functions on $\mathbb{R}_{+}$,
where $\mathcal{E}$ is a pre-Hilbert space of the representation $a\in 
\mathfrak{a}\mapsto \left( a_\nu ^\mu \right) _{\nu =+,\bullet }^{\mu
=-,\bullet }$ for the It\^{o} $\star $-algebra $\mathfrak{a}$. Assuming that 
$\mathcal{E}$ is a Fr\'{e}chet space, given by an increasing sequence of
Hilbertian norms $\left\| e^{\bullet }\right\| \left( \xi \right) >\left\|
e^{\bullet }\right\| $, $\xi \in \mathbb{N}$, we define $\mathfrak{F}$ as
the projective limit $\cap _\xi \Gamma \left( \mathfrak{E},\xi \right) $ of
the Fock spaces $\Gamma \left( \mathfrak{E},\xi \right) \subseteq \Gamma
\left( \mathfrak{E}\right) $, generated by coherent vectors $f^{\otimes }$,
with respect to the norms 
\begin{equation}
\left\| f^{\otimes }\right\| ^2\left( \xi \right) =\int_\Gamma \left\|
f^{\otimes }\left( \tau \right) \right\| ^2\left( \xi \right) \mathrm{d}\tau
:=\sum_{n=0}^\infty \frac 1{n!}\left( \int_0^\infty \left\| f^{\bullet
}\left( t\right) \right\| ^2\left( \xi \right) \mathrm{d}t\right)
^n=e^{\left\| f^{\bullet }\right\| ^2\left( \xi \right) }.  \label{2.0a}
\end{equation}
Here $f^{\otimes }\left( \tau \right) =\bigotimes_{t\in \tau }f^{\bullet
}\left( t\right) $ for each $f^{\bullet }\in \mathfrak{E}$ is represented by
tensor-functions on the space $\Gamma $ of all finite subsets $\tau =\left\{
t_1,...,t_n\right\} \subseteq \mathbb{R}_{+}$ (for a simple example of the
Fock scale see the Appendix.) Moreover, we shall assume that the It\^{o}
algebra $\mathfrak{a}$ is realized as a $\star $-subalgebra of
Hudson-Parthasarathy (HP) algebra $\mathcal{Q}\left( \mathcal{E}\right) $ of
all quadruples $\boldsymbol{a}=\left( a_\nu ^\mu \right) _{\nu =+,\bullet
}^{\mu =-,\bullet }$ with $a_{\bullet }^{\bullet }\in \mathcal{L}\left( 
\mathcal{E}\right) $, strongly representing the $\star $-semigroup $1+%
\mathfrak{a}$ on the Fr\'{e}chet space $\mathcal{E}$ by projective
contractions $\delta _{\bullet }^{\bullet }+a_{\bullet }^{\bullet }\in 
\mathcal{L}\left( \mathcal{E}\right) $ in the sense that for each $\zeta \in 
\mathbb{N}$ there exists $\xi $ such that $\left\| e^{\bullet }+a_{\bullet
}^{\bullet }e^{\bullet }\right\| \left( \zeta \right) \leq \left\|
e^{\bullet }\right\| \left( \xi \right) $ for all $e^{\bullet }\in \mathcal{E%
}$. The following theorem proves that these are natural assumptions (which
are not restrictive in the simple Fock scale for a finite dimensional $%
\mathfrak{a}$.)

\begin{proposition}
The exponential operators $W\left( t,a\right) =:\exp \left[ \Lambda \left(
t,a\right) \right] :$ defined as the solutions to the quantum It\^{o}
equation 
\begin{equation}
\mathrm{d}W_t\left( g\right) =W_t\left( g\right) \mathrm{d}\Lambda \left(
t,g\left( t\right) \right) ,\quad W_0\left( g\right) =I,g\left( t\right) \in 
\mathfrak{a}  \label{2.1a}
\end{equation}
with $g\left( t\right) =a$, are strongly continuous, $W\left( t,a\right) \in 
\mathcal{L}\left( \mathfrak{F}\right) $, if all $\widehat{a}_{\bullet
}^{\bullet }=\delta _{\bullet }^{\bullet }+a_{\bullet }^{\bullet }$ are
projective contractions on $\mathcal{E}$. They give an analytic
representation 
\begin{equation}
W\left( t,a\star a\right) =W\left( t,a\right) ^{\dagger }W\left( t,a\right)
,\quad W\left( t,0\right) =I,\quad W\left( t,d\right) =e^tI  \label{2.1g}
\end{equation}
of the unital $\star $-semigroup $1+\mathfrak{a}$ for the It\^{o} $\star $%
-algebra $\mathfrak{a}$ with respect to the $\star $-product $a\star
a=a+a^{\star }a+a^{\star }$.
\end{proposition}

\proof%
The solutions $W\left( t,a\right) $ are uniquely defined on the coherent
vectors as analytic functions 
\begin{equation}
W\left( t,a\right) f^{\otimes }\left( \tau \right) =\otimes _{r\in \tau
}^{r<t}\left( \widehat{a}_{\bullet }^{\bullet }f^{\bullet }\left( r\right)
+a_{+}^{\bullet }\right) \exp \left[ \int_0^t\left( a_{\bullet
}^{-}f^{\bullet }\left( r\right) +a_{+}^{-}\right) \mathrm{d}r\right]
\otimes _{r\in \tau }^{r\geq t}f^{\bullet }\left( r\right) ,  \label{2.1b}
\end{equation}
which obey the properties (\ref{2.1g}), see for example \cite{Bcs}. Thus the
span of coherent vectors is invariant, and it is also invariant under $%
W\left( t,a\right) ^{\dagger }=W\left( t,a^{\star }\right) $. They can be
extended on $\mathfrak{F}$ by continuity which follows from the continuity
of Wick exponentials $\otimes \widehat{a}_{\bullet }^{\bullet }$ for the
projective contractions $\widehat{a}_{\bullet }^{\bullet }\in \mathcal{E}$ ,
and boundedness of $a_{+}^{\bullet }\in \mathcal{E}$, $a_{\bullet }^{-}\in 
\mathcal{E}^{\prime }$. 
\endproof%

3. Let $\mathfrak{D}$ denote the Fr\'{e}chet space $\mathcal{D}\otimes 
\mathfrak{F}$, generated by $\psi =\eta \otimes f^{\otimes }$, $\eta \in 
\mathcal{D}$, $f^{\bullet }\in \mathfrak{E}$. Assuming for simplicity the
separability of the It\^{o} algebra in the sense $\mathcal{E}\subseteq \ell
^2$ such that $f^{\bullet }=\left( f^m\right) ^{m\in \mathbb{N}}$, one can
identify each $\psi ^{\prime }\in \mathfrak{D}^{\prime }$ with a sequence of 
$\mathcal{D}^{\prime }$-valued symmetric tensor-functions $\psi
_{m_1,...,m_n}^{\prime }\left( t_1,...t_n\right) $, $n=0,1,2,...$ . Let $%
\left( \mathfrak{D}_t\right) _{t>0}$ be the natural filtration and $\left( 
\mathfrak{D}_{[t}\right) _{t>0}$ be the backward filtration of the subspaces 
$\mathfrak{D}_t=\mathcal{D}\otimes \mathfrak{F}_t$, $\mathfrak{D}_{[t}=%
\mathcal{D}\otimes \mathfrak{F}_{[t}$ generated by $\eta \otimes f^{\otimes
} $ with $f^{\bullet }\in \mathfrak{E}_t$ and $f^{\bullet }\in \mathfrak{E}%
_{[t}$ respectively, where $\mathfrak{E}_t$ $=L_{\mathcal{E}}^2[0,t) $, $%
\mathfrak{E}_{[t}$ $=L_{\mathcal{E}}^2[t,\infty )$ are embedded into $%
\mathfrak{E}$. The spaces $\mathfrak{D}_t$, $\mathfrak{D}_{[t}$ of the
restrictions $E_t\psi =\psi |\Gamma _t$, $E_{[t}\psi =\psi |\Gamma _{[t}$
onto $\Gamma _t=\left\{ \tau _t=\tau \cap [0,t)\right\} $, $\Gamma
_{[t}=\left\{ \tau _{[t}=\tau \cap [t,\infty )\right\} $ are embedded into $%
\mathfrak{D}$ by the isometries $E_t^{\dagger }:\psi \mapsto \psi _t$, $%
E_{[t}^{\dagger }:\psi \mapsto \psi _{[t}$ as $\psi _t\left( \tau \right)
=\psi \left( \tau _t\right) \delta _\emptyset \left( \tau _{[t}\right) $, $%
\psi _{[t}\left( \tau \right) =\delta _\emptyset \left( \tau _t\right) \psi
\left( \tau _{[t}\right) $, where $\delta _\emptyset \left( \tau \right) =1$
if $\tau =\emptyset $, otherwise $\delta _\emptyset \left( \tau \right) =0$.
The projectors $E_t,E_{[t}$ onto $\mathfrak{D}_t,\mathfrak{D}^t$ are
extended onto $\mathfrak{D}^{\prime }$ as the adjoints to $E_t^{\dagger },$ $%
E_{[t}^{\dagger }$. The time shift on $\mathfrak{D}^{\prime }$ is defined by
the semigroup $\left( T^t\right) _{t>0} $ of adjoint operators $T^t=T_t^{*}$
to $T_t\psi \left( \tau \right) =\psi \left( \tau +t\right) $, where $\tau
+t=\left\{ t_1+t,...,t_n+t\right\} $, $\emptyset +t=\emptyset $, such that $%
T^t\psi \left( \tau \right) =\delta _\emptyset \left( \tau _t\right) \psi
\left( \tau _{[t}-t\right) $ are isometries for $\psi \in \mathfrak{D}$ onto 
$\mathfrak{D}_{[t}$. A family $\left( Z_t\right) _{t>0}$ of sesquilinear
forms $\left\langle \psi |Z_t\psi \right\rangle $ given by linear operators $%
Z_t:\mathfrak{D}\rightarrow \mathfrak{D}^{\prime }$ is called \emph{adapted}
(and $\left( Z^t\right) _{t>0}$ is called \emph{backward adapted}) if 
\begin{equation}
Z_t\left( \eta \otimes f^{\otimes }\right) =\psi ^{\prime }\otimes
E_{[t}f^{\otimes }\quad \left( Z^t\left( \eta \otimes f^{\otimes }\right)
=\psi ^{\prime }\otimes E_tf^{\otimes }\right) ,\quad \forall \eta \in 
\mathcal{D},f^{\bullet }\in \mathfrak{E},  \label{adp}
\end{equation}
where $\psi ^{\prime }\in \mathfrak{D}_t^{\prime }$ ($\mathfrak{D}%
_{[t}^{\prime }$) and $E_{[t}\;$($E_t$) are the projectors onto $\mathfrak{F}%
_{[t}$ ($\mathfrak{F}_t$) correspondingly.

The (vacuum) \emph{conditional expectation} on $\mathcal{B}\left( \mathfrak{D%
}\right) $ with respect to the past up to a time $t\in \mathbb{R}_{+}$ is
defined as a positive projector, $\epsilon _t\left( Z\right) \geq 0$, if $%
Z\geq 0$, $\epsilon _t=\epsilon _t\circ \epsilon _s,\forall s>t$, giving an
adapted sesquilinear form $Z_t=\epsilon _t\left( Z\right) $ in (\ref{adp})
for each $Z\in \mathcal{B}\left( \mathfrak{D}\right) $ by $\psi ^{\prime
}=E_tZE_t^{\dagger }\psi $, where $\psi =\eta \otimes E_tf^{\otimes }$. The
time shift $\left( \theta ^t\right) _{t>0}$ on $\mathcal{B}\left( \mathfrak{D%
}\right) $ is uniquely defined by the covariance condition $\theta ^t\left(
Z\right) T^t=T^tZ$ as a backward adapted family $Z^t=\theta ^t\left(
Z\right) ,t>0$ for each $Z\in \mathcal{B}\left( \mathfrak{D}\right) $. As in
the bounded case \cite{Mey} between the maps $\epsilon _t$ and $\theta ^t$
we have the relation $\theta ^r\circ \epsilon _s=\epsilon _{r+s}\circ \theta
^r$ which follows from the operator relation $T^rE_s=E_{r+s}T^r$. An adapted
family $\left( M_t\right) _{t>0}$ of positive $\left\langle \psi |M_t\psi
\right\rangle \geq 0,\forall \psi \in \mathfrak{D}$ Hermitian $M_t^{\dagger
}=M_t $ forms $M_t\in \mathcal{B}\left( \mathfrak{D}\right) $ is called 
\emph{martingale} (\emph{submartingale}) if $\epsilon _t\left( M_s\right)
=M_t$ ($\epsilon _t\left( M_s\right) \leq M_t$) for all $s\geq t\geq 0$. The
bounded operator-valued martingales $M_t$ were introduced in the case of the
finite-dimensional HP-algebra in \cite{PaS}.

4. Let $\mathfrak{B}$ denote the space of all $Y\in \mathcal{B}\left( 
\mathfrak{D}\right) $, commuting with all $X=\left\{ X\left( t\right)
\right\} $ in the sense 
\begin{equation*}
AY=YA,\quad \forall A\in \mathcal{A},\quad YW\left( t,a\right) =W\left(
t,a\right) Y,\quad \forall t>0,a\in \mathfrak{a},
\end{equation*}
where $A\left( \eta \otimes \varphi \right) =A\eta \otimes \varphi $, $%
W\left( \eta \otimes \varphi \right) =\eta \otimes W\varphi $, and the
unital $*$-algebra $\mathcal{B}\subseteq \mathcal{L}\left( \mathcal{H}%
\right) $ be weakly dense in the commutant $\mathcal{A}^c$. The quantum
filtration $\left( \mathfrak{B}_t\right) _{t>0}$ is defined as the
increasing family of subspaces $\mathfrak{B}_t\subseteq \mathfrak{B}_s,t\leq
s$ of the adapted sesquilinear forms $Y_t\in \mathfrak{B}$. The covariant
shifts $\theta ^t:Y\mapsto Y^t$ leave the space $\mathfrak{B}$ invariant,
mapping it onto the subspaces of backward adapted sesquilinear forms $%
Y^t=\theta ^t\left( Y\right) $.

The \emph{quantum stochastic positive flow} over $\mathcal{B\ }$is described
by a one parameter family $\phi =\left( \phi _t\right) _{t>0}$ of linear
w*-continuous maps $\phi _t:$ $\mathcal{B}\rightarrow \mathfrak{B}$
satisfying

\begin{enumerate}
\item the causality condition $\phi _t\left( B\right) \subseteq \mathfrak{B}%
_t,\quad \forall B\in \mathcal{B},t\in \mathbb{R}_{+}$,

\item the complete positivity condition $\left[ \phi _t\left( B_{kl}\right) %
\right] \geq 0$ for each $t>0$ and for any positive definite matrix $\left[
B_{kl}\right] \geq 0$ with $B_{kl}\in \mathcal{B}$,

\item the cocycle condition $\phi _r\circ \phi _s^r=\phi _{r+s},\forall
t,s>0 $ with respect to the covariant shift $\phi _s^r=\theta ^r\circ \phi _s
$.
\end{enumerate}

Here the composition $\circ $ is understood as $\phi _r\left[ \phi _s\left(
B\right) \right] =\phi _{r+s}\left( B\right) $ in terms of the linear normal
extensions of $\phi _t\left[ B\otimes Z\right] =\overline{\phi }_t\left(
B\right) Z^t$ to the CP maps $\mathfrak{B}\rightarrow $ $\mathfrak{B}$,
forming a one-parameter semigroup, where $B\in \overline{\mathcal{B}}$, $%
\overline{\phi }_t$ are the normal extensions of $\phi _t$ onto $\overline{%
\mathcal{B}} $, and $Z^t=\theta ^t\left( Z\right) $, $Z\in \mathcal{B}\left( 
\mathfrak{F}\right) $. These can be defined like in the classical case as $%
\phi _t\left[ Y\right] \left( \bar{f}^{\bullet },f^{\bullet }\right) =%
\overline{\phi }_t\left( \bar{f}^{\bullet },Y\left( \bar{f}_t^{\bullet
},f_t^{\bullet }\right) ,f^{\bullet }\right) $ with $f_t^{\bullet }\left(
r\right) =f^{\bullet }\left( t+r\right) $ by the coherent matrix elements $%
Y\left( \bar{f}^{\bullet },f^{\bullet }\right) =F^{*}YF$ for $Y\in \mathfrak{%
B}$ given by the continuous operators $F:\eta \mapsto \psi _f=\eta \otimes
f^{\otimes } $ , $\eta \in \mathcal{D}$ for each $f^{\bullet }\in \mathfrak{E%
}_t$ with the adjoints $F^{*}\psi ^{\prime }=\int_{\tau <t}f^{\otimes
}\left( \tau \right) ^{*}\psi ^{\prime }\left( \tau \right) \mathrm{d}\tau $
for $\psi ^{\prime }\in \mathfrak{D}^{\prime }$.

The flow is called \emph{(sub)-filtering}, if $R_t=\phi _t\left( I\right) $
is a (sub)-martingale with $R_0=I$, and is called contractive, if $I\geq
R_t\geq R_s$ for all $0\leq t\leq s\in \mathbb{R}_{+}$.

\begin{proposition}
The complete positivity for adapted linear maps $\phi _t:\mathcal{B}%
\rightarrow \mathcal{B}\left( \mathfrak{D}\right) $ can be written as 
\begin{equation}
\sum_{f,h\in \mathfrak{E}_t}\sum_{B,C\in \mathcal{B}}\langle \xi _B^f|\phi
_t\left( \bar{f}^{\bullet },B^{*}C,h^{\bullet }\right) \xi _C^h\rangle
:=\langle \eta ^k|\phi _t\left( \bar{f}_k^{\bullet },B_k^{*}B_l,h_l^{\bullet
}\right) \eta ^l\rangle \geq 0,\quad \forall t>0  \label{cpd}
\end{equation}
(the usual summation rule over repeated cross-level indices is understood),
where $\xi _B^f=\eta ^k$ if $f^{\bullet }=f_k^{\bullet }$ and $B=B_k$ with $%
f_k^{\bullet }\in \mathfrak{E}_t,B_k\in \mathcal{B}$, $k=1,2,...,$ otherwise 
$\xi _B^f=0$, and $\phi _t\left( B,f^{\bullet }\right) =\phi _t\left(
B\right) F$, $\phi _t\left( \bar{f}^{\bullet },B\right) =F^{*}\phi _t\left(
B\right) $.
\end{proposition}

\proof%
By definition the map $\phi $ into the sesquilinear forms is completely
positive on $\mathcal{B}$ if $\left\langle \psi ^k|\phi \left( B_{kl}\right)
\psi ^l\right\rangle \geq 0$ whenever $\left\langle \eta ^k|B_{kl}\eta
^l\right\rangle \geq 0$, where $\eta ^k,\psi ^k$ are arbitrary finite
sequences. Approximating from below the latter positive forms by sums of the
forms $\sum_{kl}\left\langle \eta ^k|B_{ik}^{*}B_{il}\eta ^l\right\rangle
\geq 0$, the complete positivity can be tested only for the forms $%
\sum_{kl}\left\langle \eta ^k|B_k^{*}B_l\eta ^l\right\rangle \geq 0$ due to
the additivity $\phi \left( \sum_iB_{ik}^{*}B_{il}\right) =\sum_i\phi \left(
B_{ik}^{*}B_{il}\right) $. If $\phi _t$ is adapted, this can be written as 
\begin{equation*}
\sum_{B,C\in \mathcal{B}}\left\langle \chi _B|\phi \left( B^{*}C\right) \chi
_C\right\rangle =\left\langle \psi ^k|\phi \left( B_k^{*}B_l\right) \psi
^l\right\rangle :=\sum_{k,l}\left\langle \psi ^k|\phi \left(
B_k^{*}B_l\right) \psi ^l\right\rangle \geq 0,
\end{equation*}
where $\chi _B=\psi ^k\in \mathfrak{D}_t$ if $B=B_k\in \mathcal{B}$,
otherwise $\chi _B=0$. Because any $\psi \in \mathfrak{D}_t$ can be
approximated by a $\mathcal{D}$-span $\sum_f\eta ^f\otimes f^{\otimes }$of
coherent vectors over $f_k^{\bullet }\in \mathfrak{E}_t$, it is sufficient
to define the CP property only for such spans as 
\begin{equation*}
0\leq \sum_{f,h}\sum_{B,C}\left\langle \xi _B^f\otimes f^{\otimes }|\phi
\left( B^{*}C\right) \left( \xi _C^h\otimes h^{\otimes }\right)
\right\rangle =\sum_{f,h}\sum_{B,C}\left\langle \xi _B^f|\phi \left( \bar{f}%
^{\bullet },B^{*}C,h^{\bullet }\right) \xi _C^h\right\rangle .
\end{equation*}
\endproof%

5. Note that the subfiltering (filtering) flows can be considered as quantum
stochastic CP dilations of the quantum sub-Markov (Markov) semigroups $%
\theta =\left( \theta _t\right) _{t>0}$, $\theta _r\circ \theta _s=\theta
_{r+s}$ in the sense $\theta _t=\epsilon \circ \phi _t$, where $\epsilon
\left( Y\right) \eta =EY\psi _0$, $E\psi ^{\prime }=\psi ^{\prime }\left(
\emptyset \right) ,\forall \psi ^{\prime }\in \mathfrak{D}^{\prime }$, with $%
\theta _s\left( I\right) \leq \theta _t\left( I\right) \leq I$ ($\theta
_t\left( I\right) =I$), $\forall t\leq s$. The contraction $C_t=\theta
_t\left( I\right) $ with $C_0=I$ defines the probability $\left\langle \eta
|C_t\eta \right\rangle \leq 1$, $\forall \eta \in \mathcal{H},\left\| \eta
\right\| =1$ for an unstable system not to be demolished by a time $t\in 
\mathbb{R}_{+}$, and the conditional expectations $\left\langle \eta
|AC_t\eta \right\rangle /\left\langle \eta |C_t\eta \right\rangle $ of the
initial nondemolition observables $A\in \mathcal{A}$ in any state $\eta \in 
\mathcal{D}$, and thus in any initial state $\psi _0\in \mathcal{\eta }%
\otimes \delta _\emptyset $. The following theorem shows that the
submartingale (or the contraction) $R_t=\phi _t\left( I\right) $ is also the
density operator with respect to $\psi _0=\eta \otimes \delta _\emptyset $, $%
\eta \in \mathcal{H}$ (or with respect to any $\psi \in \mathcal{H}\otimes 
\mathfrak{F}$) for the conditional state of the restricted nondemolition
process $X_t=\left\{ r\mapsto X\left( r\right) :r<t\right\} $.

\begin{theorem}
Let $t\mapsto R_t\in \mathfrak{B}_t$ be a positive (sub)-martingale and $%
\left( \mathfrak{g}_t\right) _{t>0}$ be the increasing family of $\star $%
-semigroups $\mathfrak{g}_t$ of step functions $g:\mathbb{R}_{+}\rightarrow 
\mathfrak{a}$, $g\left( s\right) =0$, $\forall s\geq t$ under the $\star $%
-product 
\begin{equation}
\left( g_k\star g_l\right) \left( t\right) =g_l\left( t\right) +g_k\left(
t\right) ^{\star }g_l\left( t\right) +g_k\left( t\right) ^{\star }
\label{2.1c}
\end{equation}
of $g_k^{\star }=g_k\star 0$ and $g_l=0\star g_l$. The generating function $%
\vartheta _t\left( g\right) =\epsilon \left[ R_tW_t\left( g\right) \right] $
of the output state for the process $\Lambda \left( t\right) $, defined for
any $g\in \mathfrak{g}_t$ and each $t>0$ as 
\begin{equation}
\left\langle \eta |\vartheta _t\left( g\right) \eta \right\rangle
=\left\langle \psi _0|R_tW_t\left( g\right) \psi _0\right\rangle ,\quad \psi
_0=\eta \otimes \delta _\emptyset ,  \label{sgf}
\end{equation}
is $\mathcal{B}^c$-valued, positive, $\vartheta _t\geq 0$ in the sense of
positive definiteness of the kernel 
\begin{equation}
\left\langle \eta ^k|\vartheta _t\left( g_k\star g_l\right) \eta
^l\right\rangle \geq 0,\quad \forall g_k\in \mathfrak{g}_t;\eta ^k\in 
\mathcal{D},  \label{2.1d}
\end{equation}
and $\vartheta _t\geq \vartheta _s|\mathfrak{g}_t$ in this sense for any $%
s\geq t $. If $R_0=I$, then $\vartheta _0\left( 0\right) =I\geq \vartheta
_t\left( 0\right) $ , and if $R_t$ is a martingale, then $\vartheta
_t=\vartheta _s|\mathfrak{g}_t$ for any $s\geq t$, and $\vartheta _t\left(
0\right) =I$ for all $t\in \mathbb{R}_{+}$. Any family $\vartheta =\left(
\vartheta _t\right) _{t\geq 0}$ of positive-definite functions $\vartheta _t:%
\mathfrak{g}_t\rightarrow \mathcal{B}^c$, satisfying the above consistency
and normalization properties, is the state generating function of the form (%
\ref{sgf}) iff it is absolutely continuous in the following sense 
\begin{equation}
\lim_{n\rightarrow \infty }\sum_{g\in \mathfrak{g}_t}\eta _n^g\otimes
g_{+}^{\otimes }=0\Rightarrow \lim_{n\rightarrow \infty }\sum_{g,h\in 
\mathfrak{g}_t}\left\langle \eta _n^g|\vartheta _t\left( g\star h\right)
\eta _n^h\right\rangle =0,  \label{2.1e}
\end{equation}
where $g_{+}^{\otimes }\left( \tau \right) =\otimes _{t\in \tau
}g_{+}^{\bullet }\left( t\right) $ and $\eta _n^g=0$ for almost all $g$
(i.e. except for a finite number of $g\in \mathfrak{g}_t$).
\end{theorem}

\proof%
Because the solutions $W_t\left( g\right) $ to the quantum stochastic
equation (\ref{2.1a}) for a step function $g$ are given by finite products
of commuting exponential operators $W\left( t,a\right) $, they are
multiplicative, $W_t\left( g_k\right) ^{*}W_t\left( g_l\right) =W_t\left(
g_k\star g_l\right) $, as the operators in (\ref{2.1b}) are. Then the
positive definiteness of $\vartheta _t$ follows from their commutativity (%
\ref{2.1c}) with positive $R_t$: 
\begin{equation*}
\left\langle \eta ^k|\vartheta _t\left( g_k\star g_l\right) \eta
^l\right\rangle =\left\langle W_t\left( g_k\right) \left( \eta ^k\otimes
\delta _\emptyset \right) |R_tW_t\left( g_l\right) \left( \eta ^l\otimes
\delta _\emptyset \right) \right\rangle =\left\langle \psi _t|R_t\psi
_t\right\rangle \geq 0\text{.}
\end{equation*}
It is $\mathcal{B}^c$-valued as 
\begin{equation*}
\left\langle \eta |\vartheta _t\left( g\right) B\eta \right\rangle
=\left\langle \psi _0|R_tW_t\left( g\right) B\psi _0\right\rangle
=\left\langle B\psi _0|R_tW_t\left( g\right) \psi _0\right\rangle
=\left\langle B\eta |\vartheta _t\left( g\right) \eta \right\rangle ,\forall
B\in \mathcal{B}\text{.}
\end{equation*}
From $W_t\left( g_r\right) =W_r\left( g\right) $, $r<t$ and $W_t\left(
0\right) =I$ as the case $g_0=0$ it follows that 
\begin{equation*}
\vartheta _t\left( g_r\right) =\left\langle \psi _0|R_tW_t\left( g_r\right)
\psi _0\right\rangle =\left\langle \psi _0|\epsilon _r\left( R_t\right)
W_r\left( g\right) \psi _0\right\rangle \leq \left\langle \psi
_0|R_rW_r\left( g\right) \psi _0\right\rangle =\vartheta _r\left( g\right)
\end{equation*}
for any finite matrix $g=\left[ g_k\star g_l\right] $, and $\vartheta
_t\left( 0\right) \leq 1=\vartheta _0\left( 0\right) $ if $R_t$ is a
submartingale with $R_0=I$. This implies the normalization and compatibility
conditions if $R_t$ is martingale. The continuity condition follows from the
continuity of the forms $R_t\in \mathcal{B}\left( \mathfrak{D}\right) $: if $%
\sum_g\left( \eta _n^g\otimes g_{+}^{\otimes }\right) \rightarrow 0$, then 
\begin{eqnarray*}
\sum_{g,h}\left\langle \eta _n^g|\vartheta _t\left( g\star h\right) \eta
_n^h\right\rangle &=&\sum_{g,h}\left\langle W_t\left( g\right) \left( \eta
_n^g\otimes \delta _\emptyset \right) |R_tW_t\left( h\right) \left( \eta
_n^h\otimes \delta _\emptyset \right) \right\rangle \\
&=&\left\langle \sum_g\eta _n^g\otimes g_{+}^{\otimes }|R_t\sum_g\eta
_n^g\otimes g_{+}^{\otimes }\right\rangle \longrightarrow 0.
\end{eqnarray*}

Conversely, let $\left( \mathcal{E},V_t,L\right) $ be the GNS triple,
describing the decomposition $\vartheta _t\left( g\right) =L^{*}V_t\left(
g\right) L$ for a positive-definite kernel-function $\vartheta _t$. It is
defined by the multiplicative $*$-representation $V_t\left( g\star h\right)
=V_t\left( g\right) ^{*}V_t\left( h\right) $ of $\mathfrak{g}_t$ on a
pre-Hilbert space $\mathcal{E}\subseteq \mathcal{H}$ and by a linear
operator $L:\mathcal{D}\rightarrow \mathcal{E}$. The correspondence $\pi
_t\left( B\right) :$ $V_t\left( g\right) L\eta \mapsto V_t\left( g\right)
LB\eta $ for $B\in \mathcal{B}$ is extended to a $*$-representation $\pi _t:%
\mathcal{B}\rightarrow V_t\left( \mathfrak{g}_t\right) ^{\prime }$ on the
linear combinations $\mathcal{E}_t^{\circ }=\left\{ \sum_kV_t\left(
g_k\right) L_t\eta _k:g_k\in \mathfrak{g}_t,\eta _k\in \mathcal{D}\right\} $
by virtue of the commutativity of $\vartheta _t\left( g\right) $ with $%
\mathcal{B}$: 
\begin{eqnarray*}
\left\langle B^{*}\eta ^g|\vartheta _t\left( g\star h\right) \eta
^h\right\rangle &=&\left\langle \pi _t\left( B^{*}\right) V_t\left( g\right)
L_t\eta ^g|V_t\left( h\right) L_t\eta ^h\right\rangle \\
&=&\left\langle V_t\left( g\right) L_t\eta ^g|\pi _t\left( B\right)
V_t\left( h\right) L_t\eta ^h\right\rangle =\left\langle \eta ^g|\vartheta
_t\left( g\star h\right) B\eta ^h\right\rangle .
\end{eqnarray*}
The linear correspondence $F_t:\sum_g\eta ^g\otimes g_{+}^{\otimes }\mapsto
\sum_gV_t\left( g\right) L\eta ^g$ obviously intertwines this representation
with $B\mapsto B\otimes I$ as well as the representation $V_t$ with $W_t$ on 
$\mathfrak{D}_t^{\circ }=\left\{ \sum_k\eta _k\otimes W_t\left( g_k\right)
\delta _\emptyset :g_k\in \mathfrak{g}_t,\eta _k\in \mathcal{D}\right\}
\subseteq \mathfrak{D}_t$: 
\begin{equation*}
F_t\left( \eta \otimes W_t\left( f^{\star }\right) g_{+}^{\otimes }\right)
=V_t\left( f\star g\right) L\eta =V_t\left( f^{\star }\right) V_t\left(
g\right) L\eta =V_t\left( f^{\star }\right) F_t\left( \eta \otimes
g_{+}^{\otimes }\right) ,
\end{equation*}
where $g_{+}^{\otimes }=W\left( g\right) \delta _\emptyset $. It is
continuous operator with respect to the Hilbert space norm in $\mathcal{H}_t$
because if $\sum_g\eta _n^g\otimes g_{+}^{\otimes }\rightarrow 0$, then 
\begin{equation*}
\left\| F_t\sum_g\eta _n^g\otimes g_{+}^{\otimes }\right\| ^2=\left\|
\sum_gV_t\left( g\right) L\eta _n^g\right\| ^2=\sum_{f,h}\left\langle \eta
_n^f|\vartheta _t\left( f\star h\right) \eta _n^h\right\rangle \rightarrow 0
\end{equation*}
due to the strong absolute continuity of $\vartheta _t$. Hence $F_t$ can be
continued to an intertwining operator $\mathfrak{D}_t^{\circ }\rightarrow 
\mathcal{H}$, and there exists the adjoint intertwining operator $F_t^{*}:%
\mathcal{E}\rightarrow \mathfrak{D}_t^{\prime }$, $F_t^{*}V_t\left( g\right)
=W_t\left( g\right) F_t^{*}$ such that 
\begin{equation*}
\left\langle \psi _0|F_t^{*}F_tW_t\left( g\right) \psi _0\right\rangle
=\left\langle F_t\psi _0|V_t\left( g\right) F_t\psi _0\right\rangle
=\left\langle L_t\eta |V_t\left( g\right) L_t\eta \right\rangle
=\left\langle \eta |L_t^{*}V_t\left( g\right) L_t\eta \right\rangle .
\end{equation*}
The positive operators $F_t^{*}F_t\in \mathcal{B}\left( \mathfrak{D}%
_t\right) $ uniquely extended to the adapted ones $R_t:\mathfrak{D}%
\rightarrow \mathfrak{D}^{\prime }$, commute with all $W_t\left( g\right)
,g\in \mathfrak{g}_t.$ They define a submartingale (martingale) $R_t\in 
\mathfrak{B}_t$ due to the property $W_t=W_s|\mathfrak{g}_t$ for all $s\geq
t $ and 
\begin{equation*}
\left\langle W_t\left( g_k\right) \psi _0^k|\epsilon _t\left( R_s\right)
W_t\left( g_l\right) \psi _0^l\right\rangle =\left\langle \psi
_0^k|R_sW_s\left( g_k\star g_l\right) \psi _0^l\right\rangle =\left\langle
\eta ^k|\vartheta _s\left( g_k\star g_l\right) \eta ^l\right\rangle
\end{equation*}
\begin{equation*}
\leq \left( =\right) \left\langle \eta ^k|\vartheta _t\left( g_k\star
g_l\right) \eta ^l\right\rangle =\left\langle \psi _0^k|R_tW_t\left(
g_k\star g_l\right) \psi _0^l\right\rangle =\left\langle W_t\left(
g_k\right) \psi _0^k|R_tW_t\left( g_l\right) \psi _0^l\right\rangle
\end{equation*}
if $\vartheta _s|\mathfrak{g}_t\leq \left( =\right) \vartheta _t$. It is
normalized, $R_0=F_0^{*}F_0=I$, as $F_0=I$ if $\vartheta _0\left( 0\right)
=I $. 
\endproof%

\section{Generators of quantum CP dynamics}

The quantum stochastically differentiable positive flow $\phi $ is defined
as a weakly continuous function $t\mapsto \phi _t$ with CP values $\phi _t:%
\mathcal{B}\rightarrow \mathfrak{B}_t$, $\phi _0\left( B\right) =B\otimes
I,\forall B\in \mathcal{B}$ such that for any product-vector $\psi _f=\eta
\otimes f^{\otimes }$ given by $\eta \in \mathcal{D}$ and $f^{\bullet }\in 
\mathfrak{E}$, 
\begin{equation}
\frac{\mathrm{d}}{\mathrm{d}t}\left\langle \psi _f|\phi _t\left( B\right)
\psi _f\right\rangle =\left\langle \psi _f|\phi _t\left( \lambda \left( \bar{%
f}^{\bullet }\left( t\right) ,B,f^{\bullet }\left( t\right) \right) \right)
\psi _f\right\rangle ,\qquad B\in \mathcal{B},  \label{2.1}
\end{equation}
where $\lambda \left( \bar{e}^{\bullet },B,e^{\bullet }\right) =$ $\lambda
\left( B\right) +e_{\bullet }\lambda ^{\bullet }\left( B\right) +\lambda
_{\bullet }\left( B\right) e^{\bullet }+e_{\bullet }\lambda _{\bullet
}^{\bullet }\left( B\right) e^{\bullet }$, $e_{\bullet }=\bar{e}^{\bullet }$
is the linear form on $\mathcal{E}$ with $e_{\bullet }^{*}=e^{\bullet }\in 
\mathcal{E}$ and $\left\langle \psi _f|\phi _0\left( B\right) \psi
_f\right\rangle =\left\langle \eta |B\eta \right\rangle \exp \left\|
f^{\bullet }\right\| ^2$. The generator $\lambda \left( B\right) =\lambda
\left( 0,B,0\right) $ of the quantum dynamical semigroup $\theta _t=\epsilon
\circ \phi _t\,$ is a linear w*-continuous map $B\mapsto \lambda \left(
B\right) \in \mathcal{A}^c$, $\lambda ^{\bullet }=\lambda _{\bullet
}^{\dagger }$ is a linear w*-continuous map given by the Hermitian adjoint
values $\lambda _{\bullet }\left( B^{*}\right) =\lambda ^{\bullet }\left(
B\right) ^{\dagger }$ in the continuous operators $\mathcal{E}\rightarrow 
\mathcal{A}^c$, and $\lambda _{\bullet }^{\bullet }:\mathcal{B}\rightarrow 
\mathcal{B}\left( \mathcal{D}\otimes \mathcal{E}\right) $is a w*-continuous
map with the values $\lambda _{\bullet }^{\bullet }\left( B\right) $ given
by continuous operators $\mathcal{E}\otimes \mathcal{E}\rightarrow \mathcal{A%
}^c$.

1. The differential evolution equation (\ref{2.1}) for the coherent vector
matrix elements $\left\langle \psi _f|\phi _t\left( B\right) \psi
_f\right\rangle $ corresponds to the It\^{o} form \cite{16} of the quantum
stochastic equation 
\begin{equation}
\mathrm{d}\phi _t\left( B\right) =\phi _t\circ \lambda _\nu ^\mu \left(
B\right) \mathrm{d}\Lambda _\mu ^\nu :=\sum_{\mu ,\nu }\phi _t\left( \lambda
_\nu ^\mu \left( B\right) \right) \mathrm{d}\Lambda _\mu ^\nu ,\qquad \text{ 
}B\in \mathcal{B}\qquad  \label{2.2}
\end{equation}
with the initial condition $\phi _0\left( B\right) =B$, for all $B\in 
\mathcal{B}$. Here $\lambda _\nu ^\mu $ are the flow generators $\lambda
_{+}^{-}=\lambda $, $\lambda _{+}^{\bullet }=\lambda ^{\bullet }$, $\lambda
_{\bullet }^{-}=\lambda _{\bullet }$, $\lambda _{\bullet }^{\bullet }$,
called the structural maps, and the summation is taken over the indices $\mu
=-,\bullet $, $\nu =+,\bullet $ of the standard quantum stochastic
integrators $\Lambda _\mu ^\nu $. For simplicity we shall assume that the
pre-Hilbert Fr\'{e}chet space $\mathcal{E}$ is separable, $\mathcal{E}%
\subseteq $ $\ell ^2$. Then the index $\bullet $ can take any value in $%
\left\{ 1,2,...\right\} $ and $\Lambda _\mu ^\nu \left( t\right) $ are
indexed with $\mu \in \left\{ -,1,2,...\right\} $, $\nu \in \left\{
+,1,2,...\right\} $ as the standard time $\Lambda _{-}^{+}\left( t\right) =t%
\mathrm{I}$, annihilation $\Lambda _{-}^m\left( t\right) $, creation $%
\Lambda _n^{+}\left( t\right) $ and exchange-number $\Lambda _n^m\left(
t\right) $ operator integrators with $m,n\in \mathbb{N}$. The infinitesimal
increments $\mathrm{d}\Lambda _\nu ^\mu \left( t\right) =\Lambda _\nu ^{t\mu
}\left( \mathrm{d}t\right) $ are formally defined by the HP multiplication
table \cite{16} and the $\star $ -property \cite{3}, 
\begin{equation}
\mathrm{d}\Lambda _\mu ^\alpha \mathrm{d}\Lambda _\beta ^\nu =\delta _\beta
^\alpha \mathrm{d}\Lambda _\mu ^\nu ,\qquad \text{ }\Lambda ^{\star
}=\Lambda ,\qquad  \label{2.3}
\end{equation}
where $\delta _\beta ^\alpha $ is the usual Kronecker delta restricted to
the indices $\alpha \in \left\{ -,1,2,...\right\} $, $\beta \in \left\{
+,1,2,...\right\} $ and $\Lambda _{-\nu }^{\star \mu }=\Lambda _{-\mu }^{\nu
*}$ with respect to the reflection $-(-)=+,$ $-(+)=-$ of the indices $\left(
-,+\right) $ only.

The linear equation (\ref{2.2}) of a particular type, (quantum Langevin
equation) with bounded finite-dimensional structural maps $\lambda _\nu ^\mu 
$ was introduced by Evans and Hudson \cite{EvH} in order to describe the $*$%
-homomorphic quantum stochastic evolutions. The constructed quantum
stochastic $*$-homomorphic flow (EH-flow) is identity preserving and is
obviously completely positive, but it is hard to prove these algebraic
properties for the unbounded case. However the typical quantum filtering
dynamics is not homomorphic or identity preserving, but it is completely
positive and in the most interesting cases is described by unbounded
generators $\lambda _\nu ^\mu $. In the general content Eq. (\ref{2.2}) was
studied in \cite{20}, and the correspondent quantum stochastic, not
necessarily homomorphic and normalized flow was constructed even for the
infinitely-dimensional non-adapted case under the natural integrability
condition for the chronological products of the generators $\lambda _\nu
^\mu $ in the norm scale (\ref{A.2}). The EH flows with unbounded $\lambda
_\nu ^\mu $, satisfying certain analyticity conditions, have been recently
constructed in the strong sense by Fagnola-Sinha in \cite{18} for the
non-Hilbert class $L^\infty $ of test functions $f^{\bullet }$. Another type
of sufficient analyticity conditions, which is related to the Hilbert scales
of the test functions, is given in the Appendix. Here we will formulate the
necessary differential conditions which follow from the complete positivity,
causality, and martingale properties of the filtering flows, and which are
sufficient for the construction of the quantum stochastic flows obeying
these properties in the case of the bounded $\lambda _\nu ^\mu $. As we
showed in \cite{15}, the found properties are sufficient to define the
general structure of the bounded generators, and this structure will help us
in construction of the minimal completely positive weak solutions for the
quantum filtering equations also with unbounded $\lambda _\nu ^\mu $.

2. Obviously the linear w*-continuous generators $\lambda _\nu ^\mu :%
\mathcal{B}\rightarrow \mathcal{A}^c$ for CP flows $\phi _t^{*}=\phi _t$,
where $\phi _t^{*}\left( B\right) =\phi _t\left( B^{*}\right) ^{\dagger }$,
must satisfy the $\star $ -property $\lambda ^{\star }=\lambda $, where $%
\lambda _{-\mu }^{\star \nu }=\lambda _{-\nu }^{\mu *}$, $\lambda _\nu ^{\mu
*}\left( B\right) =\lambda _\nu ^\mu \left( B^{*}\right) ^{*}$ and are
independent of $t$, corresponding to cocycle property $\phi _s\circ \phi
_r^s=\phi _{s+r}$, where $\phi _t^s$ is the solution to (\ref{2.2}) with $%
\Lambda _\nu ^\mu \left( t\right) $ replaced by $\Lambda _\nu ^{s\mu }\left(
t\right) $, and $\lambda _{+}^{-}\left( I\right) =0$ if $\phi $ is a
filtering flow, $\phi _t\left( I\right) =I$, as it is in the multiplicative
case \cite{EvH}. We shall assume that $\boldsymbol{\lambda }=\left( \lambda
_\nu ^\mu \right) _{\nu =+,\bullet }^{\mu =-,\bullet }$ for each $B^{*}=B$
defines a continuous Hermitian form $\boldsymbol{b}=\boldsymbol{\lambda }%
\left( B\right) $ on the Fr\'{e}chet space $\mathcal{D}\oplus \mathcal{D}%
_{\bullet }$, 
\begin{equation*}
\left\langle \boldsymbol{\eta }\right| \boldsymbol{b}\left. \boldsymbol{\eta 
}\right\rangle =\sum_{m,n}\left\langle \eta ^m|b_n^m\eta ^n\right\rangle
+\sum_m\left\langle \eta ^m|b_{+}^m\eta \right\rangle +\sum_n\left\langle
\eta |b_n^{-}\eta ^n\right\rangle +\left\langle \eta |b_{+}^{-}\eta
\right\rangle ,
\end{equation*}
where $\eta \in \mathcal{D}$, $\eta ^{\bullet }=\left( \eta ^m\right) ^{m\in 
\mathbb{N}}\in \mathcal{D}_{\bullet }=\mathcal{D}\otimes \mathcal{E}$. We
say that an It\^{o} algebra $\mathfrak{a}$ , represented on $\mathcal{E}$,
commutes in HP sense with a $\boldsymbol{b}$, given by the form-generator $%
\boldsymbol{\lambda }$ if $\left( I\otimes a_{\bullet }^\mu \right) b_\nu
^{\bullet }=b_{\bullet }^\mu \left( I\otimes a_\nu ^{\bullet }\right) $ (For
simplicity the ampliation $I\otimes a_\nu ^\mu $ will be written again as $%
a_\nu ^\mu $.) Note that if we define the matrix elements $a_\nu ^\mu $, $%
b_\nu ^\mu $ also for $\mu =+$ and $\nu =-$, by the extension 
\begin{equation*}
a_\nu ^{+}=0=a_{-}^\mu ,\qquad \lambda _\nu ^{+}\left( B\right) =0=\lambda
_{-}^\mu \left( B\right) ,\quad \forall a\in \mathfrak{a},B\in \mathcal{B},
\end{equation*}
the HP product (\ref{0.4}) of $\boldsymbol{a}$ and $\boldsymbol{b}$ can be
written in terms of the usual matrix product $\mathbf{ab}=\left[ a_\lambda
^\mu b_\nu ^\lambda \right] $ of the extended quadratic matrices $\mathbf{a}=%
\left[ a_\nu ^\mu \right] _{\nu =-,\bullet ,+}^{\mu =-,\bullet ,+}$ and $%
\mathbf{b=}\boldsymbol{b}\mathbf{g}$, where $\mathbf{g}=\left[ \delta _{-\nu
}^\mu \right] $. Then one can extend the summation in (\ref{2.2}) so it is
also over $\mu =+$, and $\nu =-$, such that $b_\nu ^\mu \mathrm{d}\Lambda
_\mu ^\nu $ is written as the trace $\mathbf{b\cdot }\mathrm{d}\boldsymbol{%
\Lambda }$ over all $\mu ,\nu $. By such an extension the multiplication
table for $\mathrm{d}\Lambda \left( a\right) =\mathbf{a\cdot }\mathrm{d}%
\boldsymbol{\Lambda }$ , $\mathrm{d}\Lambda \left( b\right) =\mathbf{b\cdot }%
\mathrm{d}\boldsymbol{\Lambda }$ can be represented as $\mathrm{d}\Lambda
\left( a\right) \mathrm{d}\Lambda \left( b\right) =\mathbf{ab\cdot }\mathrm{d%
}\boldsymbol{\Lambda }$, and the involution $\mathbf{b\mapsto b}^{\star }$,
defining $\mathrm{d}\Lambda \left( b\right) ^{\dagger }=\mathbf{b}^{\star }%
\mathbf{\cdot }\mathrm{d}\boldsymbol{\Lambda }$, can be obtained by the
pseudo-Hermitian conjugation $b_\alpha ^{\star \nu }=g_{\alpha \mu }b_\beta
^{\mu *}g^{\beta \nu }$ respectively to the indefinite Minkowski metric
tensor $\mathbf{g}=\left[ g_{\mu \nu }\right] $ and its inverse $\mathbf{g}%
^{-1}=\left[ g^{\mu \nu }\right] $, given by $g^{\mu \nu }=\delta _{-\nu
}^\mu I=g_{\mu \nu }$.

Now let us find the differential form of the normalization and causality
conditions with respect to the quantum stationary process, with independent
increments $\mathrm{d}X\left( t\right) =X\left( t+\Delta \right) -X\left(
s\right) $ generated by an It\^{o} algebra $\mathfrak{a}$ on the separable
space $\mathcal{E}$.

\begin{proposition}
Let $\phi $ be a flow, satisfying the quantum stochastic equation (\ref{2.2}%
), and $\left[ W_t\left( g\right) ,\phi _t\left( B\right) \right] =0$ for
all $g\in \mathfrak{g},B\in \mathcal{B}$. Then the coefficients $b_\nu ^\mu
=\lambda _\nu ^\mu \left( B\right) $, $\mu =-,\bullet $, $\nu =+,\bullet $,
where $\bullet =1,2,...$, written in the matrix form $\boldsymbol{b}=\left(
b_\nu ^\mu \right) _{\nu =+,\bullet }^{\mu =-,\bullet }$, commute in the
sense of the HP product with $\boldsymbol{a}=\left( a_\nu ^\mu \right) _{\nu
=+,\bullet }^{\mu =-,\bullet }$ for all $a\in \mathfrak{a}$ and $B\in 
\mathcal{B} $: 
\begin{equation}
\left[ \boldsymbol{a},\boldsymbol{b}\right] :=\left( a_{\bullet }^\mu b_\nu
^{\bullet }-b_{\bullet }^\mu a_\nu ^{\bullet }\right) _{\nu =+,\bullet
}^{\mu =-,\bullet }=0.  \label{2.4}
\end{equation}
\end{proposition}

\proof%
Since $\epsilon _t\left( \phi _s\left( I\right) -\phi _t\left( I\right)
\right) $ is a negative Hermitian form, 
\begin{equation*}
\epsilon _t\left( \mathrm{d}\phi _t\left( I\right) \right) =\epsilon
_t\left( \phi _t\left( \lambda _\nu ^\mu \left( I\right) \right) \mathrm{d}%
\Lambda _\mu ^\nu \right) =\phi _t\left( \lambda _{+}^{-}\left( I\right)
\right) \mathrm{d}t\leq 0.
\end{equation*}
Since $Y_t=\phi _t\left( B\right) $ commutes with $W_t\left( g\right) $ for
all $B$ and $g\left( t\right) =a$, we have by virtue of quantum It\^{o}'s
formula 
\begin{equation*}
\mathrm{d}\left[ Y_t,W_t\right] =\left[ \mathrm{d}Y_t,W_t\right] +\left[ Y_t,%
\mathrm{d}W_t\right] +\left[ \mathrm{d}Y_t,\mathrm{d}W_t\right] =0.
\end{equation*}
Equations (\ref{2.1a}), (\ref{2.2}) and commutativity of $a_\nu ^\mu $ with $%
Y_t$ and $W_t$ imply 
\begin{eqnarray*}
&&\left( \left[ \phi _t\left( b_\nu ^\mu \right) ,W_t\right] +\left[
Y_t,a_\nu ^\mu W_t\right] +\phi _t\left( b_{\bullet }^\mu \right) a_\nu
^{\bullet }W_t-a_{\bullet }^\mu W_t\phi _t\left( b_\nu ^{\bullet }\right)
\right) \mathrm{d}\Lambda _\mu ^\nu \\
&=&W_t\left( \phi _t\left( b_{\bullet }^\mu \right) a_\nu ^{\bullet
}-a_{\bullet }^\mu \phi _t\left( b_\nu ^{\bullet }\right) \right) \mathrm{d}%
\Lambda _\nu ^\mu =W_t\phi _t\left( b_{\bullet }^\mu a_\nu ^{\bullet
}-a_{\bullet }^\mu b_\nu ^{\bullet }\right) \mathrm{d}\Lambda _{\mu .}^\nu
=0.
\end{eqnarray*}
Thus $\boldsymbol{a}\bullet \boldsymbol{b}=\boldsymbol{b}\bullet \boldsymbol{%
a}$ by the argument \cite{Par}of independence of the integrators $\mathrm{d}%
\Lambda _\mu ^\nu $. 
\endproof%

3. In order to formulate the CP differential condition we need the notion of 
\emph{quantum stochastic germ} for the CP flow $\phi $ at $t=0$. It was
defined in \cite{20, Bge}, for a quantum stochastic differential (\ref{2.2})
with $\phi _0\left( B\right) =B,\forall B\in \mathcal{B}$ as $\gamma _\nu
^\mu =\lambda _\nu ^\mu +\imath _\nu ^\mu $, where $\lambda _\nu ^\mu $ are
the structural maps $B\mapsto \lambda _\nu ^\mu \left( B\right) $ given by
the generators of the quantum It\^{o} equation (\ref{2.2}) and $\imath _\nu
^\mu :B\mapsto B\delta _\nu ^\mu $ is the ampliation of $\mathcal{B}$. Let
us prove that the germ-maps $\gamma _\nu ^\mu $ of a CP flow $\phi $ must be
conditionally completely positive (CCP) in a degenerated sense as it was
found for the finite-dimensional bounded case in \cite{15, Be96}.

\begin{theorem}
If $\phi $ is a completely positive flow satisfying the quantum stochastic
equation (\ref{2.2}) with $\phi _0\left( B\right) =B$, then the germ-matrix $%
\boldsymbol{\gamma }=\left( \lambda _\nu ^\mu +\imath _\nu ^\mu \right)
_{\nu =+,\bullet }^{\mu =-,\bullet }$ is conditionally completely positive
in the sense 
\begin{equation*}
\sum_{B\in \mathcal{B}}\boldsymbol{\iota }\left( B\right) \boldsymbol{\zeta }%
_B=0\Rightarrow \sum_{B,C\in \mathcal{B}}\langle \boldsymbol{\zeta }_B|%
\boldsymbol{\gamma }\left( B^{*}C\right) \boldsymbol{\zeta }_C\rangle \geq 0.
\end{equation*}
Here $\boldsymbol{\zeta }\in \mathcal{D}\oplus \mathcal{D}_{\bullet },%
\mathcal{D}_{\bullet }=\mathcal{D}\otimes \mathcal{E}$, and $\boldsymbol{%
\iota }=\left( \iota _\nu ^\mu \right) _{\nu =+,\bullet }^{\mu =-,\bullet }$
is the degenerate representation $\iota _\nu ^\mu \left( B\right) =B\delta
_\nu ^{+}\delta _{-}^\mu $, written both with $\boldsymbol{\gamma }$ in the
matrix form as 
\begin{equation}
\boldsymbol{\gamma }=\left( 
\begin{array}{cc}
\gamma & \gamma _{\bullet } \\ 
\gamma ^{\bullet } & \gamma _{\bullet }^{\bullet }%
\end{array}
\right) ,\qquad \text{ }\boldsymbol{\iota }\left( B\right) =\left( 
\begin{array}{cc}
B & 0 \\ 
0 & 0%
\end{array}
\right) ,\qquad  \label{2.5}
\end{equation}
where $\gamma =\lambda _{+}^{-},\quad $ $\gamma ^m=\lambda _{+}^m,\quad $ $%
\gamma _n=\lambda _n^{-},\quad \gamma _n^m=\imath _n^m+\lambda _n^m$ with $%
\imath _n^m\left( B\right) =B\delta _n^m$ such that 
\begin{equation}
\gamma \left( B^{*}\right) =\gamma \left( B\right) ^{*},\qquad \text{ }%
\gamma ^n\left( B^{*}\right) =\gamma _n\left( B\right) ^{*},\qquad \text{ }%
\gamma _n^m\left( B^{*}\right) =\gamma _m^n\left( B\right) ^{*}.  \label{2.6}
\end{equation}
If $\phi $ is subfiltering, then $D=-\lambda _{+}^{-}\left( I\right) $ is a
positive Hermitian form, $\left\langle \eta |D\eta \right\rangle \geq 0$,
for all $\eta \in \mathcal{D}$, and if $\phi $ is contractive, then $%
\boldsymbol{D}=-\boldsymbol{\lambda }\left( I\right) $ is positive in the
sense $\langle \boldsymbol{\eta }|\boldsymbol{D}\boldsymbol{\eta }\rangle
\geq 0$ for all $\boldsymbol{\eta }\in \mathcal{D}\oplus \mathcal{D}%
_{\bullet }$.
\end{theorem}

\proof%
The CP condition in the form (\ref{cpd}) for the adapted map $\phi _t$ can
be obviously extended on all $f^{\bullet }\in \mathfrak{E}$ if the
sesquianalytical function $f^{\bullet }\mapsto \phi _t\left( \bar{f}%
^{\bullet },B,f^{\bullet }\right) $ is defined as the $\mathfrak{E}$%
-function 
\begin{equation}
\left\langle \eta |\phi _t\left( \bar{f}^{\bullet },B,f^{\bullet }\right)
\eta \right\rangle =\left\langle \eta \otimes f^{\otimes }|\phi _t\left(
B\right) \eta \otimes f^{\otimes }\right\rangle \exp \left[ -\int_t^\infty
\left\| f^{\bullet }\left( s\right) \right\| ^2\mathrm{d}s\right] ,
\label{2.7}
\end{equation}
where $\left\| f^{\bullet }\left( t\right) \right\| ^2=\sum_{n=1}^\infty
\left| f^n\left( t\right) \right| ^2$. It coincides with the former
definition on $\mathfrak{E}_t$ and does not depend on $f^{\bullet }\left(
s\right) $, $s>t$ due to the adaptiveness (\ref{adp}) of $Y_t=\phi _t\left(
B\right) $. If the $\mathfrak{D}$-form $\phi _t\left( B\right) $ satisfies
the stochastic equation (\ref{2.2}), the $\mathcal{D}$-form $\phi _t\left( 
\bar{f}^{\bullet },B,f^{\bullet }\right) $ satisfies the differential
equation \cite{16} 
\begin{equation*}
\frac{\mathrm{d}}{\mathrm{d}t}\phi _t\left( \bar{f}^{\bullet },B,f^{\bullet
}\right) =\left\| f^{\bullet }\left( t\right) \right\| ^2\phi _t\left( \bar{f%
}^{\bullet },B,f^{\bullet }\right) +\phi _t\left( \bar{f}^{\bullet },\lambda
_{+}^{-}\left( B\right) ,f^{\bullet }\right)
\end{equation*}
\begin{eqnarray*}
&&+\sum_{m=1}^\infty \overline{f}^m\left( t\right) \phi _t\left( \bar{f}%
^{\bullet },\lambda _{+}^m\left( B\right) ,f^{\bullet }\right)
+\sum_{n=1}^\infty \phi _t\left( \bar{f}^{\bullet },\lambda _n^{-}\left(
B\right) ,f^{\bullet }\right) f^n\left( t\right) \\
&&+\sum_{m,n=1}^\infty \overline{f}^m\left( t\right) \phi _t\left( \bar{f}%
^{\bullet },\lambda _n^m\left( B\right) ,f^{\bullet }\right) f^n\left(
t\right) =\phi _t\left( \bar{f}^{\bullet },\gamma \left( \bar{f}^{\bullet
}\left( t\right) ,B,f^{\bullet }\left( t\right) \right) ,f^{\bullet }\right)
.
\end{eqnarray*}
The positive definiteness of (\ref{2.7}) ensures the conditional positive
definiteness\newline
$\sum_f\sum_BB\xi _B^f=0\Rightarrow $ 
\begin{equation*}
\sum_{B,C}\sum_{f,h}\left\langle \xi _B^f\right| \left. \gamma _t\left( \bar{%
f}^{\bullet },B^{*}C,h^{\bullet }\right) \xi _C^h\right\rangle =\frac
1t\sum_{B,C}\sum_{f,h}\left\langle \xi _B^f\right| \left. \phi _t\left( \bar{%
f}^{\bullet },B^{*}C,h^{\bullet }\right) \xi _C^h\right\rangle \geq 0
\end{equation*}
of the form, given by $\gamma _t\left( \bar{f}^{\bullet },B,f^{\bullet
}\right) =\frac 1t\left( \phi _t\left( \bar{f}^{\bullet },B,f^{\bullet
}\right) -B\right) $ for each $t>0$. This holds also at the limit $\gamma
_0\left( \bar{f}^{\bullet },B,f^{\bullet }\right) =\gamma \left( \bar{f}%
^{\bullet }\left( 0\right) ,B,f^{\bullet }\left( 0\right) \right) $, given
at $t\downarrow 0$ by the $\mathcal{E}$-form 
\begin{equation*}
\gamma \left( \bar{e}^{\bullet },B,e^{\bullet }\right) =\sum_{m,n}\bar{e}%
^m\gamma _n^m\left( B\right) e^n+\sum_m\bar{e}^m\gamma ^m\left( B\right)
+\sum_n\gamma _n\left( B\right) e^n+\gamma \left( B\right) ,
\end{equation*}
where $e^{\bullet }=f^{\bullet }\left( 0\right) \in \mathcal{E}$, $\bar{e}%
^{\bullet }=e_{\bullet }$ and the $\gamma $'s are defined in (\ref{2.5}).
Hence the form 
\begin{equation*}
\sum_{B,C}\sum_{\mu ,\nu }\left\langle \zeta _B^\mu \right| \gamma _\nu ^\mu
\left( B^{*}C\right) \zeta _C^\nu \rangle :=\sum_{B,C}\sum_{m,n}\left\langle
\zeta _B^m\right| \left. \gamma _n^m\left( B^{*}C\right) \zeta
_C^n\right\rangle
\end{equation*}
\begin{equation*}
+\sum_{B,C}\left( \sum_n\left\langle \zeta _B\right| \left. \gamma _n\left(
B^{*}C\right) \zeta _C^n\right\rangle +\sum_m\left\langle \zeta _B^m\right|
\left. \gamma ^m\left( B^{*}C\right) \zeta _C\right\rangle +\left\langle
\zeta _B\right| \gamma \left( B^{*}C\right) \left| \zeta _C\right\rangle
\right)
\end{equation*}
with $\zeta =\sum_f\xi ^f,\quad \zeta ^{\bullet }=\sum_f\xi ^f\otimes
e_f^{\bullet }$, where $e_f^{\bullet }=f^{\bullet }\left( 0\right) $, is
positive if $\sum_BB\zeta _B=0$. The components $\zeta $ and $\zeta
^{\bullet }$ of these vectors are independent because for any $\zeta \in 
\mathcal{D}$ and $\zeta ^{\bullet }=\left( \zeta ^1,\zeta ^2,...\right) \in 
\mathcal{D}\otimes \mathcal{E}$ there exists such a function $e^{\bullet
}\mapsto \xi ^e$ on $\mathcal{E}$ with a countable support, that $\sum_e\xi
^e=\zeta ,\quad \sum_e\xi ^e\otimes e^{\bullet }=\zeta ^{\bullet }$, namely, 
$\xi ^e=0$ for all $e^{\bullet }\in \mathcal{E}$ except $e^{\bullet }=0$
with $\xi ^0=\zeta -\sum_{n=1}^\infty \zeta ^n$ and $e^{\bullet
}=e_n^{\bullet }$, the $n$-th basis element in $\mathcal{\ell }^2$, for
which $\xi ^e=\zeta ^n$. This proves the complete positivity of the matrix
form $\boldsymbol{\gamma }$, with respect to the matrix representation $%
\boldsymbol{\iota }$ defined in (\ref{2.5}) on the ket-vectors $\boldsymbol{%
\zeta }=\left( \zeta ^\mu \right) $.

If $\epsilon \left( R_t\right) \leq I$, then $D=-\lambda \left( I\right)
=\lim \frac 1t\epsilon \left( I-R_t\right) \geq 0$, and we also conclude the
dissipativity $\sum_{k,l}\left\langle \xi ^k|D\left( \bar{k}^{\bullet
},l^{\bullet }\right) \xi ^l\right\rangle \geq 0$ from 
\begin{equation*}
0\leq \lim \frac 1t\sum_{f,h}\left\langle \xi ^f|e^{\int_0^tf_{\bullet
}h^{\bullet }}I-\phi _t\left( \bar{f}^{\bullet },I,h^{\bullet }\right) \xi
^h\right\rangle =-\left\langle \xi ^f|\lambda \left( \bar{f}^{\bullet
}\left( 0\right) ,I,h^{\bullet }\left( 0\right) \right) \xi ^h\right\rangle
\end{equation*}
if $\phi _t\left( I\right) \leq I$, where $\lambda \left( \bar{e}^{\bullet
},I,e^{\bullet }\right) =\gamma \left( \bar{e}^{\bullet },I,e^{\bullet
}\right) -\left\| e^{\bullet }\right\| ^2I=$ $D\left( \bar{e}^{\bullet
},e^{\bullet }\right) $. 
\endproof%

4. Obviously the CCP property for the germ-matrix $\boldsymbol{\gamma }$ is
invariant under the transformation $\boldsymbol{\gamma }\mapsto \boldsymbol{%
\varphi }$ given by 
\begin{equation}
\boldsymbol{\varphi }\left( B\right) =\boldsymbol{\gamma }\left( B\right) +%
\boldsymbol{\iota }\left( B\right) \boldsymbol{K}+\boldsymbol{K}^{*}%
\boldsymbol{\iota }\left( B\right) ,  \label{2.8}
\end{equation}
where $\boldsymbol{K}=\left( K_\nu ^\mu \right) _{\nu =+,\bullet }^{\mu
=-,\bullet }$ is an arbitrary matrix of $K_\nu ^\mu \in \mathcal{L}\left( 
\mathcal{D}\right) $ with $K_{-\nu }^{*\mu }=K_{-\mu }^{\nu *}$. As was
proven in \cite{15, Be96} for the case of a finite-dimensional matrix $%
\boldsymbol{\gamma }$ of bounded $\gamma _\nu ^\mu $, see also \cite{LiP},
the matrix elements $K_\nu ^{-}$ can be chosen in such way that the matrix
map $\boldsymbol{\varphi }=\left( \varphi _\nu ^\mu \right) _{\nu =+,\bullet
}^{\mu =-,\bullet }$ becomes CP from $\mathcal{B}$ into the quadratic
matrices of $\varphi _\nu ^\mu \left( B\right) .$ (The other elements can be
chosen arbitrarily, say as $K_{+}^{\bullet }=0$, $K_{\bullet }^{\bullet
}=\frac 12I_{\bullet }^{\bullet }$, because (\ref{2.8}) does not depend on $%
K_{+}^{\bullet },K_{\bullet }^{\bullet }$.) Thus the generator $\boldsymbol{%
\lambda }=\boldsymbol{\gamma }-\boldsymbol{\imath }$ for a quantum
stochastic CP flow $\phi $ can be written (at least in the bounded case) as $%
\boldsymbol{\varphi }-\imath \boldsymbol{K}-\boldsymbol{K}^{*}\imath $: 
\begin{equation}
\lambda _\nu ^\mu \left( B\right) =\varphi _\nu ^\mu \left( B\right)
-B\left( \tfrac 12\delta _\nu ^\mu I+\delta _{-}^\mu K_\nu \right) -\left(
\tfrac 12\delta _\nu ^\mu I+K^\mu \delta _\nu ^{+}\right) B,  \label{2.10}
\end{equation}
where $\varphi _\nu ^\mu :\mathcal{B}\rightarrow \mathcal{B}\left( \mathcal{D%
}\right) $ are matrix elements of the CP map $\boldsymbol{\varphi }$ and $%
K_\nu \in \mathcal{L}\left( \mathcal{D}\right) $, $K^{-}=K_{+}^{*}$, $%
K^m=K_m^{*}$. Now we show that the germ-matrix of this form obeys the CCP
property even in the general case of unbounded $K_\nu ^{-},$ $\varphi _\nu
^\mu \left( B\right) \in \mathcal{B}\left( \mathcal{D}\right) $.

\begin{proposition}
The matrix map $\boldsymbol{\gamma }=\left( \gamma _\nu ^\mu \right) _{\nu
=+,\bullet }^{\mu =-,\bullet }$ given in (\ref{2.8}) by $\quad $%
\begin{equation}
\boldsymbol{\varphi }=\left( 
\begin{array}{cc}
\varphi & \varphi _{\bullet } \\ 
\varphi ^{\bullet } & \varphi _{\bullet }^{\bullet }%
\end{array}
\right) ,\quad \mathrm{and\quad }\boldsymbol{K}=\left( 
\begin{array}{cc}
K & K_{\bullet } \\ 
0 & \frac 12I_{\bullet }^{\bullet }%
\end{array}
\right) ,\;\boldsymbol{K}^{*}=\left( 
\begin{array}{cc}
K^{*} & 0 \\ 
K_{\bullet }^{*} & \frac 12I_{\bullet }^{\bullet }%
\end{array}
\right) ,  \label{2.9}
\end{equation}
with $\varphi =\varphi _{+}^{-},\quad \varphi ^m=\varphi _{+}^m,\quad
\varphi _n=\varphi _n^{-}$ and $\varphi _n^m=\gamma _n^m$ is CCP with
respect to the degenerate representation $\boldsymbol{\iota }=\left( \delta
_{-}^\mu \delta _\nu ^{+}\iota \right) _{\nu =+,\bullet }^{\mu =-,\bullet }$%
, where $\iota \left( B\right) =B$, if $\boldsymbol{\varphi }$ is a CP map.
\end{proposition}

\proof%
If $\boldsymbol{\iota }\left( B_k\right) \boldsymbol{\eta }^k=0$, then 
\begin{equation*}
\langle \boldsymbol{\eta }^k|\boldsymbol{\iota }\left( B_k^{*}B_l\right) 
\boldsymbol{K}+\boldsymbol{K}^{*}\boldsymbol{\iota }\left( B_k^{*}B_l\right) 
\boldsymbol{\eta }^l\rangle
\end{equation*}
\begin{equation*}
=2\func{Re}\left\langle \boldsymbol{\iota }\left( B_k\right) \boldsymbol{%
\eta }^k|\boldsymbol{\iota }\left( B_l\right) \boldsymbol{K}\boldsymbol{\eta 
}^l\right\rangle =0.
\end{equation*}
Hence the CCP for $\boldsymbol{\gamma }$ is equivalent to the CCP property
for (\ref{2.8}) and follows from its CP property: 
\begin{equation*}
\left\langle \boldsymbol{\eta }^k|\boldsymbol{\gamma }\left(
B_k^{*}B_l\right) \boldsymbol{\eta }^l\right\rangle =\left\langle 
\boldsymbol{\eta }^k|\boldsymbol{\varphi }\left( B_k^{*}B_l\right) 
\boldsymbol{\eta }^l\right\rangle \geq 0
\end{equation*}
for such sequences $\boldsymbol{\eta }^k\in \mathcal{D}\oplus \mathcal{D}%
_{\bullet }$.

\section{Construction of quantum CP flows}

The necessary conditions for the stochastic generator $\boldsymbol{\lambda }%
=\left( \lambda _\nu ^\mu \right) _{\nu =+,\bullet }^{\mu =-,\bullet }$ of a
CP flow $\phi $ at $t=0$ are found in the previous section in the form of a
CCP property for the corresponding germ $\boldsymbol{\gamma }=\left( \gamma
_\nu ^\mu \right) _{\nu =+,\bullet }^{\mu =-,\bullet }$. In the next section
we shall show, these conditions are essentially equivalent to the assumption
(\ref{2.10}), corresponding to 
\begin{equation}
\gamma ^m\left( B\right) =\varphi ^m\left( B\right) -K_m^{*}B=\gamma
_m^{*}\left( B\right) ,\quad \gamma \left( B\right) =\varphi \left( B\right)
-K^{*}B-BK,  \label{3.0'}
\end{equation}
where $\boldsymbol{\varphi }=\left( \varphi _\nu ^\mu \right) _{\nu
=+,\bullet }^{\mu =-,\bullet }$ is a CP map with $\varphi _n^m=\gamma _n^m$.
Here we are going to prove under the following conditions for the operators $%
K,K_{\bullet }$ and the maps $\varphi _\nu ^\mu $ that this general form is
also sufficient for the existence of the CP solutions to the quantum
stochastic equation (\ref{2.2}). We are going to construct the minimal
quantum stochastic positive flow $B\mapsto \phi _t\left( B\right) $ for a
given w*-continuous unbounded germ-matrix map of the above form, satisfying
the following conditions.

\begin{enumerate}
\item First, we suppose that the operator $K\in \mathcal{B}\left( \mathcal{D}%
\right) $ generates the one parametric semigroup $\left( e^{-Kt}\right)
_{t>0}$, $e^{-Kr}e^{-Ks}=e^{-K\left( r+s\right) }$ of continuous operators $%
e^{-Kt}\in \mathcal{L}\left( \mathcal{D}\right) $ in the strong sense 
\begin{equation*}
\lim_{t\searrow 0}\frac 1t\left( I-e^{-Kt}\right) \eta =K\eta ,\quad \forall
\eta \in \mathcal{D}.
\end{equation*}
(A contraction semigroup on the Hilbert space $\mathcal{H}$ if $K$ defines
an accretive $K+K^{\dagger }\geq 0$ and so maximal accretive form.)

\item Second, we suppose that the solution $S_t^n,n\in \mathbb{N}$ to the
recurrence 
\begin{equation*}
S_t^{n+1}=S_t^{\circ }-\int_0^tS_{t-r}^{\circ }\sum_{m=1}^\infty K_mS_r^n%
\mathrm{d}\Lambda _{-}^m,\quad S_t^0=S_t^{\circ },
\end{equation*}
where $S_t^{\circ }=e^{-Kt}\otimes T_t\in \mathcal{L}\left( \mathfrak{D}%
\right) $ is the contraction given by the shift co-isometries $T_t:\mathfrak{%
F}\rightarrow \mathfrak{F}$, strongly converges to a continuous operator $%
S_t\in \mathcal{L}\left( \mathfrak{D}\right) $ at $n\longrightarrow \infty $
for each $t>0$.

\item Third, we suppose that the solution $R_t^n,n\in \mathbb{N}$ to the
recurrence 
\begin{equation*}
R_t^{n+1}=S_t^{*}S_t+\int_0^t\mathrm{d}\Lambda _\mu ^\nu \left(
r,S_r^{*}\varphi _\nu ^\mu \left( R_{t-r}^n\right) S_r\right) ,\quad
R_t^0=S_t^{*}S_t,
\end{equation*}
where the quantum stochastic non-adapted integral is understood in the sense 
\cite{20} (see the Appendix), weakly converges to a continuous form $R_t\in 
\mathcal{B}\left( \mathfrak{D}\right) $ at $n\longrightarrow \infty $ for
each $t>0$.
\end{enumerate}

The first and second assumptions are necessary to define the existence of
the free evolution semigroup $S^{\circ }=\left( S_t^{\circ }\right) _{t>0}$
and its perturbation $S=\left( S_t\right) _{t>0}$ on the product space $%
\mathfrak{D}=\mathcal{D}\otimes \mathfrak{F}$ in the form of multiple
quantum stochastic integral 
\begin{equation}
S_t=S_t^{\circ }+\sum_{n=1}^\infty \left( -1\right) ^n\underset{0<t_1<\ldots
<t_n<t}{\int \cdots \int }K_{m_n}\left( t-t_n\right) \cdot \cdot \cdot
K_{m_1}\left( t_2-t_1\right) S_{t_1}^{\circ }\mathrm{d}\Lambda
_{-}^{m_1}\cdot \cdot \cdot \mathrm{d}\Lambda _{-}^{m_n},  \label{3.5'}
\end{equation}
iterating the quantum stochastic integral equation 
\begin{equation}
S_t=S_t^{\circ }-\int_0^t\sum_{m=1}^\infty K_m\left( t-r\right) S_r\mathrm{d}%
\Lambda _{-}^m,\quad S_0=I,  \label{3.1'}
\end{equation}
where $K_m\left( t\right) =S_t^{\circ }\left( K_m\otimes I\right) $. A
sufficient analyticity condition under which this iteration strongly
coverges in $\mathfrak{D}$ is given in the Appendix. The third assumption
supplies the weak convergence for the series 
\begin{equation}
R_t=S_t^{*}S_t+\sum_{n=1}^\infty \underset{0<t_1<\ldots <t_n<t}{\int \cdots
\int }\mathrm{d}\Lambda _{\mu _1\ldots \mu _n}^{\nu _1\ldots \nu _n}\left(
t_1,\ldots ,t_n,\varphi _{\nu _1...\nu _n}^{\mu _1...\mu _n}\left(
t_1,\ldots ,t_n,S_{t-t_n}^{*}S_{t-t_n}\right) \right)  \label{3.6'}
\end{equation}
of non-adapted n-tuple integrals, i.e. for the multiple quantum stochastic
integral (see the definition in the Appendix) with 
\begin{equation}
\varphi _{\nu _1\ldots \nu _n}^{\mu _1\ldots \mu _n}\left( t_1,\ldots
,t_n\right) =\varphi _{\nu _1\ldots \nu _{n-1}}^{\mu _1\ldots \mu
_{n-1}}\left( t_1,\ldots ,t_{n-1}\right) \circ \varphi _{\nu _n}^{\mu
_n}\left( t_n-t_{n-1}\right) ,  \label{3.9'}
\end{equation}
where$\quad \varphi _\nu ^\mu \left( t,B\right) =S_t^{*}\varphi _\nu ^\mu
\left( B\right) S_t$. A sufficient analyticity condition for this
convergence is also given in the Appendix.

The following theorem gives a characterization of the evolution semigroup $S$
in terms of cocycles with unbounded coefficients, characterized by Fagnola 
\cite{Fgn} in the isometric and unitary case.

\begin{proposition}
Let the family $V^{\circ }=\left( V_t^{\circ }\right) _{t>0}$ be a quantum
stochastic adapted cocycle, $V_r^{\circ }T_sV_s^{\circ }=T_sV_{r+s}^{\circ }$%
, satisfying the HP differential equation 
\begin{equation}
\mathrm{d}V_t^{\circ }+KV_t^{\circ }\mathrm{d}t+\sum_{m=1}^\infty
K_mV_t^{\circ }\mathrm{d}\Lambda _{-}^m+\sum_{n=1}^\infty V_t^{\circ }%
\mathrm{d}\Lambda _n^n=0,\quad V_0^{\circ }=I.  \label{3.2'}
\end{equation}
Then $S_t=T_tV_t^{\circ }$ is a semigroup solution, $S_rS_s=S_{r+s}$ to the
non-adapted integral equation (\ref{3.1'}) such that $S_t\psi _f=S_t\left(
f^{\bullet }\right) \eta \otimes \delta _{\varnothing },\forall \eta \in 
\mathcal{D}$ on $\psi _f=\eta \otimes f^{\otimes }$ with $f^{\bullet }\in 
\mathfrak{E}_t$. Conversely, if $S=\left( S_t\right) _{t>0}$ is the
non-adapted solution (\ref{3.5'}) to the integral equation (\ref{3.1'}),
then $V_t^{\circ }=T_t^{*}S_t$ is the adapted solution to (\ref{3.2'}),
defined as $V_t^{\circ }\psi _f=S_t\left( f^{\bullet }\right) \eta \otimes
f^{\otimes },\forall \eta \in \mathcal{D}$, where $S_t\left( f^{\bullet
}\right) =F^{*}S_tF$ is given by $F\eta =\eta \otimes f^{\otimes }$ with $%
f^{\bullet }\in \mathfrak{E}_t$.
\end{proposition}

\proof%
First let us show that Eq. (\ref{3.2'}) is equivalent to the integral one 
\begin{equation*}
V_t^{\circ }=e^{-Kt}\otimes I_t-\int_0^t\sum_{m=1}^\infty \left( e^{-K\left(
t-r\right) }\otimes I_t\left( r\right) \right) K_mV_r^{\circ }\mathrm{d}%
\Lambda _{-}^m,\quad V_0^{\circ }=I,
\end{equation*}
where $I_t=T_t^{\dagger }T_t$ is the orthoprojector onto $\mathfrak{F}_{[t}$
and $I_t\left( r\right) =\theta ^r\left( I_{t-r}\right) $. Indeed,
multiplying both parts of the integral equation from the left by $e^{K\left(
t-s\right) } $ and differentiating the product $e^{K\left( t-s\right)
}V_t^{\circ }$ at $t=s$, we obtain (\ref{3.2'}) by taking into account that $%
\mathrm{d}I_t+\sum_{n=1}^\infty I_t\mathrm{d}\Lambda _n^n=0$ and $\mathrm{d}%
\Lambda _n^n\mathrm{d}\Lambda _{-}^m=0$. Conversely, the integral equation
can be obtained from (\ref{3.2'}) by the integration: 
\begin{eqnarray*}
V_t^{\circ }-e^{-Kt}\otimes I_t &=&\int_0^t\mathrm{d}\left( \left(
e^{-K\left( t-r\right) }\otimes I_t\left( r\right) \right) V_r^{\circ
}\right) \\
&=&\int_0^t\left( e^{-K\left( t-r\right) }\otimes I_t\left( r\right) \right)
\left( \mathrm{d}V_r^{\circ }+KV_r^{\circ }\mathrm{d}r+V_r^{\circ }\mathrm{d}%
\Lambda _{\bullet }^{\bullet }\right) \\
&=&-\int_0^t\left( e^{-K\left( t-r\right) }\otimes I_t\left( r\right)
\right) K_{\bullet }V_r^{\circ }\mathrm{d}\Lambda _{-}^{\bullet },
\end{eqnarray*}
where we used that $\mathrm{d}I\left( r\right) =I\left( r\right) \mathrm{d}%
\Lambda _{\bullet }^{\bullet }$ and $\mathrm{d}\left( I\left( r\right)
V_r\right) =\mathrm{d}I\left( r\right) V_r+I\left( r\right) \mathrm{d}V_r$
due to the non-adapted It\^{o} formula \cite{20}. The non-adapted equation (%
\ref{3.1'}) is obtained by applying the operator $T_t=T_{t-r}T_r$ to both
parts of this integral equation and taking into account the commutativity of 
$e^{K\left( r-t\right) }K_m$ with $T_r$. Moreover, due to the adaptiveness
of $V_t^{\circ }$, $S_t\psi _f=T_t\left( E_tV_t^{\circ }\psi _f\otimes
E_{[t}f^{\otimes }\right) =S_t\left( f^{\bullet }\right) \eta \otimes
f_t^{\otimes }$, where $f_t^{\otimes }=T_tf^{\otimes }$, and $S_t\left(
f^{\bullet }\right) =EV_t^{\circ }F$ is the solution to the equation 
\begin{equation*}
S_t\left( f^{\bullet }\right) =e^{-Kt}+\int_0^te^{-K\left( t-r\right)
}K_{\bullet }f^{\bullet }\left( r\right) S_r\left( f^{\bullet }\right) 
\mathrm{d}r,\ \qquad S_0\left( f^{\bullet }\right) =I.
\end{equation*}
Hence $S_tF=E^{*}S_t\left( f^{\bullet }\right) $ if $f^{\bullet }\in 
\mathfrak{E}_t$, and $F^{*}S_tF=S_t\left( f^{\bullet }\right) $ as $EF=I$.
Since this equation is equivalent to the differential one 
\begin{equation}
\frac{\mathrm{d}}{\mathrm{d}t}S_t\left( f^{\bullet }\right) \eta +\left(
K_{\bullet }f^{\bullet }\left( t\right) +K\right) S_t\left( f^{\bullet
}\right) \eta =0,\quad S_0\left( f^{\bullet }\right) \eta =\eta ,\qquad
\forall \eta \in \mathcal{D},  \label{3.7'}
\end{equation}
the function $t\mapsto S_t\left( f^{\bullet }\right) ,$ $f^{\bullet }\in 
\mathfrak{E}$ is a strongly continuous cocycle, 
\begin{equation*}
S_r\left( f_s^{\bullet }\right) S_s\left( f^{\bullet }\right) =S_{r+s}\left(
f^{\bullet }\right) ,\ \forall r,s>0,\qquad f_s^{\bullet }\left( t\right)
=f^{\bullet }\left( t+s\right) ,\quad S_0\left( f^{\bullet }\right) =I.
\end{equation*}
As was proved in \cite{20}, the multiple integral (\ref{3.5'}) gives a
solution to the integral equation (\ref{3.1'}), and so the multiple integral
for $V_t^{\circ }\psi _f=S_t\left( f^{\bullet }\right) \eta \otimes
f^{\otimes },$ 
\begin{equation*}
S_t\left( f^{\bullet }\right) =e^{-Kt}+\sum_{n=1}^\infty \left( -1\right) ^n%
\underset{0<t_1<\ldots <t_n<t}{\int \cdots \int }K\left( t,t_n\right) \cdot
\cdot \cdot K\left( t_2,t_1\right) e^{-Kt_1}\mathrm{d}t_1\cdot \cdot \cdot 
\mathrm{d}t_n,
\end{equation*}
where $K\left( t,r\right) =e^{-K\left( t-r\right) }K_{\bullet }f^{\bullet
}\left( r\right) $, corresponding to the iteration of the integral equation
for $V_t^{\circ }$ on $\psi _f$, satisfies the HP equation (\ref{3.2'}). 
\endproof%

The following theorem reduces the problem of solving differential evolution
equations to the problem of iteration of integral equations similar to the
nonstochastic case \cite{Be89, Chb}.

\begin{proposition}
Let $S_t=T_tV_t^{\circ }$, where $V_t^{\circ }\in \mathcal{L}\left( 
\mathfrak{D}\right) $ are continuous operators defining the adapted cocycle
solution to Eq. (\ref{3.2'}). Then the linear stochastic evolution equation (%
\ref{2.2}) is equivalent to the quantum non-adapted (in the sense of \cite%
{20}) integral equation 
\begin{equation}
\phi _t\left( B\right) =S_t^{*}BS_t+\int_0^t\mathrm{d}\Lambda _\mu ^\nu
\left( r,\phi _r\left[ \varphi _\nu ^\mu \left( S_{t-r}^{*}BS_{t-r}\right) %
\right] \right)  \label{3.3'}
\end{equation}
with $\phi _0\left( B\right) =B\in \mathcal{B}$, where $\varphi _\nu ^\mu $
are extended onto $\mathfrak{B}$ by w*-continuity and linearity as $\varphi
_\nu ^\mu \left( B\otimes Z\right) =\varphi _\nu ^\mu \left( B\right)
\otimes Z$ for $B\in \mathcal{B}$, $Z\in \mathcal{B}\left( \mathfrak{F}%
\right) $.
\end{proposition}

\proof%
The non-adapted equation (\ref{3.3'}) is understood in the coherent form
sense as 
\begin{equation*}
\left\langle \psi _f|\phi _t\left( B\right) \psi _f\right\rangle
=\left\langle S_t\psi _f|BS_t\psi _f\right\rangle +\int_0^t\left\langle \psi
_f|\phi _r\left[ \varphi \left( \bar{f}^{\bullet }\left( r\right)
,S_{t-r}^{*}BS_{t-r},f^{\bullet }\left( r\right) \right) \right] \psi
_f\right\rangle \mathrm{d}r,
\end{equation*}
where $\varphi \left( \bar{e}^{\bullet },B,e^{\bullet }\right) =\sum_{m,n}%
\bar{e}^m\varphi _n^m\left( B\right) e^n+\sum_m\bar{e}^m\varphi ^m\left(
B\right) +\sum_n\varphi _n\left( B\right) e^n+\varphi \left( B\right) $. Due
to the adaptiveness of $\phi _t$ this can be written for $\psi _f=\eta
\otimes f^{\otimes }=F\eta $ with $f^{\bullet }\in \mathfrak{E}_t$ as 
\begin{eqnarray}
\phi _t\left( \bar{f}^{\bullet },B,f^{\bullet }\right) &=&S_t^{*}\left( \bar{%
f}^{\bullet }\right) BS_t\left( f^{\bullet }\right)  \label{3.4'} \\
&&+\int_0^t\phi _r\left( \bar{f}^{\bullet },\varphi \left(
r,S_{t-r}^{*}\left( \bar{f}_r^{\bullet }\right) BS_{t-r}\left( f_r^{\bullet
}\right) \right) ,f^{\bullet }\right) \mathrm{d}r,  \notag
\end{eqnarray}
where $S_t^{*}\left( \bar{f}^{\bullet }\right) =S_t\left( f^{\bullet
}\right) ^{*}$, $f_r^{\bullet }\left( t\right) =f^{\bullet }\left(
t+r\right) $. Here we take into account that due to adaptiveness $F^{*}\phi
_r\left[ Y\right] F=\phi _r\left( \bar{f}^{\bullet },F_r^{*}YF_r,f^{\bullet
}\right) $, where $F_r=T_rF$, and therefore 
\begin{equation*}
F^{*}\phi _r\left[ \varphi \left( r,S_{t-r}^{*}BS_{t-r}\right) \right]
F=\phi _r\left( \bar{f}^{\bullet },F_r^{*}\varphi \left(
r,S_{t-r}^{*}BS_{t-r}\right) F_r,f^{\bullet }\right) =
\end{equation*}
\begin{equation*}
\phi _r\left( \bar{f}^{\bullet },\varphi \left(
r,F_r^{*}S_{t-r}^{*}BS_{t-r}F_r\right) ,f^{\bullet }\right) =\phi _r\left( 
\bar{f}^{\bullet },\varphi \left( r,S_{t-r}^{*}\left( \bar{f}_r^{\bullet
}\right) BS_{t-r}\left( f_r^{\bullet }\right) \right) ,f^{\bullet }\right)
\end{equation*}
as $F_t^{*}\varphi \left( t,B\right) F_t=\varphi \left( t,F_t^{*}BF_t\right) 
$ for $\varphi \left( t,B\right) =\varphi \left( \bar{f}^{\bullet }\left(
t\right) ,B,f^{\bullet }\left( t\right) \right) $ and $S_{t-r}F_r=F_tS_{t-r}%
\left( f_r^{\bullet }\right) $, where $F_t\eta =\eta \otimes \delta
_\emptyset $ for any $f^{\bullet }\in \mathfrak{E}_t$.

Let us prove that the operator-valued function $t\mapsto S_s\left(
t,f^{\bullet }\right) :=S_{s-t}\left( f_t^{\bullet }\right) $ satisfies the
backward evolution equation 
\begin{equation*}
\frac{\mathrm{d}}{\mathrm{d}t}S_s\left( t,f^{\bullet }\right) \eta
=S_s\left( t,f^{\bullet }\right) \left( K_{\bullet }f^{\bullet }\left(
t\right) +K\right) \eta ,\quad S_0\left( f_s^{\bullet }\right) \eta =\eta
\quad \forall t\in [0,s).
\end{equation*}
Indeed, taking into account the forward equation (\ref{3.7'}), we obtain it
at $r=t$ from the cocycle property $S_s\left( t,f^{\bullet }\right)
S_t\left( r,f^{\bullet }\right) =S_s\left( r,f^{\bullet }\right) $: 
\begin{equation*}
0=\frac{\mathrm{d}}{\mathrm{d}t}\left( S_s\left( r,f^{\bullet }\right) \eta
\right) =\left( \frac{\mathrm{d}}{\mathrm{d}t}S_s\left( t,f^{\bullet
}\right) -S_s\left( t,f^{\bullet }\right) \left( K_{\bullet }f^{\bullet
}\left( t\right) +K\right) \right) S_t\left( r,f^{\bullet }\right) \eta .
\end{equation*}

Now, replacing $B$ in (\ref{3.4'}) by $Y_s\left( \bar{f}^{\bullet
},t,f^{\bullet }\right) =S_s^{*}\left( t,\bar{f}^{\bullet }\right)
BS_s\left( t,f^{\bullet }\right) $, we can write 
\begin{eqnarray*}
&&\phi _t\left( \vec{f}^{\bullet },S_s^{*}\left( t,\bar{f}^{\bullet }\right)
BS_s\left( t,f^{\bullet }\right) ,f^{\bullet }\right) \\
&=&S_s^{*}\left( \bar{f}^{\bullet }\right) BS_s\left( f^{\bullet }\right)
+\int_0^t\phi _r\left( \vec{f}^{\bullet },\varphi \left( r,S_s^{*}\left( r,%
\bar{f}^{\bullet }\right) BS_s\left( r,f^{\bullet }\right) \right)
,f^{\bullet }\right) \mathrm{d}r.
\end{eqnarray*}
Calculating the total derivative $\frac{\mathrm{d}}{\mathrm{d}t}\phi
_t\left( \bar{f}^{\bullet },S_s^{*}\left( t,\bar{f}^{\bullet }\right)
BS_s\left( t,f^{\bullet }\right) ,f^{\bullet }\right) $ by taking into
account the backward equation, we obtain the differential equation at $s=t$: 
\begin{equation*}
\frac{\mathrm{d}}{\mathrm{d}t}\phi _t\left( \bar{f}^{\bullet },B,f^{\bullet
}\right) +\phi _t\left( \bar{f}^{\bullet },K\left( t\right) ^{*}B+BK\left(
t\right) ,f^{\bullet }\right) =\phi _t\left( \bar{f}^{\bullet },\varphi
\left( \bar{f}^{\bullet }\left( t\right) ,B,f^{\bullet }\left( t\right)
\right) ,f^{\bullet }\right)
\end{equation*}
where $K\left( t\right) =K+K_{\bullet }f^{\bullet }\left( t\right) $. This
equation written for $\left\langle \eta |\phi _t\left( \bar{f}^{\bullet
},B,f^{\bullet }\right) \eta \right\rangle $ coincides with the coherent
matrix form (\ref{2.1}) for the quantum stochastic equation (\ref{2.2}) with 
$\psi _f=F\eta $.

The converse is easy to show by integrating the equation for $\phi _t\left( 
\bar{f}^{\bullet },B,f^{\bullet }\right) $ with $B$ replaced by $Y\left(
t\right) =S_s^{*}\left( t,\bar{f}^{\bullet }\right) BS_s\left( t,f^{\bullet
}\right) $: 
\begin{equation*}
\phi _s\left( \bar{f}^{\bullet },B,f^{\bullet }\right) -S_s^{*}\left( \bar{f}%
^{\bullet }\right) BS_s\left( f^{\bullet }\right) =\int_0^s\frac{\mathrm{d}}{%
\mathrm{d}t}\phi _t\left( \bar{f}^{\bullet },S_s^{*}\left( t,\bar{f}%
^{\bullet }\right) BS_s\left( t,f^{\bullet }\right) ,f^{\bullet }\right) 
\mathrm{d}t
\end{equation*}
\begin{eqnarray*}
&=&\int_0^s\left( \frac{\mathrm{d}}{\mathrm{d}t}\phi _t\left( \bar{f}%
^{\bullet },Y\left( r\right) ,f^{\bullet }\right) +\phi _r\left( \bar{f}%
^{\bullet },\frac{\mathrm{d}}{\mathrm{d}t}Y\left( t\right) ,f^{\bullet
}\right) \right) _{r=t}\mathrm{d}t \\
&=&\int_0^s\phi _t\left( \bar{f}^{\bullet },\varphi \left( t,S_s^{*}\left( t,%
\bar{f}^{\bullet }\right) BS_s\left( t,f^{\bullet }\right) \right)
,f^{\bullet }\right) \mathrm{d}t,
\end{eqnarray*}
whereas $\frac{\mathrm{d}}{\mathrm{d}t}Y\left( t\right) =\left( K_{\bullet
}f^{\bullet }\left( t\right) +K\right) ^{*}Y\left( t\right) +Y\left(
t\right) \left( K_{\bullet }f^{\bullet }\left( t\right) +K\right) $. 
\endproof%

\begin{theorem}
Let $\boldsymbol{\varphi }$ be a w*-continuous CP-map, and $%
S_t=T_tV_t^{\circ } $ be given by the solution to the quantum stochastic
equation (\ref{3.2'}). Then the solutions to the evolution equation (\ref%
{2.2}) with the generators, corresponding to (\ref{3.0'}), have the CP
property, and satisfy the submartingale (contractivity) condition $\phi
_t\left( I\right) \leq \epsilon _t\left[ \phi _s\left( I\right) \right] $
for all $t<s$ if $\varphi \left( I\right) \leq K+K^{\dagger }$ ($\phi
_t(I)\leq \phi _s(I)$ if $\boldsymbol{\varphi }(I)\leq \boldsymbol{K}+%
\boldsymbol{K}^{\dagger }$). The minimal solution can be constructed in the
form of a multiple quantum stochastic integral in the sense \cite{20} as the
series 
\begin{equation}
\phi _t\left( B\right) =\sum_{n=0}^\infty \underset{0<t_1<\ldots <t_n<t}{%
\int \cdots \int }\mathrm{d}\Lambda _{\mu _1\ldots \mu _n}^{\nu _1\ldots \nu
_n}\left( t_1,\ldots ,t_n,\varphi _{\nu _1\ldots \nu _n}^{\mu _1\ldots \mu
_n}\left( t_1,\ldots ,t_n,S_{t-t_n}^{*}BS_{t-t_n}\right) \right)
\label{3.8'}
\end{equation}
of non-adapted n-tuple CP integrals with $S_t^{*}BS_t$ at $n=0$ and 
\begin{equation*}
\varphi _{\nu _1\ldots \nu _n}^{\mu _1\ldots \mu _n}\left( t_1,\ldots
,t_n\right) =\varphi _{\nu _1}^{\mu _1}\left( t_1\right) \circ \varphi _{\nu
_2}^{\mu _2}\left( t_2-t_1\right) \circ \ldots \circ \varphi _{\nu _n}^{\mu
_n}\left( t_n-t_{n-1}\right) ,
\end{equation*}
where $\varphi _\nu ^\mu \left( t,B\right) =S_t^{*}\varphi _\nu ^\mu \left(
B\right) S_t$. If $\boldsymbol{\varphi }$ is bounded, then the solution to
the equation is unique, and $\phi _t\left( I\right) =\epsilon _t\left[ \phi
_s\left( I\right) \right] $ for all $t<s$ if $K+K^{\dagger }=\varphi \left(
I\right) $ ($\phi _t(I)=I$ if $\boldsymbol{K}+\boldsymbol{K}^{\dagger }=%
\boldsymbol{\varphi }(I)$).
\end{theorem}

\proof%
The existence and uniqueness of the solutions $\phi _t\left( B\right) $ to
the quantum stochastic equations (\ref{2.2}) with the bounded generators $%
\lambda _\nu ^\mu \left( B\right) =\gamma _\nu ^\mu \left( B\right) -B\delta
_\nu ^\mu $ and the initial conditions $\phi _0\left( B\right) =B$ in an
operator algebra $\mathcal{B\subseteq L}\left( \mathcal{H}\right) $ was
proved in \cite{20}. The CP property of the solution to this equation with
the generators, corresponding to the conditionally positive germ-matrix (\ref%
{3.0'}), can be proven in the form (\ref{3.8'}), which is obtained by the
iteration 
\begin{equation*}
\phi _t^{n+1}\left( B\right) =S_t^{*}BS_t+\int_0^t\mathrm{d}\Lambda _\mu
^\nu \left( r,\phi _r^n\left( \varphi _\nu ^\mu \left(
S_{t-r}^{*}BS_{t-r}\right) \right) \right) ,\quad \phi _t^0\left( B\right)
=S_t^{*}BS_t
\end{equation*}
of the equivalent non-adapted integral equation (\ref{3.3'}). Indeed, in
order to prove the complete positivity of the solution, written in this
form, one should prove the positive definiteness of the iteration 
\begin{eqnarray*}
\phi _t^{n+1}\left( \bar{f}^{\bullet },B,f^{\bullet }\right)
&=&S_t^{*}\left( \bar{f}^{\bullet }\right) BS_t\left( f^{\bullet }\right) \\
&&+\int_0^t\phi _r^n\left( \bar{f}^{\bullet },\varphi \left( \bar{f}%
^{\bullet }\left( r\right) ,S_{t-r}^{*}\left( \bar{f}_r^{\bullet }\right)
BS_{t-r}\left( f_r^{\bullet }\right) ,f^{\bullet }\left( r\right) \right)
,f^{\bullet }\right) \mathrm{d}r
\end{eqnarray*}
of the integral equation (\ref{3.4'}) with the CP $\phi _t^0\left( B\right)
=S_t^{*}BS_t$. Thus, we have to test the positive definiteness of the forms 
\begin{equation*}
\sum_{B,C}\sum_{f,h}\left\langle \xi _B^f\right| \left. \phi _t^{n+1}\left( 
\bar{f}^{\bullet },B^{*}C,h^{\bullet }\right) \xi _C^h\right\rangle
=\sum_{B,C}\sum_{f,h}\left\langle BS_t\left( f^{\bullet }\right) \xi
_B^f|CS_t\left( h^{\bullet }\right) \xi _C^h\right\rangle
\end{equation*}
\begin{equation*}
+\int_0^t\sum_{B,C}\sum_{f,h}\left\langle \boldsymbol{\eta }_B^f\left(
r\right) |\phi _s^n\left( \bar{f}^{\bullet },\boldsymbol{\varphi }\left(
S_{t-r}^{*}\left( \bar{f}_r^{\bullet }\right) B^{*}CS_{t-r}\left(
h_r^{\bullet }\right) \right) ,h^{\bullet }\right) \boldsymbol{\eta }%
_C^h\left( r\right) \right\rangle ,
\end{equation*}
where $\boldsymbol{\eta }_B^f\left( r\right) =\sum_{f\left( r\right) }\xi
_B^f\otimes \boldsymbol{f}\left( r\right) $, and $\boldsymbol{f}\left(
r\right) =1\oplus f^{\bullet }\left( r\right) $. It is a consequence of the
CP condition for $\boldsymbol{\varphi }$ and the CP property for $\phi
_{t_n}^n,\forall t_n<t$, which obviously follows from the positive
definiteness of $\phi _r^{n-1},r<t_n$, and so on up to $\phi _r^0,r<t_1$.
The direct iteration of this integral recursion with the initial CP
condition $\phi _t^0\left( B\right) =S_t^{*}BS_t$ gives at the limit $%
n\rightarrow \infty $ the solution in the form of a series 
\begin{equation*}
\phi _t\left( \bar{f}^{\bullet },B,f^{\bullet }\right) =\sum_{n=0}^\infty 
\underset{0<t_1<\ldots <t_n<t}{\int \cdots \int }\varphi \left( t_1,\ldots
,t_n;S_{t-t_n}^{*}\left( \bar{f}_{t_n}^{\bullet }\right) BS_{t-t_n}\left(
f_{t_n}^{\bullet }\right) \right) \mathrm{d}t_1\cdot \cdot \cdot \mathrm{d}%
t_n
\end{equation*}
of n-tuple integrals on the interval $[0,t)$ with $S_t^{*}\left( \bar{f}%
^{\bullet }\right) BS_t\left( f^{\bullet }\right) $ at $n=0$. The positive
definite kernels 
\begin{equation*}
\quad \varphi \left( t_1,\ldots ,t_n\right) =\varphi ^0\left( t_1\right)
\circ \varphi ^{t_1}\left( t_2\right) \circ \ldots \circ \varphi
^{t_{n-1}}\left( t_n\right) ,
\end{equation*}
where $\varphi ^r\left( t,B\right) =S_{t-r}^{*}\left( \bar{f}_r^{\bullet
}\right) \varphi \left( \bar{f}^{\bullet }\left( t\right) ,B,f^{\bullet
}\left( t\right) \right) S_{t-r}\left( f_r^{\bullet }\right) $, are obtained
by the recurrence 
\begin{equation*}
\varphi \left( t_1,\ldots ,t_n\right) =\varphi \left( t_1,\ldots
,t_{n-1}\right) \circ \varphi ^{t_{n-1}}\left( t_n\right) ,\qquad \varphi
\left( t\right) =\varphi ^0\left( t\right) ,
\end{equation*}
corresponding to (\ref{3.9'}). This proves the CP property for the series (%
\ref{3.8'}), which converges to a $0\leq Y_t\leq \kappa R_t$ for any
positive bounded $0\leq B\leq \kappa I$ because of the increase $Y_t^n\leq
Y_t^{n+1}$ for $Y_t^n=\phi _t^n\left( B\right) $ and the boundedness $%
Y_t^n\leq \kappa R_t^n$, $R_t^n\leq R_t$, where $R_t$ is the continuous
sesquilinear form (\ref{3.6'}).

As follows from the exponential estimate \cite{20} for the solutions to the
quantum stochastic equations (\ref{2.2}) with the bounded generators, $%
R_t=\phi _t\left( I\right) $ might be unbounded, but strongly continuous in
the Fock scale $\mathfrak{F}$. In the case of unbounded generators the
solution to (\ref{3.6'}) might not be unique, and the iterated series (\ref%
{3.8'}) gives obviously the minimal one, which is unique among such
solutions. Let us prove the submartingale property for the sesquilinear form 
$R_t$, given by the weakly convergent series (\ref{3.6'}). $R_s$ for a $s>t$
is defined as the iterated solution $Y_s=R_s:=\lim R_s^n$ to the backward
integral equation 
\begin{equation*}
Y_s=S_s^{*}BS_s+\int_0^s\mathrm{d}\Lambda _\mu ^\nu \left( r,S_r^{*}\varphi
_\nu ^\mu \left( Y_{s-r}\right) S_r\right)
\end{equation*}
for the series $Y_s=\phi _s\left( B\right) $ with $B=I$. It satisfies the
integral equation 
\begin{equation*}
R_s=S_t^{*}R_{s-t}S_t+\int_0^t\mathrm{d}\Lambda _\mu ^\nu \left(
r,S_r^{*}\varphi _\nu ^\mu \left( R_{s-r}\right) S_r\right) ,
\end{equation*}
where we used the semigroup property $S_{s-t}S_t=S_s$ and that 
\begin{equation*}
\int_t^s\mathrm{d}\Lambda _\mu ^\nu \left( r,S_r^{*}\varphi _\nu ^\mu \left(
Y_{s-r}\right) S_r\right) =V_t^{\circ *}\int_t^s\mathrm{d}\Lambda _\mu ^\nu
\left( r,T_t^{*}S_{r-t}^{*}\varphi _\nu ^\mu \left( Y_{s-r}\right)
S_{r-t}T_t\right) V_t^{\circ },
\end{equation*}
\begin{equation*}
\int_t^s\mathrm{d}\Lambda _\mu ^\nu \left( r-t,S_{r-t}^{*}\varphi _\nu ^\mu
\left( Y_{s-r}\right) S_{r-t}\right) =\int_0^{s-t}\mathrm{d}\Lambda _\mu
^\nu \left( r,S_r^{*}\varphi _\nu ^\mu \left( Y_{s-t-r}\right) S_r\right) .
\end{equation*}
This can be written in terms of the coherent matrix elements $R_s\left( \bar{%
f}^{\bullet },f^{\bullet }\right) =F^{*}R_sF$, $f^{\bullet }\in \mathfrak{E}%
_s$ as 
\begin{eqnarray*}
R_s\left( \bar{f}^{\bullet },f^{\bullet }\right) &=&S_t^{*}\left( \bar{f}%
^{\bullet }\right) R_{s-t}\left( \bar{f}^{\bullet },f_t^{\bullet }\right)
S_t\left( f^{\bullet }\right) \\
&&+\int_0^tS_r^{*}\left( \bar{f}^{\bullet }\right) \varphi \left( \bar{f}%
^{\bullet }\left( r\right) ,R_{s-r}\left( \bar{f}^{\bullet },f_r^{\bullet
}\right) ,f_r^{\bullet }\right) S_r\left( f^{\bullet }\right) \mathrm{d}r.
\end{eqnarray*}
The coherent matrix elements $Y_t\left( \bar{f}^{\bullet },f^{\bullet
}\right) $ of the conditional expectation $Y_s=\epsilon _t\left( R_s\right) $
coincide with $R_s\left( \bar{f}^{\bullet },f^{\bullet }\right) $ if $%
f^{\bullet }\in \mathfrak{E}_t$. Hence, they satisfy the integral equation 
\begin{eqnarray*}
Y_t\left( \bar{f}^{\bullet },f^{\bullet }\right) &=&S_t^{*}\left( \bar{f}%
^{\bullet }\right) P_{s-t}S_t\left( f^{\bullet }\right) \\
&&+\int_0^tS_r^{*}\left( \bar{f}^{\bullet }\right) \varphi \left( \bar{f}%
^{\bullet }\left( r\right) ,Y_{t-r}\left( \bar{f}^{\bullet },f_r^{\bullet
}\right) ,f_r^{\bullet }\right) S_r\left( f^{\bullet }\right) \mathrm{d}r,
\end{eqnarray*}
corresponding to the non-adapted backward equation 
\begin{equation*}
Y_t=S_t^{*}P_{s-t}^{\circ }S_t+\int_0^t\mathrm{d}\Lambda _\mu ^\nu \left(
r,S_r^{*}\varphi _\nu ^\mu \left( Y_{t-r}\right) S_r\right) ,
\end{equation*}
where $P_s^{\circ }=P_s\otimes I$, $P_s=R_s\left( 0,0\right) $, as $%
f_t^{\bullet }\left( r\right) =f^{\bullet }\left( r+t\right) =0,\forall r\in 
\mathbb{R}_{+}$ if $f^{\bullet }\in \mathfrak{E}_t$. The operators $%
P_s=\epsilon \left( R_s\right) =\theta _s\left( I\right) $ are given by the
Markov semigroup $\theta _s=\epsilon \circ \phi _s$ as the decreasing
solutions to the integral equation 
\begin{equation*}
P_s=e^{-Ks*}e^{-Ks}+\int_0^se^{-Kr*}\varphi \left( P_{s-r}\right) e^{-Kr}%
\mathrm{d}r,
\end{equation*}
and $P_t\leq I$ if $K+K^{\dagger }\leq 0$. (See, for example, \cite{Be89}.)
Thus, the difference $\widetilde{R}_t=R_t-\epsilon _t\left( R_s\right)
=R_t-Y_t$ satisfies the same equation 
\begin{equation*}
\widetilde{R}_t=S_t^{*}\widetilde{I}_{s-t}S_t+\int_0^t\mathrm{d}\Lambda _\mu
^\nu \left( r,S_r^{*}\varphi _\nu ^\mu \left( \widetilde{R}_{t-r}\right)
S_r\right)
\end{equation*}
as $R_t$ with $\widetilde{I}_s=I-P_s^{\circ }$ instead of $I$. The iteration
of this equation defines it as the weak limit $\widetilde{R}_t=\lim 
\widetilde{R}_t^n$ in the form of the series (\ref{3.8'}) with $%
B=I-P_{s-t}\geq 0$. Hence $\widetilde{R}_t=\phi _t\left( I-P_{s-t}\right) $
is a positive sesquilinear form on $\mathfrak{D}$ for any $s\geq t$ due to
the positivity of $\phi _t$. The proof of contractivity $\phi _t(I)\leq \phi
_s(I)$ for $t<s$ is similar to that one, without the vacuum averaging of $%
R_t $. 
\endproof%

\section{The structure of the generators and flows}

First, let us prove the structure (\ref{2.10}) for the (unbounded)
form-generator of CP flows over the algebra $\mathcal{B}=\mathcal{L}\left( 
\mathcal{H}\right) $ of all bounded operators. This algebra contains the
one-dimensional operators $|\eta ^{\prime }\rangle \langle \eta ^0|:\eta
\mapsto \left\langle \eta ^0|\eta \right\rangle \eta ^{\prime }$ given by
the vectors $\eta ^0,\eta ^{\prime }\in \mathcal{H}$.

1. Let us fix a vector $\boldsymbol{\eta }^0\in \mathcal{D}\oplus \mathcal{D}%
_{\bullet }$ with the unit projection $\eta ^0\in \mathcal{D}$, $\left\|
\eta ^0\right\| =1$, and make the following assumption of the weak
continuity for the linear operator $\eta ^{\prime }\mapsto $ $\boldsymbol{%
\gamma }\left( |\eta ^{\prime }\rangle \langle \eta ^0|\right) \boldsymbol{%
\eta }^0$.

\begin{enumerate}
\item[0)] The sequence $\boldsymbol{\eta }_n^{\prime }=\boldsymbol{\gamma }%
\left( |\eta _n^{\prime }\rangle \langle \eta ^0|\right) \boldsymbol{\eta }%
^0\in \mathcal{D}^{\prime }\oplus \mathcal{D}_{\bullet }^{\prime }$ of
anti-linear forms 
\begin{equation*}
\boldsymbol{\eta }\in \mathcal{D}\oplus \mathcal{D}_{\bullet }\mapsto
\left\langle \boldsymbol{\eta }|\boldsymbol{\eta }_n^{\prime }\right\rangle
:=\left\langle \boldsymbol{\eta }|\boldsymbol{\gamma }\left( |\eta
_n^{\prime }\rangle \langle \eta ^0|\right) \boldsymbol{\eta }^0\right\rangle
\end{equation*}
converges for each sequence $\eta _n^{\prime }\in \mathcal{H}$ converging in 
$\mathcal{D}^{\prime }\supseteq \mathcal{H}$.
\end{enumerate}

\begin{proposition}
Let the CCP germ-matrix $\boldsymbol{\gamma }$ satisfy the above continuity
condition for a given $\boldsymbol{\eta }^0$. Then there exist strongly
continuous operators $K\in \mathcal{L}\left( \mathcal{D}\right) ,K_{\bullet
}:\mathcal{D}_{\bullet }\rightarrow \mathcal{D}$ defining the matrix
operator $\boldsymbol{K}$ in (\ref{2.10}), such that the matrix map (\ref%
{2.8}) is CP, and there exists a Hilbert space $\mathcal{K}$, a $*$%
-representation $\jmath :B\mapsto B\otimes J$ of $\mathcal{B}=\mathcal{L}%
\left( \mathcal{H}\right) $ on the Hilbert product $\mathcal{G}=\mathcal{H}%
\otimes \mathcal{K}$, given by an orthoprojector $J$ in $\mathcal{K}$, such
that 
\begin{equation}
\boldsymbol{\varphi }\left( B\right) =\left( L^\mu \jmath \left( B\right)
L_\nu \right) _{\nu =+,\bullet }^{\mu =-,\bullet }=\boldsymbol{L}^{*}\jmath
\left( B\right) \boldsymbol{L}.  \label{4.0}
\end{equation}
Here $\boldsymbol{L}=\left( L,L_{\bullet }\right) $ is a strongly continuous
operator $\mathcal{D}\oplus \mathcal{D}_{\bullet }\rightarrow \mathcal{G}$
with $L=L_{+}$, $L^{-}=L^{*}$, $L^{\bullet }=L_{\bullet }^{*}$ which is
always possible to make 
\begin{equation}
\left\langle \eta ^0\otimes e|\boldsymbol{L}\boldsymbol{\eta }%
^0\right\rangle =0,\quad \quad \forall e\in \mathcal{K}_1,  \label{4.1}
\end{equation}
where $\mathcal{K}_1=J\mathcal{K}$. If $D=-\lambda \left( I\right) \geq 0$,
then one can make $L^{*}L=K+K^{\dagger }$ in a canonical way, and in
addition $L^{*}L_{\bullet }=K_{\bullet }$, $L_{\bullet }^{*}L_{\bullet
}=I_{\bullet }^{\bullet }$, where $I_{\bullet }^{\bullet }=I\delta _{\bullet
}^{\bullet }$, if $\boldsymbol{D}=-\boldsymbol{\lambda }\left( I\right) \geq
0$.
\end{proposition}

\proof%
Define the linear operator $\boldsymbol{A}:\mathcal{H}\rightarrow \mathcal{D}%
^{\prime }\oplus \mathcal{D}_{\bullet }^{\prime }$ by the relation 
\begin{equation*}
\left\langle \boldsymbol{\eta }|\boldsymbol{A}\eta ^{\prime }\right\rangle
=\left\langle \boldsymbol{\eta }|\boldsymbol{\gamma }\left( |\eta ^{\prime
}\rangle \langle \eta ^0|\right) \boldsymbol{\eta }^0\right\rangle
\end{equation*}
for all $\boldsymbol{\eta }\in \mathcal{D}\oplus \mathcal{D}_{\bullet }$ and 
$\eta ^{\prime }\in \mathcal{H}$. By the weak continuity it can be extended
on $\mathcal{D}^{\prime }$ and its dual operator $\boldsymbol{A}^{*}=\left(
A^{*},A_{\bullet }^{*}\right) $ into $\mathcal{D}$ is strongly continuous on 
$\mathcal{D}\oplus \mathcal{D}_{\bullet }$. The operators 
\begin{equation*}
K=\tfrac 12\left\langle \boldsymbol{\eta }^0|\boldsymbol{\gamma }\left(
|\eta ^0\rangle \langle \eta ^0|\right) \boldsymbol{\eta }^0\right\rangle
I-A^{*},\quad K_{\bullet }=-A_{\bullet }^{*}
\end{equation*}
define the matrix-map (\ref{2.8}) in the form 
\begin{eqnarray*}
\left\langle \boldsymbol{\eta }|\boldsymbol{\varphi }\left( B\right) 
\boldsymbol{\eta }\right\rangle &=&\left\langle \boldsymbol{\eta }|%
\boldsymbol{\gamma }\left( B\right) \boldsymbol{\eta }\right\rangle
+\left\langle \boldsymbol{\eta }^0|\boldsymbol{\gamma }\left( |\eta
^0\rangle \langle \eta ^0|\right) \boldsymbol{\eta }^0\right\rangle
\left\langle \eta |B\eta \right\rangle - \\
&&\left\langle \boldsymbol{\eta }|\boldsymbol{\gamma }\left( B|\eta \rangle
\langle \eta ^0|\right) \boldsymbol{\eta }^0\right\rangle -\left\langle 
\boldsymbol{\eta }^0|\boldsymbol{\gamma }\left( |\eta ^0\rangle \langle \eta
|B\right) \boldsymbol{\eta }\right\rangle ,
\end{eqnarray*}
where $\eta \in \mathcal{D}$ is the natural projection of $\boldsymbol{\eta }%
\in \mathcal{D}\oplus \mathcal{D}_{\bullet }$ onto $\mathcal{D}$. Let us
prove that this is a CP map, i.e.

\begin{equation*}
\sum_{B,C\in \mathcal{B}}\left\langle \boldsymbol{\xi }_B|\boldsymbol{%
\varphi }\left( B^{*}C\right) \boldsymbol{\xi }_C\right\rangle \geq 0
\end{equation*}
for all $\boldsymbol{\xi }_B=0$ except for a finite number of $B=B_k\in 
\mathcal{B},k=1,2,\ldots $, for which $\boldsymbol{\xi }_B=\boldsymbol{\eta }%
^k $. Indeed,

\begin{eqnarray*}
\left\langle \boldsymbol{\eta }^k|\boldsymbol{\varphi }\left(
B_k^{*}B_l\right) \boldsymbol{\eta }^l\right\rangle &=&\left\langle 
\boldsymbol{\eta }^k|\boldsymbol{\gamma }\left( B_k^{*}B_l\right) 
\boldsymbol{\eta }^l\right\rangle +\left\langle \boldsymbol{\eta }^0|%
\boldsymbol{\gamma }\left( |\eta ^0\rangle \langle \eta ^0|\right) 
\boldsymbol{\eta }^0\right\rangle \left\langle \eta ^k|B_k^{*}B_l\eta
^l\right\rangle - \\
&&\left\langle \boldsymbol{\eta }^k|\boldsymbol{\gamma }\left(
B_k^{*}B_l|\eta ^l\rangle \langle \eta ^0|\right) \boldsymbol{\eta }%
^0\right\rangle -\left\langle \boldsymbol{\eta }^0|\boldsymbol{\gamma }%
\left( |\eta ^0\rangle \langle \eta ^k|B_k^{*}B_l\right) \boldsymbol{\eta }%
^l\right\rangle \\
&=&\sum_{k,l\geq 0}\left\langle \boldsymbol{\eta }^k|\boldsymbol{\gamma }%
\left( B_k^{*}B_l\right) \boldsymbol{\eta }^l\right\rangle ,
\end{eqnarray*}
where $B_0=-\sum_{B\in \mathcal{B}}B|\xi _B\rangle \langle \eta ^0|$, and $%
\boldsymbol{\eta }^k=\boldsymbol{\xi }_B$ for $B=B_k,k=1,2,\ldots $. Because 
$\sum_{k\geq 0}B_k\eta ^k=B_0\eta ^0+\sum_{B\in \mathcal{B}}B\xi _B=0$, this
form is positive, as it is written as a conditionally positive form 
\begin{equation*}
\sum_{B,C\in \mathcal{B}}\left\langle \boldsymbol{\zeta }_B|\boldsymbol{%
\gamma }\left( B^{*}C\right) \boldsymbol{\zeta }_C\right\rangle \geq 0,
\end{equation*}
with $\sum_{B\in \mathcal{B}}\boldsymbol{\iota }\left( B\right) \boldsymbol{%
\zeta }_B=0$, where $\boldsymbol{\zeta }_B=\boldsymbol{\eta }^k=\boldsymbol{%
\xi }_B$ if $B=B_k\neq B_0$, and $\boldsymbol{\zeta }_B=\boldsymbol{\xi }_B+%
\boldsymbol{\eta }^0$ for $B=B_0$, otherwise $\boldsymbol{\zeta }_B=0$.
Moreover, 
\begin{eqnarray*}
\left\langle \boldsymbol{\eta }|\boldsymbol{\varphi }\left( |\eta ^{\prime
}\rangle \langle \eta ^0|\right) \boldsymbol{\eta }^0\right\rangle
&=&\left\langle \boldsymbol{\eta }|\boldsymbol{\gamma }\left( |\eta ^{\prime
}\rangle \langle \eta ^0|\right) \boldsymbol{\eta }^0\right\rangle
+\left\langle \boldsymbol{\eta }^0|\boldsymbol{\gamma }\left( |\eta
^0\rangle \langle \eta ^0|\right) \boldsymbol{\eta }^0\right\rangle
\left\langle \eta |\eta ^{\prime }\right\rangle - \\
&&\left\langle \boldsymbol{\eta }|\boldsymbol{\gamma }\left( |\eta ^{\prime
}\rangle \langle \eta ^0|\right) \boldsymbol{\eta }^0\right\rangle
-\left\langle \boldsymbol{\eta }^0|\boldsymbol{\gamma }\left( |\eta
^0\rangle \left\langle \eta |\eta ^{\prime }\right\rangle \langle \eta
^0|\right) \boldsymbol{\eta }^0\right\rangle =0.
\end{eqnarray*}
Thus, the form-generator over $\mathcal{B}=\mathcal{L}\left( \mathcal{H}%
\right) $ has the form (\ref{2.10}), where the CP map $\boldsymbol{\varphi }$
can always be chosen to satisfy $\boldsymbol{\varphi }\left( |\eta ^{\prime
}\rangle \langle \eta ^0|\right) \boldsymbol{\eta }^0=0$ for all $\eta
^{\prime }\in \mathcal{H}$ and a given vector $\eta ^0\in \mathcal{D}$. The
Steinspring dilation (\ref{4.0}) of the CP map $\boldsymbol{\varphi }$ into
the continuous forms $\boldsymbol{\varphi }\left( B\right) \in \mathcal{B}%
\left( \mathcal{D}\oplus \mathcal{D}_{\bullet }\right) $ is given by a
continuous operator $\boldsymbol{L}:\mathcal{D}\oplus \mathcal{D}_{\bullet
}\rightarrow \mathcal{G}$ with the dual $\boldsymbol{L}^{*}:\mathcal{G}%
\rightarrow \mathcal{D}^{\prime }\oplus \mathcal{D}_{\bullet }^{\prime }$
because 
\begin{equation*}
\left\| \boldsymbol{L}\boldsymbol{\eta }_n\right\| ^2=\left\langle 
\boldsymbol{\eta }_n|\boldsymbol{\varphi }\left( I\right) \boldsymbol{\eta }%
_n\right\rangle \longrightarrow 0
\end{equation*}
if $\boldsymbol{\eta }_n\longrightarrow 0$ strongly in $\mathcal{D}\oplus 
\mathcal{D}_{\bullet }$. The w*-representation $\jmath :\mathcal{B}%
\rightarrow \mathcal{L}\left( \mathcal{G}\right) $ of $\mathcal{B}=\mathcal{L%
}\left( \mathcal{H}\right) $ is always an ampliation $\jmath \left( B\right)
=B\otimes J$, where $J$ is an orthoprojector onto a subspace $\mathcal{K}%
_1\subseteq \mathcal{K}$, corresponding to the minimal dilation in $\mathcal{%
G}_1=\mathcal{H}\otimes \mathcal{K}_1$. The property (\ref{4.1}) follows
from the inequality $|e\rangle \langle e|\leq J$ if $Je=e$: 
\begin{eqnarray*}
\left| \left\langle \eta ^0\otimes e|\boldsymbol{L}\boldsymbol{\eta }%
^0\right\rangle \right| ^2 &\leq &\left\langle \boldsymbol{L}\boldsymbol{%
\eta }^0|\jmath \left( |\eta ^0\rangle \langle \eta ^0|\right) \boldsymbol{L}%
\boldsymbol{\eta }^0\right\rangle \\
&=&\left\langle \boldsymbol{\eta }^0|\boldsymbol{\varphi }\left( |\eta
^0\rangle \langle \eta ^0|\right) \boldsymbol{\eta }^0\right\rangle =0.
\end{eqnarray*}
If $K+K^{\dagger }\geq \varphi \left( I\right) $, $L^1=\left( I\otimes
J\right) L$ is the operator of the minimal dilation $\varphi \left( B\right)
=L_1\jmath \left( B\right) L^1$, so that $\varphi \left( I\right) =L_1L^1$
with respect to the adjoint $L_1:$ $\mathcal{G}_1\rightarrow \mathcal{D}%
^{\prime }$, and $L^0$ is an operator on $\mathcal{D}$ into a Hilbert
product $\mathcal{G}_0=\mathcal{H}\otimes \mathcal{K}_0$, satisfying the
condition $L_0L^0=D$ with respect to the adjoint $L_0:\mathcal{G}%
_0\rightarrow \mathcal{D}^{\prime }$, then $L^{\circ }:\eta \mapsto L^0\eta
\oplus L^1\eta $ defines the canonical dilation in $\mathcal{G}_{\circ }=%
\mathcal{H}\otimes \mathcal{K}_{\circ }$ having the property $L_{\circ
}L^{\circ }=\varphi \left( I\right) +D=K+K^{\dagger }$, where $\mathcal{K}%
_{\circ }=\mathcal{K}_0\oplus \mathcal{K}_1$ and $L_{\circ }:\mathcal{G}%
_{\circ }\rightarrow \mathcal{D}^{\prime }$ is the adjoint to $L^{\circ }:%
\mathcal{D}\rightarrow \mathcal{G}_{\circ }$. Moreover, if 
\begin{equation*}
\boldsymbol{D}=\left( \delta _\nu ^\mu I-\varphi _\nu ^\mu \left( I\right)
+\delta _{-}^\mu K_\nu +K^\mu \delta _\nu ^{+}\right) _{\nu =+,\bullet
}^{\mu =-,\bullet }=\boldsymbol{K}+\boldsymbol{K}^{\dagger }-\boldsymbol{%
\varphi }(I)\geq 0,
\end{equation*}
the operator $\boldsymbol{L}^{\circ }:\eta \mapsto \boldsymbol{L}^0\eta
\oplus \boldsymbol{L}^1\eta $ defines the canonical dilation with the
property $\boldsymbol{L}^{*}\boldsymbol{L}=\boldsymbol{K}+\boldsymbol{K}%
^{\dagger }$: 
\begin{equation*}
L_{\circ }^\mu L_\nu ^{\circ }=L_0^\mu L_\nu ^0+L_1^\mu L_\nu ^1=D_\nu ^\mu
+\varphi _\nu ^\mu \left( I\right) =\delta _{-}^\mu K_\nu +K^\mu \delta _\nu
^{+}+\delta _\nu ^\mu I,
\end{equation*}
where $\boldsymbol{L}^0:\mbox{\boldmath$\mathcal D$}\rightarrow \mathcal{G}%
_0 $ are operators $\left( L^0,L_{\bullet }^0\right) $ with the adjoints $%
L_0^\mu =L_{-\mu }^{0*}$, satisfying the conditions $L_0^\mu L_\nu ^0=D_\nu
^\mu $, and $\boldsymbol{L}_\nu ^1=\left( I\otimes J\right) \boldsymbol{L}%
_\nu $ are the operators of the minimal dilation $\varphi _\nu ^\mu \left(
B\right) =L_1^\mu \jmath \left( B\right) L_\nu ^1$. 
\endproof%

2. Thus we have proved that Eq. (\ref{2.2}) for completely positive quantum
stochastic flows over $\mathcal{B}=\mathcal{L}\left( \mathcal{H}\right) $
has the following general form: 
\begin{equation*}
\mathrm{d}\phi _{t}\left( B\right) +\phi _{t}\left( K^{\ast }B+BK-L^{\ast
}\jmath \left( B\right) L\right) \mathrm{d}t=\sum_{m,n=1}^{\infty }\phi
_{t}\left( L_{m}^{\ast }\jmath \left( B\right) L_{n}-B\delta _{n}^{m}\right) 
\mathrm{d}\Lambda _{m}^{n}
\end{equation*}
\begin{equation*}
+\sum_{m=1}^{\infty }\phi _{t}\left( L_{m}^{\ast }\jmath \left( B\right)
L-K_{m}^{\ast }B\right) \mathrm{d}\Lambda _{m}^{+}+\sum_{n=1}^{\infty }\phi
_{t}\left( L^{\ast }\jmath \left( B\right) L_{n}-BK_{n}\right) \mathrm{d}%
\Lambda _{-}^{n},
\end{equation*}
generalizing the Lindblad form \cite{19} for the semigroups of completely
positive maps. This can be written in the tensor notation form as 
\begin{equation}
\mathrm{d}\phi _{t}\left( B\right) =\phi _{t}\left( L_{\alpha }^{\star \mu
}\jmath _{\beta }^{\alpha }\left( B\right) L_{\nu }^{\beta }-\imath _{\nu
}^{\mu }\left( B\right) \right) \mathrm{d}\Lambda _{\mu }^{\nu }=\phi
_{t}\left( \mathbf{L}^{\star }\text{\textbf{\j }}\left( B\right) \mathbf{L}-%
\text{\textbf{\i }}\left( B\right) \right) \cdot \mathrm{d}\boldsymbol{%
\Lambda },  \label{4.3}
\end{equation}
where the summation is taken over all $\alpha ,\beta =-,\circ ,+$ and $\mu
,\nu =-,\bullet ,+$, $\jmath _{-}^{-}\left( B\right) =B=\jmath
_{+}^{+}\left( B\right) $, $\jmath _{\circ }^{\circ }\left( B\right) =\jmath
\left( B\right) $, $\jmath _{\beta }^{\alpha }\left( B\right) =0$ if $\alpha
\neq \beta $, $\imath _{\nu }^{\mu }\left( B\right) =B\delta _{\nu }^{\mu }$%
, and $\mathbf{L}^{\star }\mathbf{=}\left[ L_{\alpha }^{\star \mu }\right]
_{\alpha =-,\circ ,+}^{\mu =-,\bullet ,+}$ is the triangular matrix,
pseudoadjoint to $\mathbf{L=}\left[ L_{\nu }^{\beta }\right] _{\nu
=-,\bullet ,+}^{\beta =-,\circ ,+}$ with $L_{-}^{-}=I=L_{+}^{+}$, 
\begin{equation*}
L_{\bullet }^{\circ }=L_{\bullet },\quad L_{+}^{\circ }=L,\quad L_{\bullet
}^{-}=-K_{\bullet },\quad L_{+}^{-}=-K.
\end{equation*}
(All other $L_{\nu }^{\beta }$ are zero.) If the Hilbert space $\mathcal{H}%
\otimes \mathcal{G}$ is embedded into the direct sum $\mathcal{H}\oplus 
\mathcal{H}\oplus ...$ of copies of the initial Hilbert space $\mathcal{H}$
such that $J=\left[ \delta _{l}^{i}\right] $ for a subset $i,l\notin \mathbb{%
N}_{0}\subseteq \mathbb{N}$, this equation can be resolved as $\phi
_{t}\left( B\right) =V_{t}^{\ast }\left( B\otimes I_{t}\right) V_{t}$, where 
$V=\left( V_{t}\right) _{t>0}$ is an (unbounded) cocycle on the product $%
\mathcal{D}\otimes \mathfrak{F}$ with Fock space $\mathfrak{F}$ over the
Hilbert space $L^{2}\left( \mathbb{N\times R}_{+}\right) $ of the quantum
noise, and $I_{t}$ is the solution to the stochastic equation $\mathrm{d}%
I_{t}+\sum_{n\in \mathbb{N}_{0}}I_{t}\mathrm{d}\Lambda _{n}^{n}$ with $%
I_{0}=I$ in $\mathfrak{F}$. The cocycle $V$ satisfies the quantum stochastic
equation $\mathrm{d}V_{t}=(L_{\nu }^{\mu }-I\delta _{\nu }^{\mu })V_{t}%
\mathrm{d}\Lambda _{\mu }^{\nu }$ of the form 
\begin{equation}
\mathrm{d}V_{t}+KV_{t}\mathrm{d}t+\sum_{n=1}^{\infty }K_{n}V_{t}\mathrm{d}%
\Lambda _{-}^{n}=\sum_{m,n=1}^{\infty }\left( L_{n}^{m}-I\delta
_{n}^{m}\right) V_{t}\mathrm{d}\Lambda _{m}^{n}+\sum_{m=1}^{\infty
}L^{m}V_{t}\mathrm{d}\Lambda _{m}^{+},\qquad  \label{4.4}
\end{equation}
where $L_{n}^{i\text{ }}$ and $L^{i}$ are the operators in $\mathcal{D}$,
defining 
\begin{eqnarray}
\varphi _{n}^{m}\left( B\right) &=&\sum_{l\notin \mathbb{N}_{0}}L_{m}^{l\ast
}BL_{n}^{l},\qquad \varphi \left( B\right) =\sum_{l\notin \mathbb{N}%
_{0}}L^{l\ast }BL^{l}  \label{4.5} \\
\varphi ^{m}\left( B\right) &=&\sum_{l\notin \mathbb{N}_{0}}L_{m}^{l\ast
}BL^{l},\qquad \varphi _{n}\left( B\right) =\sum_{l\notin \mathbb{N}%
_{0}}L^{l\ast }BL_{n}^{l}\qquad  \notag
\end{eqnarray}
with $\sum_{i=1}^{\infty }L^{i\ast }L^{i}=K+K^{\dagger }$ if $K+K^{\dagger
}\geq \varphi \left( I\right) =\sum_{l\notin \mathbb{N}_{0}}L^{i\ast }L^{i}$%
. \allowbreak The formal derivation of Eq. (\ref{4.4}) from (\ref{4.3}) is
obtained by a simple application of the HP It\^{o} formula. The martingale $%
M_{t}$, describing the density operator for the output state of $\Lambda
\left( t,a\right) $, is then defined as $M_{t}=V_{t}^{\ast }V_{t}$.

3. The following theorem ensures the existence of a $*$-representation $%
\iota :\Lambda \left( t,a\right) \mapsto \Lambda \left( t,i\left( a\right)
\right) :=i_\beta ^\alpha \left( a\right) \Lambda _\alpha ^\beta \left(
t\right) $ of the quantum stochastic process (\ref{0.2}), commuting with $%
Y_t=\phi _t\left( B\right) $ for all $a\in \mathfrak{a},B\in \mathcal{L}%
\left( \mathcal{H}\right) $, in the form 
\begin{equation*}
\Lambda \left( t,i\left( a\right) \right) =i_{\circ }^{\circ }\left(
a\right) \Lambda _{\circ }^{\circ }\left( t\right) +i_{+}^{\circ }\left(
a\right) \Lambda _{\circ }^{+}\left( t\right) +i_{\circ }^{-}\left( a\right)
\Lambda _{-}^{\circ }\left( t\right) +i_{+}^{-}\left( a\right) \Lambda
_{-}^{+}\left( t\right) \text{.}
\end{equation*}
Here $\boldsymbol{i}\mathbf{=}\left( i_\beta ^\alpha \right) _{\beta
=+,\circ }^{\alpha =-,\circ }$ is a $\star $-representation 
\begin{equation*}
i_\beta ^\alpha \left( a^{\star }a\right) =i_{\circ }^\alpha \left( a^{\star
}\right) i_\beta ^{\circ }\left( a\right) ,\quad i_{-\beta }^\alpha \left(
a^{\star }\right) =i_{-\alpha }^\beta \left( a\right) ^{*}
\end{equation*}
of the It\^{o} algebra $\mathfrak{a}$ in the operators $i_\beta ^\alpha
\left( a\right) :\mathcal{K}_\beta \rightarrow \mathcal{K}_\alpha $, with a
domain $\mathcal{K}_{\circ }\subseteq \mathcal{K}$, $\mathcal{K}_{-}=\mathbb{%
C=}\mathcal{K}_{+}$, and $\Lambda _\alpha ^\beta \left( t\right) $ are the
canonical quantum stochastic integrators in the Fock space $\Gamma \left( 
\mathfrak{K}\right) $ over $\mathfrak{K}=L_{\mathcal{K}}^2\left( \mathbb{R}%
_{+}\right) $, the space of $\mathcal{K}$-valued square-integrable functions
on $\mathbb{R}_{+}$.We shall extend $\boldsymbol{i}$ to the triangular
matrix representation $\mathbf{i=}\left[ i_\beta ^\alpha \right] _{\beta
=-,\circ ,+}^{\alpha =-,\circ ,+}$ on the pseudo-Hilbert space $\mathbb{%
C\oplus }\mathcal{K}\oplus \mathbb{C}$ with the Minkowski metrics tensor $%
\mathbf{g=}\left[ \delta _{-\beta }^\alpha \right] =\mathbf{g}^{-1}$, by $%
i_\beta ^{+}\left( a\right) =0=i_{-}^\alpha \left( a\right) $, for all $a\in 
\mathfrak{a} $, as it was done for $\mathbf{a=}\left[ a_\nu ^\mu \right]
_{\nu =-,\bullet ,+}^{\mu =-,\bullet ,+}$, and denote the ampliation $%
I\otimes i_\beta ^\alpha \left( a\right) $ again as $i_\beta ^\alpha \left(
a\right) $ by omitting the index $\circ $. Note that if the stochastic
generator of the form (\ref{2.10}) is restricted onto an operator algebra $%
\mathcal{B}\subseteq \mathcal{L}\left( \mathcal{H}\right) $ with the weak
closure $\mathcal{\bar{B}}=\mathcal{A}^c$, and all the sesquilinear forms $%
\gamma _\nu ^\mu \left( B\right) $, $B\in \mathcal{B}$ commute with the $*$%
-algebra $\mathcal{A}\subset \mathcal{L}\left( \mathcal{D}\right) $, then $%
\lambda _\nu ^\mu \left( B\right) \in \mathcal{\bar{B}}$.

\begin{proposition}
Let $\boldsymbol{b}=\boldsymbol{\gamma }\left( B\right) -\boldsymbol{\imath }%
\left( B\right) $ satisfy the commutativity conditions (\ref{2.4}) for all $%
a\in \mathfrak{a}$, $B\in \mathcal{L}\left( \mathcal{H}\right) $. Then there
exists a $\star $-representation $a\mapsto \boldsymbol{i}\left( a\right) $
of the It\^{o} algebra $\mathfrak{a}$, defining the operators $i_\beta
^\alpha \left( a\right) :\mathcal{K}_\beta \rightarrow \mathcal{K}_\alpha $,
with $i_\beta ^\alpha \left( a\right) ^{*}\mathcal{K}_\alpha \subseteq 
\mathcal{K}_\beta $, where $\mathcal{K}_{-}=\mathbb{C=}\mathcal{K}_{+}$,
such that $L_\mu ^\alpha \left( I\otimes a_\nu ^\mu \right) =\left( I\otimes
i_\beta ^\alpha \left( a\right) \right) L_\nu ^\beta $ for all $a\in 
\mathfrak{a}:$ 
\begin{eqnarray}
L_{\bullet }a_{\bullet }^{\bullet } &=&i\left( a\right) L_{\bullet },\quad
\quad a_{+}^{-}-K_{\bullet }a_{+}^{\bullet }=i^{-}\left( a\right)
L+i_{+}^{-}\left( a\right) ,  \label{4.6} \\
L_{\bullet }a_{+}^{\bullet } &=&i\left( a\right) L+i_{+}\left( a\right)
,\quad \quad a_{\bullet }^{-}-K_{\bullet }a_{\bullet }^{\bullet
}=i^{-}\left( a\right) L_{\bullet }.  \notag
\end{eqnarray}
If $\left[ A,\gamma _\nu ^\mu \left( B\right) \right] =0$ for all $A\in 
\mathcal{A}$ and $B\in \mathcal{B}$, where $\mathcal{B}\subseteq \mathcal{L}%
\left( \mathcal{H}\right) $ is a $*$-algebra of bounded operators, and $%
\mathcal{\bar{B}}=\mathcal{A}^c$, then there exists a triangular $\star $%
-representation $\mathbf{j=}\left[ j_\beta ^\alpha \right] _{\beta =-,\circ
,+}^{\alpha =-,\circ ,+}$ of the operator algebra $\mathcal{A}$ with $%
j_{\circ }^{\circ }\left( I\right) =J$ such that 
\begin{equation}
\mathbf{J\QTR{mathbf}{LA}=\QTR{mathbf}{j}}\left( A\right) \mathbf{%
\QTR{mathbf}{L,\quad }}\left[ \mathbf{j}\left( A\right) ,\mathbf{i}\left(
a\right) \right] =0\mathbf{,\quad }\left[ \mathbf{j}\left( A\right) ,\text{%
\textbf{\emph{\j }}}\left( B\right) \right] =0,\quad \forall A\in \mathcal{A}%
,a\in \mathfrak{a},B\in \mathcal{B}.  \label{4.7}
\end{equation}
\end{proposition}

\proof%
Let $\mbox{\boldmath$\mathcal G$}=\mathcal{G}_{-}\oplus \mathcal{G}\oplus 
\mathcal{G}_{+}$ be the pseudo-Hilbert space, where $\mathcal{G}_{+}=%
\mathcal{D}=\mathcal{G}_{-}$, $\mathcal{G}\subseteq \mathcal{H}\otimes 
\mathcal{K}$ is the linear span of $\left\{ \jmath \left( B\right) 
\boldsymbol{L}\boldsymbol{\eta }|B\in \mathcal{B},\boldsymbol{\eta }\in 
\mathcal{D}\oplus \mathcal{D}_{\bullet }\right\} $ and the indefinite
metrics is defined by 
\begin{equation*}
\left\langle \xi ^\alpha |g_{\alpha \beta }\xi ^\beta \right\rangle =\left\|
\xi \right\| ^2+2\func{Re}\left\langle \xi ^{+}|\xi ^{-}\right\rangle ,\quad
\xi ^\alpha \in \mathcal{G}_\alpha ,\xi ^{\circ }=\xi \in \mathcal{G}.
\end{equation*}
The algebra $\mathcal{B}=\mathcal{L}\left( \mathcal{H}\right) $ is
represented on $\mbox{\boldmath$\mathcal G$}$ by the ampliation \textbf{\j }$%
\left( B\right) =B\otimes \mathbf{J}$, where $\mathbf{J}=1\oplus J\oplus 1$,
and \textbf{\j }$\left( B\right) \mathbf{Lg}\boldsymbol{\eta }\in %
\mbox{\boldmath$\mathcal G$}_{\circ }$, where the pre-Hilbert space $%
\mathcal{D}\oplus \mathcal{D}_{\bullet }$ is isometrically embedded into $%
\mathcal{D}\oplus \mathcal{D}_{\bullet }\oplus \mathcal{D}$ as $\mathbf{g}%
\left( \eta \oplus \eta ^{\bullet }\right) =0\oplus \eta ^{\bullet }\oplus
\eta $. We define the representations $\mathbf{i}$ and $\mathbf{j}$ on $%
\mbox{\boldmath$\mathcal G$}$ by intertwining 
\begin{equation*}
\mathbf{i}\left( a\right) \mathbf{L=La},\quad \mathbf{i}\left( a\right) 
\text{\textbf{\j }}\left( B\right) \mathbf{L}=\text{\textbf{\j }}\left(
B\right) \mathbf{La,\quad j}\left( A\right) \text{\textbf{\j }}\left(
B\right) \mathbf{L}=\text{\textbf{\j }}\left( B\right) \mathbf{LA},
\end{equation*}
the operators $\mathbf{a=}I\otimes \boldsymbol{a}\mathbf{g}$ and $\mathbf{A}%
=A\otimes \mathbf{I}$. Such a definition is correct, because if \textbf{\j }$%
\left( B_k\right) \mathbf{L}\boldsymbol{\zeta }^k=0$ for a finite family of
non-zero $\boldsymbol{\zeta }^k\in \mathcal{D}\oplus \mathcal{D}_{\bullet }$%
, then 
\begin{equation*}
\left( \text{\textbf{\j }}\left( B\right) \mathbf{Lg}\boldsymbol{\eta }|%
\mathbf{i}\left( a\right) \text{\textbf{\j }}\left( B_k\right) \mathbf{Lg}%
\boldsymbol{\zeta }^k\right) =\left( \text{\textbf{\j }}\left( B\right) 
\mathbf{Lg}\boldsymbol{\eta }|\text{\textbf{\j }}\left( B_k\right) \mathbf{%
Lag}\boldsymbol{\zeta }^k\right) =
\end{equation*}
\begin{equation*}
\left\langle \boldsymbol{\eta }|\boldsymbol{\gamma }\left( B^{*}B_k\right) 
\mathbf{g}\boldsymbol{a}\boldsymbol{\zeta }^k\right\rangle =\left\langle 
\boldsymbol{\eta }|\boldsymbol{a}\mathbf{g}\boldsymbol{\gamma }\left(
B^{*}B_k\right) \boldsymbol{\zeta }^k\right\rangle =
\end{equation*}
\begin{equation*}
\left\langle \boldsymbol{a}^{\star }\boldsymbol{\eta }|\mathbf{g}\boldsymbol{%
\gamma }\left( B^{*}B_k\right) \boldsymbol{\zeta }^k\right\rangle =\left( 
\text{\textbf{\j }}\left( B\right) \mathbf{La}^{\star }\mathbf{g}\boldsymbol{%
\eta }|\text{\textbf{\j }}\left( B_k\right) \mathbf{Lg}\boldsymbol{\zeta }%
^k\right) =0
\end{equation*}
for all $\boldsymbol{\eta }\in \mathcal{D}\oplus \mathcal{D}_{\bullet }$ and 
$B\in \mathcal{L}\left( \mathcal{H}\right) $, and so $\mathbf{i}\left(
a\right) $\textbf{\j }$\left( B_k\right) \mathbf{L}\boldsymbol{\zeta }^k=0$.
Here we used the condition 
\begin{equation*}
\boldsymbol{\gamma }\left( B\right) \mathbf{g}\boldsymbol{a}=\left( \text{%
\textbf{\i }}\left( B\right) +\mathbf{b}\right) \mathbf{ag=a}\left( \text{%
\textbf{\i }}\left( B\right) +\mathbf{b}\right) \mathbf{g}=\boldsymbol{a}%
\mathbf{g}\boldsymbol{\gamma }\left( B\right) ,
\end{equation*}
as $\mathbf{ab}=\mathbf{ba}$ due to the HP commutativity $\boldsymbol{a}%
\mathbf{g}\boldsymbol{b}=\boldsymbol{b}\mathbf{g}\boldsymbol{a}$ of $%
\boldsymbol{b}=\boldsymbol{\lambda }\left( B\right) $ where $\boldsymbol{%
\gamma }=(\gamma _\nu ^\mu )$ is extended to all indexes as $\gamma _\nu
^\mu (B)=B\delta _\nu ^\mu +C_\nu ^\mu $ with $C_\nu ^\mu =0$ if $\mu =+$ or 
$\nu =-$. In the same way, the operators $\mathbf{j}\left( A\right) $ are
correctly defined for $B\in \mathcal{A}^c$ if $\boldsymbol{\gamma }\left(
B\right) $\textbf{\i }$\left( A\right) =$\textbf{\i }$\left( A\right) 
\boldsymbol{\gamma }\left( B\right) $. This also proves that $\mathbf{i}%
\left( a\right) ^{\star }=\mathbf{i}\left( a^{\star }\right) $, and $\mathbf{%
j}\left( A\right) ^{\star }=\mathbf{j}\left( A^{*}\right) $. (The
multiplicativity of $\mathbf{i,j}$ as well as the commutativity properties (%
\ref{4.7}) directly follow from the definition of these operators.) Note
that $\mathbf{j}\left( I\right) =I\otimes \mathbf{J}$, and if the dilation
is minimal, $\mathbf{j}\left( I\right) =I\otimes \mathbf{I}$. If it is not,
the unital property can still be achieved for the canonical dilations in $%
\mathcal{K}_{\circ }$, by adding $j_{\circ }^{\circ }\left( A\right) \left(
\eta \otimes e_0\right) =A\eta \otimes e_0$ for all $e_0\in \mathcal{K}$
with $Je_0=0$. 
\endproof%

4. Now we are going to construct the quantum stochastic dilation for the
flow $\phi _t\left( B\right) $ and the quantum state generating function $%
\vartheta _t^a=\epsilon \left[ R_tW\left( t,a\right) \right] $ of the output
process $\boldsymbol{\Lambda }\left( t,a\right) $ in the form 
\begin{equation*}
\phi _t\left( B\right) =V_t^{*}\left( I_t\otimes B\right) V_t,\quad
\vartheta _t\left( g\right) =\epsilon \left[ V_t^{*}\left( W_t^a\otimes
I\right) V_t\right] ,\quad \forall B\in \mathcal{L}\left( \mathcal{H}\right)
,a\in \mathfrak{a},
\end{equation*}
where $V_t$ is an operator on $\mathcal{D}\otimes \mathfrak{F}$ into $\Gamma
\left( \mathfrak{K}\right) \otimes \mathcal{D}\otimes \mathfrak{F}$,
intertwining the Weyl operators $W\left( t,a\right) $ with the operators $%
W_t^a=W\left( t,i\left( a\right) \right) I_t$ in the Fock space $\Gamma
\left( \mathfrak{K}\right) $, 
\begin{equation*}
\mathrm{d}W\left( t,i\left( a\right) \right) =W\left( t,i\left( a\right)
\right) \mathrm{d}\Lambda \left( t,i\left( a\right) \right) ,\quad W\left(
0,i\left( a\right) \right) =I,
\end{equation*}
and $I_t\geq I_s,\forall t\leq s$ is a decreasing family of orthoprojectors.

In order to prove the existence of the Fock space dilation we need the
following assumptions in addition to the continuity assumptions of this and
previous sections.

\begin{enumerate}
\item[1)] The minimal quantum stochastic CP flow over the algebra $\mathcal{A%
}$, resolving the quantum Langevin equation 
\begin{equation}
\mathrm{d}\tau _t\left( A\right) =\tau _t\left( \mathbf{j}\left( A\right) -%
\text{\textbf{\i }}\left( A\right) \right) \cdot \mathrm{d}\boldsymbol{%
\Lambda },\quad \tau _0\left( A\right) =I\otimes A,\quad A\in \mathcal{A},
\label{4.2}
\end{equation}
where $\mathbf{j}\left( I\right) =I\otimes \mathbf{J}$, \textbf{\i }$%
(A)=A\otimes $ $\mathbf{I}$, is the multiplicative flow, satisfying the
condition $\tau _t\left( I\right) =I_t\otimes I$, where $I_t$ is the
solution to the stochastic equation $\mathrm{d}I_t=\left( J-I\right) _{\circ
}^{\circ }I_t\mathrm{d}\Lambda _{\circ }^{\circ }$ with $I_0=I$.

\item[2)] Let us assume the strong continuity of the operators $L\left( \bar{%
e}\right) :\mathcal{D}\rightarrow \mathcal{D}$, $L_{\bullet }\left( \bar{e}%
\right) :\mathcal{D}_{\bullet }\rightarrow \mathcal{D}$, given for all $e\in 
\mathcal{K}$ as $L_\nu \left( \bar{e}\right) =\left( I\otimes e^{*}\right)
L_\nu $ by $\left\langle L_\nu \left( \bar{e}\right) \eta |\eta ^{\prime
}\right\rangle =\left\langle L_\nu \eta |\eta ^{\prime }\otimes
e\right\rangle \quad \forall \eta \in \mathcal{D},\eta ^{\prime }\in 
\mathcal{D}^{\prime }$. This is necessary for the definition of the
operators $V_t\left( \sigma \right) $ for each subset $\sigma \subset [0,t)$
of a finite cardinality $\left| \sigma \right| \in \mathbb{N}$ by the
recurrence 
\begin{equation*}
V_t\left( \sigma \right) \psi =V_t^{\circ }\left( s\right) \left( LV_s\left(
\sigma \backslash s\right) \psi +\sum_mL_mV_r\left( \sigma \backslash
s\right) \psi ^m\left( s\right) \right) ,\quad s=\max \sigma ,
\end{equation*}
with $V_t(\emptyset )=V_t^{\circ }$. Here $V_t^{\circ }\left( s\right)
=T_t^{*}S_{t-s}T_s$, $V_t^{\circ }$ is the solution to Eq. (\ref{3.1'}) in $%
\mathcal{D}\otimes \mathfrak{F}$ acting as $I_{\circ }^{\otimes \left|
\sigma \right| }\otimes V_t^{\circ }$ on $\mathcal{K}^{\otimes \left| \sigma
\right| }\otimes \mathcal{D}\otimes \mathfrak{F}$, the operators $L_\nu :%
\mathcal{D}\rightarrow \mathcal{K}\otimes \mathcal{H}$ act on $\mathcal{K}%
^{\otimes \left| \sigma \backslash s\right| }\otimes \mathcal{H}\otimes 
\mathfrak{F}$ as $I_{\circ }^{\otimes \left| \sigma \backslash s\right|
}\otimes L\otimes I$ ($s$ is identified with the single point subset $%
\left\{ s\right\} $ such that $\sigma \backslash \max \sigma $ is the $%
\sigma $ without its maximum,) and $\psi ^{\bullet }\left( s\right) \in 
\mathcal{K}\otimes \mathcal{H}\otimes \mathfrak{F}$ is given as $\psi
^{\bullet }\left( \tau ,s\right) =\psi \left( \tau \sqcup s\right) $ of $%
\psi \in \mathcal{H}\otimes \mathfrak{F}$, where $\tau \sqcup s$ is defined
for almost all $s$ ($s\notin \sigma $) as the disjoint union of the single
point $\left\{ s\right\} $ with a finite subset $\tau \in \mathbb{R}_{+}$.

\item[3)] The operator-valued function $\sigma \mapsto $ $V_{t}\left( \sigma
\right) $, defined for all such $\sigma \in \Gamma _{t}$, is weakly square
integrable for each $t$ with respect to the measure $\mathrm{d}\sigma
=\prod_{s\in \sigma }\mathrm{d}s$ in the sense 
\begin{equation*}
\int_{\Gamma _{t}}\left\| V_{t}\left( \sigma \right) \psi \right\| ^{2}%
\mathrm{d}\sigma :=\sum_{n=0}^{\infty }\underset{0<s_{1}\ldots s_{n}<t}{\int
\ldots \int }\left\| V_{t}\left( s_{1,\ldots ,}s_{n}\right) \psi \right\|
^{2}\mathrm{d}s_{1}\ldots \mathrm{d}s_{1}<\infty ,
\end{equation*}
for all $\psi \in \mathcal{D}\otimes \mathfrak{F}.$ Thus the operators $%
V_{t} $ can be extended to the Fock space ones $V_{t}:$ $\mathcal{D}\otimes 
\mathfrak{F}\rightarrow \Gamma \left( \mathfrak{K}\right) \otimes \mathcal{D}%
\otimes \mathfrak{F} $, say, by letting $V_{t}\left( \sigma \right)
=V_{t}\left( \sigma _{t}\right) \otimes \delta _{\emptyset }\left( \sigma
_{\lbrack t}\right) \quad $for all finite $\sigma \subset \mathbb{R}_{+}$ if 
$\sigma _{\lbrack t}=\sigma \cap \lbrack t,\infty )\neq \emptyset $.
Obviously they form a cocycle, $V_{t-r}^{r}\left( \sigma \right) V_{r}\left(
\sigma \right) =V_{t}\left( \sigma \right) $, where $V_{s}^{r}\left( \sigma
\right) =I_{\circ }^{\otimes \left| \sigma _{r}\right| }\otimes T_{r}^{\ast
}V_{s}\left( \sigma _{\lbrack r}-r\right) T_{r}$ with $\sigma _{r}=\sigma
\cap \lbrack 0,r)$.
\end{enumerate}

\begin{theorem}
Under the given assumptions 0), 1), 2), 3) there exist:

\begin{enumerate}
\item[(i)] A cocycle dilation $V_t:\mathcal{D}\otimes \mathfrak{F}%
\rightarrow \Gamma \left( \mathfrak{K}\right) \otimes \mathcal{D}\otimes 
\mathfrak{F}$ of the minimal CP flow $\phi $, intertwining the Weyl operator 
$W\left( t,a\right) $ with $W_t^a$: 
\begin{equation}
V_t\left( I\otimes W\left( t,a\right) \right) =\left( W_t^a\otimes I\right)
V_t,\quad \phi _t\left( B\right) =V_t^{*}\left( I_t\otimes B\right) V_t\text{
},\quad \forall a\in \mathfrak{a},B\in \mathcal{L}\left( \mathcal{H}\right) ,
\label{4.8}
\end{equation}
where $I_t\leq I_s$, $\forall t<s$ are orthoprojectors in $\Gamma \left( 
\mathfrak{K}\right) $.

\item[(ii)] A $*$-multiplicative flow $\tau =\left( \tau _t\right) $ over $%
\mathcal{A}$ in $\Gamma \left( \mathfrak{K}\right) \otimes \mathcal{H}$ with
the properties $\tau _t\left( I\right) =I_t$, 
\begin{equation}
V_tA=\tau _t\left( A\right) V_t,\quad \left[ \tau _t\left( A\right) ,W_t^a%
\right] =0,\quad \left[ \tau _t\left( A\right) ,I\otimes B\right] =0,\quad
\forall A\in \mathcal{A},a\in \mathfrak{a},B\in \mathcal{B}\text{.}
\label{4.9}
\end{equation}

\item[(iii)] If $\lambda \left( I\right) \leq 0$, then one can make $%
M_t=V_t^{*}V_t$ martingale, and, if $\boldsymbol{\lambda }\left( I\right)
\leq 0$, one can make $V_t$ isometric, $V_t^{*}V_t=I$.

\item[(iv)] Moreover, let $U=\left( U_t\right) _{t>0}$ be a one parametric
weakly continuous cocycle of unitary operators on $\Gamma \left( \mathfrak{K}%
\right) \otimes \mathcal{H}\otimes \Gamma \left( \mathfrak{E}\right) $ ,
giving the unique solution to the quantum stochastic equation 
\begin{eqnarray}
&&\mathrm{d}U_t+\left( K\mathrm{d}t+K_{\bullet }^{-}\mathrm{d}\Lambda
_{-}^{\bullet }+K_{\circ }^{-}\mathrm{d}\Lambda _{-}^{\circ }\right) U_t
\label{4.12} \\
&=&\left( L_{+}^{\circ }\mathrm{d}\Lambda _{\circ }^{+}-I_{\bullet
}^{\bullet }\mathrm{d}\Lambda _{\bullet }^{\bullet }+J_{\bullet }^{\circ }%
\mathrm{d}\Lambda _{\circ }^{\bullet }+J_{\circ }^{\bullet }\mathrm{d}%
\Lambda _{\bullet }^{\circ }+\left( J_{\circ }^{\circ }-I_{\circ }^{\circ
}\right) \mathrm{d}\Lambda _{\circ }^{\circ }\right) U_t  \notag
\end{eqnarray}
with $U_0=I$ and the necessary differential unitarity conditions 
\begin{equation*}
K+K^{\dagger }=L_{\circ }^{-}L_{+}^{\circ },K_{\bullet }^{-}=L_{\circ
}^{-}J_{\bullet }^{\circ },J_{\circ }^{\bullet }J_{\bullet }^{\circ
}=I_{\bullet }^{\bullet },K_{\circ }^{-}=L_{\circ }^{-}J_{\circ }^{\circ
},J_{\circ }^{\circ }=I_{\circ }^{\circ }-J_{\bullet }^{\circ }J_{\circ
}^{\bullet },
\end{equation*}
where $L_{\circ }^{-}=L_{+}^{\circ *}$, $J_{\circ }^{\bullet }=J_{\bullet
}^{\circ *}$. If $\lambda \left( I\right) \leq 0$, and $L_{+}^{\circ
}=L^{\circ }$ is the canonical operator in the dilation (\ref{4.0}), then 
\begin{equation}
\left\langle \psi |\left( A\otimes I\right) \phi _t^a\left( B\right) \psi
\right\rangle =\left\langle U_t\left( \delta _\emptyset \otimes \psi \right)
|\left( \tau _t^a\left( A\right) \left( I\otimes B\right) \right) U_t\left(
\delta _\emptyset \otimes \psi \right) \right\rangle  \label{4.11}
\end{equation}
for all $A\in \mathcal{A},a\in \mathfrak{a},B\in \mathcal{B}$, where $\psi $
is any initial state $\psi _0=\eta \otimes \delta _\emptyset $, $\eta \in 
\mathcal{D}$, and 
\begin{equation*}
\phi _t^a\left( B\right) =\left( I\otimes W\left( t,a\right) \right) \phi
_t\left( B\right) ,\quad \tau _t^a\left( A\right) =\left( W_t^a\otimes
I\right) \tau _t\left( A\right) .
\end{equation*}
If $\boldsymbol{\lambda }\left( I\right) \leq 0$, and in addition $%
J_{\bullet }^{\circ }=L_{\bullet }^{\circ }$ is the canonical isometry in (%
\ref{4.0}), this unitary cocycle dilation is valied also for any
vector-state $\psi \in \mathcal{D}\otimes \mathfrak{F}$ .
\end{enumerate}
\end{theorem}

\proof%
(Sketch). The cocycle $V=\left( V_t\right) _{t>0}$ is recurrently
constructed due to the above assumptions 0)--3). It obviously intertwines
the Weyl operators (\ref{2.1b}) with the operators $W_t^a$, acting in the
same way in $\Gamma \left( \mathfrak{K}\right) $, by virtue of the property (%
\ref{4.6})$.$

Let us denote by $\mathfrak{K}_1=L_{\mathcal{K}}^2\left( \mathbb{R}%
_{+}\right) $ the functional Hilbert space corresponding to the minimal
dilation (\ref{4.0}) sub-space $\mathcal{K}=\mathcal{K}_1$ for the CP map $%
\boldsymbol{\varphi }$, given by the orthoprojector $J=J_1$ in the space $%
\mathcal{K}_{\circ }$ of the canonical dilation, and $\mathfrak{K}_0$ its
orthogonal compliment, corresponding to $\mathcal{K}_0=J_0\mathcal{K}_{\circ
}$, where $J_0=I-J_1$. Representing $\Gamma \left( \mathfrak{K}_0\oplus 
\mathfrak{K}_1\right) $ as $\Gamma \left( \mathfrak{K}_0\right) \otimes
\Gamma \left( \mathfrak{K}_1\right) $, let us denote by $I_t$ the survival
orthoprojectors $I_t\chi \left( \sigma ^0,\sigma ^1\right) =\delta
_\emptyset \left( \sigma _t^0\right) \chi \left( \sigma ^0,\sigma ^1\right) $%
, $\sigma _t=\sigma \cap [0,t)$, where $\chi \left( \sigma ^0,\sigma
^1\right) =\chi \left( \sigma ^0\sqcup \sigma ^1\right) \in \mathcal{K}%
^{\otimes \left| \sigma ^0\right| }\otimes \mathcal{K}^{\otimes \left|
\sigma ^1\right| }$ is the set function, representing a $\chi \in \Gamma
\left( \mathfrak{K}_0\oplus \mathfrak{K}_1\right) $. The decreasing family $%
\left( I_t\right) _{t>0}$ defines the decay orthoprojectors $E_t=I-I_t$ in $%
\Gamma \left( \mathfrak{K}_{\circ }\right) $ satisfying the quantum
stochastic equation $\mathrm{d}E_t=E_tJ_0\cdot \mathrm{d}\Lambda _{\circ
}^{\circ }$ with $E_0=0$, and $\Lambda _{\circ }^{\circ }$ is the number
integrator in the Fock space $\Gamma \left( \mathfrak{K}_{\circ }\right) $
over $\mathfrak{K}_{\circ }=\mathfrak{K}_0\oplus \mathfrak{K}_1$. Then one
easily find that the minimal CP flow (\ref{3.8'}) can be represented as $%
\phi _t\left( B\right) =V_t^{*}\left( I_t\otimes B\right) V_t$.

We may also construct the minimal quantum stochastic $*$-flow \cite{EvH}
over the operator algebra $\mathcal{A}$, resolving the quantum Langevin
equation (\ref{4.2}) by its iteration as it was done in Sect 4 for the flow $%
\phi $, and then prove its $*$-multiplicativity under certain conditions as
in \cite{18}. However, we can directly construct the representations $\tau
_t $ with the property $\tau _t\left( I\right) =I_t$ in a similar way as it
was done for the representation $\mathbf{j}$, and then prove that it
satisfies the Langevin equation. Then the properties (\ref{4.9}) follow from
the definition of the operators $V_t$, and can be checked recurrently by use
of (\ref{4.6}) and (\ref{4.7}).

The cocycle $U=\left( U_t\right) $ is constructed to satisfy the HP quantum
stochastic equation (\ref{4.12}). It can be represented in the form of the
stochastic multiple integral of the chronologically ordered products of the
coefficients of the quantum differential equation under the integrability
conditions given in the Appendix.

If $K+K^{\dagger }\geq \varphi \left( I\right) $, the HP unitarity condition 
\cite{16} is satisfied for the canonical choice $L_{+}^{\circ }=L^{\circ }$,
where $L^{\circ }=L^0+L^1$, and arbitrary isometric operator $J_{\bullet
}^{\circ }$, $J_{\circ }^{\bullet }J_{\bullet }^{\circ }=I_{\bullet
}^{\bullet }$ with $K_{\bullet }^{-}=L_{\circ }J_{\bullet }^{\circ }$, $%
K_{\circ }^{-}=L_{\circ }J_{\circ }^{\circ }$, $J_{\circ }^{\circ }=I_{\circ
}^{\circ }-J_{\bullet }^{\circ }J_{\circ }^{\bullet }$. In addition if $%
\boldsymbol{K}+\boldsymbol{K}^{\dagger }\geq \boldsymbol{\varphi }\left(
I\right) $, we make the choice $J_{\bullet }^{\circ }=L_{\bullet }^{\circ }$
from the canonical dilation, $L_{\bullet }^{\circ }=L_{\bullet
}^0+L_{\bullet }^1$, and so $K_{\bullet }^{-}=L_{\circ }L_{\bullet }^{\circ
}=K_{\bullet }$, where $L_{\circ }^{*}=L^{\circ }$, $J_{\circ }^{\bullet
}=J_{\bullet }^{\circ *}$. In the first, subfiltering case$\ \lambda \left(
I\right) \leq 0$ such a choice gives $U_t\left( \delta _\emptyset \otimes
\psi _0\right) =V_t\psi _0$ for any $\psi _0=\eta \otimes \delta _\emptyset $%
, $\eta \in \mathcal{D}$ and therefore $\left\| V_t\psi _0\right\| =\left\|
\psi _0\right\| $. Thus $M_t=V_t^{*}V_t$ is a martingale and the condition (%
\ref{4.11}) is satisfied for any initial $\psi _0$. In the second,
contractive case $\boldsymbol{\lambda }\left( I\right) \leq 0$ the canonical
choice gives $U_t\left( \delta _\emptyset \otimes \psi \right) =V_t\psi $
and therefore $\left\| V_t\psi \right\| =\left\| \psi \right\| $ for any $%
\psi \in \mathcal{D}\otimes \mathfrak{F}$. Thus $V_t^{*}V_t=I$ and the
condition (\ref{4.11}) is satisfied for any $\psi $. 
\endproof%

\section{Appendix}

Here we give a resume on the sufficient analytical conditions for the
quantum multiple integration \cite{Bcs} of stochastic linear differential
equations in Hilbert spaces, based on the noncommutative analysis in the
Fock scale \cite{20}.

1. Let $\left\| e^{\bullet }\right\| ^2\left( \xi \right) =\xi \left\|
e^{\bullet }\right\| ^2,\xi >0$ as in \cite{20}, so that the projective
limit $\mathcal{E}$ and the dual space $\mathcal{E}^{\prime }$ coincide with
the Hilbert space $\mathcal{K}$ with the norm $\left\| e\right\| ^2$. The
projective Fock space $\mathfrak{F}=\cap _\xi \Gamma \left( \mathfrak{K},\xi
\right) $ over $\mathfrak{K}=L_{\mathcal{K}}^2\left( \mathbb{R}_{+}\right) $
with respect to the exponential scale (\ref{2.0a}), where $\left\|
f^{\otimes }\right\| ^2\left( \xi \right) =\exp \left[ \xi \left\|
f^{\bullet }\right\| ^2\right] $, $f^{\bullet }\in \mathfrak{K}$, is the
natural domain for the quantum stochastic integration \cite{Bcs}, and $%
\mathfrak{F}^{\prime }=\cup _\xi \Gamma \left( \mathfrak{K},\xi ^{-1}\right) 
$. If $\mathcal{D}=\cap \mathcal{H}_p $ is the projective limit of an
increasing family of the dense Hilbert subspaces $\mathcal{H}_p\subseteq 
\mathcal{H}_{p-1}$, the $\pi $-product $\mathfrak{D}=\mathcal{D}\otimes 
\mathfrak{F}$ of the Fr\'{e}chet spaces $\mathcal{D}$ and $\mathfrak{F}$ is
the projective limit of the directed family of the spaces $\mathfrak{H}%
_p\left( \xi \right) =$ $\mathcal{H}_p\otimes \Gamma \left( \mathfrak{K},\xi
\right) $ and $\mathfrak{D}^{\prime }=\mathcal{D}^{\prime }\otimes \mathfrak{%
F}^{\prime }$ is given as $\cup \mathfrak{H}_{-p}\left( \xi ^{-1}\right) $,
where $\mathcal{H}_{-p}$ denote the duals $\mathcal{H}_p^{\prime }$ to the
Hilbert spaces $\mathcal{H}_p$, with respect to the standard pairing in the
Hilbert product $\mathfrak{H}$ of $\mathcal{H}=\mathcal{H}_0$ and $\Gamma
\left( \mathfrak{K}\right) =\Gamma \left( \mathfrak{K},1\right) $ .
Following \cite{20, Bcs}, we define the multiple quantum stochastic integral 
$Y_t=\Lambda _{[0,t)}^{\otimes }\left( B\right) $ of a function $B\left( 
\boldsymbol{\tau }\right) $ of the quadruple $\boldsymbol{\tau }=\left( \tau
_\nu ^\mu \right) _{\nu =+,\bullet }^{\mu =-,\bullet }$ of finite subsets $%
\tau _\nu ^\mu \subset [0,t)$ with values in the nonadapted kernels $%
\mathfrak{D}\otimes \mathcal{K}^{\otimes \left| \tau \cup \tau ^{-}\right|
}\rightarrow \mathfrak{D}^{\prime }\otimes \mathcal{K}^{\otimes \left| \tau
\cup \tau _{+}\right| }$, as the sesquilinear form $\left\langle \psi
|Y_t\psi \right\rangle =$ 
\begin{equation}
\int_{\Gamma _t}\int_{\Gamma _t}\int_{\Gamma _t}\int_{\Gamma _t}\left\langle 
\overset{\bullet }{\psi }\left( \tau \cup \tau _{+}\right) |B\left( 
\begin{tabular}{ll}
$\tau _{+}^{-}$ & $\tau ^{-}$ \\ 
$\tau _{+}$ & $\tau $%
\end{tabular}
\right) \overset{\bullet }{\psi }\left( \tau \cup \tau ^{-}\right)
\right\rangle \mathrm{d}\tau \mathrm{d}\tau ^{-}\mathrm{d}\tau _{+}\mathrm{d}%
\tau _{+}^{-},  \label{A.1}
\end{equation}
where $\overset{\bullet }{\psi }\left( \sigma ,\tau \right) =\psi \left(
\sigma \cup \tau \right) $, given by the quadruple of the multiple integrals 
\begin{equation*}
\int_{\Gamma _t}B\left( \tau \right) \mathrm{d}\tau =\sum_{n=0}^\infty 
\underset{0<t_1,<\ldots <t_n<t}{\idotsint }B\left( t_1,\ldots ,t_n\right) 
\mathrm{d}t_1\cdots \mathrm{d}t_n.
\end{equation*}
The function $B$ is integrable up to a $t>0$ if $Y_t\in \mathcal{B}\left( 
\mathfrak{D}\right) $, and it is strongly integrable if $Y_t\in \mathcal{L}%
\left( \mathfrak{D}\right) $. The natural criterion of multiple
integrability was formulated in \cite{20} in terms of the norms $\left\|
B\right\| _{p,q}^{\prime }\left( \xi ,\zeta \right) =$ 
\begin{equation}
\int_{\Gamma _t}\left( \int_{\Gamma _t}\int_{\Gamma _t}\left( \frac 1\zeta
\right) ^{\left| \tau ^{-}\right| }\left( \frac 1\xi \right) ^{\left| \tau
_{+}\right| }\sup_{\tau \in \Gamma _t}\left( \left( \frac 1{\xi \zeta
}\right) ^{\frac{\left| \tau \right| }2}\left\| B\left( \boldsymbol{\tau }%
\right) \right\| _{p,q}^{^{\prime }}\right) ^2\mathrm{d}\tau _{+}\mathrm{d}%
\tau ^{-}\right) ^{\frac 12}\mathrm{d}\tau _{+}^{-}  \label{A.2}
\end{equation}
as $\left\| B\right\| _{p,q}^{^{\prime }}\left( \xi ,\zeta \right) <\infty $
for some $p,q,\xi ,\zeta <\infty $. The function $B$ is strongly integrable
if $\left\| B\right\| _{p,q}^{^{\prime }}\left( \xi ,\zeta \right) <\infty $
for any $p<0,\xi <1$ and some $q,\zeta $. Here 
\begin{equation}
\left\| B\left( 
\begin{tabular}{ll}
$\tau _{+}^{-}$ & $\tau ^{-}$ \\ 
$\tau _{+}$ & $\tau $%
\end{tabular}
\right) \right\| _{p,q}^{^{\prime }}\left( \xi ,\zeta \right) =\sup_{\psi
,\chi }\frac{\left| \left\langle \psi _{n,n_{+}}|B\left( \boldsymbol{\tau }%
\right) \chi _{n,n^{-}}\right\rangle \right| }{\left\| \psi
_{n,n_{+}}\right\| _p\left( \zeta \right) \left\| \chi _{n,n^{-}}\right\|
_q\left( \zeta \right) }  \label{A.8}
\end{equation}
denotes the norm of the kernel $B\left( \boldsymbol{\tau }\right) :\mathfrak{%
H}_q\left( \zeta \right) \otimes \mathcal{K}^{\otimes n}\otimes \mathcal{K}%
^{\otimes n_{-}}\rightarrow \mathfrak{H}_{-p}\left( \xi ^{-1}\right) \otimes 
\mathcal{K}^{\otimes n}\otimes \mathcal{K}^{\otimes n_{+}}$, where $n_\nu
^\mu =\left| \tau _\nu ^\mu \right| $ are the cardinalities of $\tau _\nu
^\mu $. The norms (\ref{A.2}) define the estimate for the integral by virtue
of the inequality \cite{20} 
\begin{equation*}
\left\| \Lambda _{[0,t)}^{\otimes }\left( B\right) \right\| _{p,q}^{\prime
}\left( 3\xi ,3\zeta \right) \leq \left\| B\right\| _{p,q}^{^{\prime
}}\left( \xi ,\zeta \right)
\end{equation*}
so that $\Lambda _{[0,t)}^{\otimes }\left( B\right) \in \mathcal{B}\left( 
\mathfrak{D}\right) $ ( or $\in \mathcal{L}\left( \mathfrak{D}\right) $) is
defined as a bounded kernel $Y_t:\mathfrak{H}_q\left( \zeta \right)
\rightarrow \mathfrak{H}_{-p}\left( \xi ^{-1}\right) $ if $\left\| B\right\|
_{p,q}^{^{\prime }}\left( \frac 13\xi ,\frac 13\zeta \right) <\infty $ ( or
if for each $p,\xi $ the norm is finite for some $q,\zeta $).

2. Let us consider the case when the operator-valued function $B\left( 
\boldsymbol{\tau }\right) $ is relatively bounded in the following sense 
\begin{equation}
\left\| B\left( 
\begin{tabular}{ll}
$\tau _{+}^{-}$ & $\tau ^{-}$ \\ 
$\tau _{+}$ & $\tau $%
\end{tabular}
\right) \right\| _{p,q}^{^{\prime }}\leq \frac{\left(
n+n_{+}+n^{-}+n_{+}^{-}\right) !}{n!\sqrt{n_{+}!n^{-}!}}c_{p,q}\left(
n+n_{+}+n^{-}+n_{+}^{-}\right) ,  \label{A.3}
\end{equation}
where $c_{p,q}\left( n\right) $ are positive constants such that $%
\sum_{n=0}^\infty c_{p,q}\left( n\right) \rho ^n<\infty $ for a strictly
positive number $\rho $ and sufficiently large $\xi $ or $\zeta $.
Integrating $\underset{0<t_1,<\ldots <t_n<t}{\idotsint }\mathrm{d}t_1\cdots 
\mathrm{d}t_n=t^n/n!$ three times, one can find that $\left\| B\right\|
_{p,q}^{^{\prime }}\left( \xi ,\zeta \right) \leq $ 
\begin{eqnarray*}
&&\sum_{n_{+}^{-}}\frac{\left( t\sqrt{\xi \zeta }\right) ^{n_{+}^{-}}}{%
n_{+}^{-}!}\left( \sum_{n^{-}}\sum_{n_{+}}\sup_n\left( \frac{\left( t\xi
\right) ^{\frac{n^{-}}2}\left( t\zeta \right) ^{\frac{n_{+}}%
2}n!c_{p,q}\left( n\right) }{\left( \xi \zeta \right) ^{\frac
n2}n_{+}!n^{-}!\left( n-n_{+}-n^{-}-n_{+}^{-}\right) !}\right) ^2\right)
^{\frac 12} \\
&\leq &\sum_n\sum_{n_{+}}\sum_{n^{-}}\sum_{n_{+}^{-}}\frac{\left( t\zeta
\right) ^{\frac{n_{+}}2}\left( t\xi \right) ^{\frac{n^{-}}2}\left( t\xi
t\zeta \right) ^{\frac{n_{+}^{-}}2}n!}{\left( \xi \zeta \right) ^{\frac
n2}n_{+}!n^{-}!n_{+}^{-}!\left( n-n_{+}-n^{-}-n_{+}^{-}\right) !}%
c_{p,q}\left( n\right) ,
\end{eqnarray*}
where the supremum and summation is taken over $n\geq n_{+}+n^{-}+n_{+}^{-}$%
. Then the function $B$ is integrable up to a $t<\rho $ as it has the finite
estimate 
\begin{equation}
\left\| B\right\| _{p,q}^{^{\prime }}\left( \xi ,\zeta \right) \leq
\sum_{n=0}^\infty \left( \frac{\left( 1+\sqrt{t\xi }\right) \left( 1+\sqrt{%
t\zeta }\right) }{\sqrt{\xi \zeta }}\right) ^nc_{p,q}\left( n\right) \leq
\sum_{n=0}^\infty \rho ^nc_{p,q}\left( n\right) ,  \label{A.4}
\end{equation}
and so $\left\| Y_t\right\| _{p,q}^{\prime }\left( 3\xi ,3\zeta \right) \leq
\sum_{n=0}^\infty \rho ^nc_{p,q}\left( n\right) $ if $\sqrt{\xi \zeta }%
>1/\rho $ and 
\begin{equation*}
\sqrt{\xi \zeta t}<\left( \xi \zeta \rho +\frac 14\left( \sqrt{\xi }-\sqrt{%
\zeta }\right) ^2\right) ^{\frac 12}-\frac 12\left( \sqrt{\xi }+\sqrt{\zeta }%
\right) \text{.}
\end{equation*}
In particular, the integral $Y_t$ is defined as a continuous operator $%
\mathfrak{D}\rightarrow \mathfrak{H}$ into the Hilbert space $\mathfrak{H}$
if this analytical estimate is valid for $p=0$, $\xi =1/3$ and some $q,\zeta 
$, and it is a strongly continuous operator, $Y_t\in \mathcal{L}\left( 
\mathfrak{D}\right) $, if it is also valid for any $p<0,\xi <1/3$.

3. Let us apply this estimate to the multiple integral of the chronological
products 
\begin{equation}
B\left( \boldsymbol{\tau }\right) =L\left( \boldsymbol{t}_n\right) \cdots
L\left( \boldsymbol{t}_1\right) =L\left( \left| \boldsymbol{\tau }\right|
\right) ,  \label{A.5}
\end{equation}
defined by the unique decomposition $\boldsymbol{\tau }=\boldsymbol{t}_1\cup
\cdots \cup \boldsymbol{t}_n$ of the set table $\boldsymbol{\tau }$ into the
single point tables 
\begin{equation*}
\boldsymbol{t}_{+}^{-}=\left( 
\begin{tabular}{ll}
$t$ & $\varnothing $ \\ 
$\varnothing $ & $\varnothing $%
\end{tabular}
\right) ,\boldsymbol{t}_{\bullet }^{-}=\left( 
\begin{tabular}{ll}
$\varnothing $ & $t$ \\ 
$\varnothing $ & $\varnothing $%
\end{tabular}
\right) ,\boldsymbol{t}_{+}^{\bullet }=\left( 
\begin{tabular}{ll}
$\varnothing $ & $\varnothing $ \\ 
$t$ & $\varnothing $%
\end{tabular}
\right) ,\boldsymbol{t}_{\bullet }^{\bullet }=\left( 
\begin{tabular}{ll}
$\varnothing $ & $\varnothing $ \\ 
$\varnothing $ & $t$%
\end{tabular}
\right) ,
\end{equation*}
with $L\left( \boldsymbol{t}_\nu ^\mu \right) =L_\nu ^\mu \in \mathcal{L}%
\left( \mathcal{D}\right) $ and $t_1<\ldots <t_n$, $n=\left| \cup \tau _\nu
^\mu \right| $. If the norms (\ref{A.2}) of the chronological product (\ref%
{A.5}) with 
\begin{equation}
\left\| B\left( 
\begin{tabular}{ll}
$\tau _{+}^{-}$ & $\tau ^{-}$ \\ 
$\tau _{+}$ & $\tau $%
\end{tabular}
\right) \right\| _{p,q}^{^{\prime }}=\sup_{\eta _{n,n_{+}}\in \mathcal{H}%
_p\otimes \mathcal{K}^{\otimes n_{+}},\eta _{n,n^{-}}\in \mathcal{H}%
_q\otimes \mathcal{K}^{\otimes n^{-}}}\frac{\left| \left\langle \eta
_{n,n_{+}}|B\left( \boldsymbol{\tau }\right) \eta _{n,n^{-}}\right\rangle
\right| }{\left\| \eta _{n,n_{+}}\right\| _p\left\| \eta _{n,n^{-}}\right\|
_q}  \label{A.7}
\end{equation}
are finite for some $\xi \zeta \geq 1$ and a $t=T$, the multiple integral $%
V_t=\Lambda _{[0,t)}^{\otimes }\left( B\right) $ satisfies the quantum
linear differential equation $\mathrm{d}V_t=L_\nu ^\mu V_t\mathrm{d}\Lambda
_\mu ^\nu $, $t\leq T$ with $V_0=I$, see Theorem 1 in \cite{20}. Thus the
estimate (\ref{A.3}) for the chronological products (\ref{A.5}) with 
\begin{equation*}
L_{+}^{-}=-K,\quad L_{\bullet }^{-}=-K^{-},\quad L_{+}^{\bullet }=L,\quad
L_{\bullet }^{\bullet }=J-I
\end{equation*}
gives a sufficient condition for the existence of the unique solution to Eq.
(\ref{1.3}), (\ref{1.1}) of the type (\ref{1.6}) in the form of the
stochastic chronologically ordered operator-valued exponents $V_t$.

4. A similar estimate 
\begin{equation}
\left\| L\left( \boldsymbol{t}_n\right) \cdots L\left( \boldsymbol{t}%
_1\right) \right\| _{p,q}^{\prime }\leq \frac{n!}{\sqrt{n^{-}!}}%
c_{p,q}\left( n\right)  \label{A.6}
\end{equation}
for the chronological products $B\left( \tau _{+}^{-},\tau ^{-}\right) $ of $%
L\left( \boldsymbol{t}_{+}^{-}\right) =-K$ and $L\left( \boldsymbol{t}%
^{-}\right) =-K_{\bullet }$ gives the sufficient condition 
\begin{eqnarray}
\left\| B\right\| _{p,q}^{\prime }\left( \xi ,\zeta \right) &=&\left(
\int_{\Gamma _t}\left( \frac 1\zeta \right) ^{\left| \tau ^{-}\right|
}\left( \int_{\Gamma _t}\left\| B\left( \tau _{+}^{-},\tau ^{-}\right)
\right\| _{p,q}^{\prime }\mathrm{d}\tau _{+}^{-}\right) ^2\mathrm{d}\tau
^{-}\right) ^{\frac 12}  \label{A.9} \\
&\leq &\sum_n\left( \frac 1{\sqrt{\zeta }}+\sqrt{t}\right) ^nc_{p,q}\left(
n\right) <\infty ,\quad \forall \xi ^{-1}\leq \zeta  \notag
\end{eqnarray}
of the integrability $V_t^{\circ }=\int_{\Gamma _t}\int_{\Gamma _t}B\left(
\tau _{+}^{-},\tau ^{-}\right) \mathrm{d}\tau _{+}^{-}\Lambda ^{-}\left( 
\mathrm{d}\tau ^{-}\right) $ for the quantum stochastic equation (\ref{3.2'}%
) with $t\leq \left( \rho -1/\sqrt{\zeta }\right) ^2$. Thus the iteration $%
S_t=T_tV_t^{\circ }$ of the nonadapted integral equation \ref{3.1'} has the
estimates $\left\| S_t\right\| _{p,q}\left( \xi ,\zeta \right) \leq
\sum_{n=0}^\infty \rho ^nc_{p,q}\left( n\right) $ for all $2^{-1}\zeta \geq
\max \left\{ \rho ^{-2},2\xi ^{-1}\right\} $, and $S_t\in \mathcal{L}\left( 
\mathfrak{D}\right) $ if the chronological products $B$ satisfy the
analyticity condition (\ref{A.6}) for each $p<0$ and some $q>0$.

5. In order to formulate an analyticity condition for the weak convergence
of (\ref{3.6'}) in terms of the structural maps $\lambda _\nu ^\mu $, let us
represent this multiple integral in the equivalent form (see Theorem 2 in 
\cite{20}) as the adapted one $R_t=\Lambda _{[0,t)}^{\otimes }\left(
L\right) $, for the integrant 
\begin{equation}
L\left( \boldsymbol{\tau }\right) =\lambda \left( \boldsymbol{\tau }%
,I\right) ,\quad \boldsymbol{\tau }=\cup _{i=1}^n\boldsymbol{t}_i,
\label{A.10}
\end{equation}
giving the solution to the equivalent equation (\ref{2.2}) for $B=I$. The
integrant $L$ for such a representation is given by the chronological
composition 
\begin{equation*}
\lambda \left( \boldsymbol{\tau }\right) =\lambda \left( \boldsymbol{t}%
_1\right) \circ \ldots \circ \lambda \left( \boldsymbol{t}_n\right)
\end{equation*}
of $\lambda \left( \boldsymbol{t}_\nu ^\mu ,\cdot \right) =\lambda _\nu ^\mu
\left( \cdot \right) $, see \cite{20}. If this integrant $B=L$ has the
estimate (\ref{A.3}) for some $p,q>0$ and $\xi ,\zeta >1$, then the series (%
\ref{3.6'}) converges to the continuous sesquilinear form $R_t\in \mathcal{B}%
\left( \mathfrak{D}\right) $ with $\left\| R_t\right\| _{p,q}\left( 3\xi
,3\zeta \right) <\infty $. Another analyticity condition, corresponding to a
smaller ($L^\infty $) space of test functions $f^{\bullet }\in L^2$ and a
stronger ($L^2$) integrability, is given in \cite{18}. In the next paper,
which will be published elsewhere, we generalize the sufficient
integrability condition (\ref{A.3}) to the $L^{1+\xi }$ spaces of test
functions, $\xi \geq 1$, and will show that our method gives more precise
estimates in the limit case $\xi \longrightarrow \infty $.

\noindent \textbf{Acknowledgements.}

The author wishes to thank Professor K.\thinspace R.\thinspace Parthasarathy
for drawing his attention to the problem of studying filtering dynamics from
a general ``completely positive'' point of view, Dr.\thinspace J.\thinspace
M.\thinspace Lindsay for stimulating discussions on the subject, and
Professor R.\thinspace L.\thinspace Hudson for encouraging the author to
write this paper.

\end{document}